\documentclass[nofootinbib, prx, twocolumn, preprintnumbers, superscriptaddress]{revtex4-2}

\usepackage[acronyms, nohypertypes={acronym}, nopostdot, style=super, nonumberlist, toc]{glossaries}
\glsenableentrycount
\setacronymstyle{long-sc-short}

\newcommand\ac[1]{\gls{#1}}

\def\acresetall{\glsresetall[acronym]\glsunset{VUMPS}}

\newacronym{QCD}{qcd}{quantum chromodynamics}
\newacronym{IR}{ir}{infrared}
\newacronym{GW}{gw}{Ginsparg-Wilson}
\newacronym{SMG}{smg}{symmetric mass generation}

\newacronym{VUMPS}{vumps}{variational uniform matrix product state}
\newacronym{MPS}{mps}{matrix product states}

\usepackage[T1]{fontenc}
\usepackage{lmodern}
\usepackage[scaled=0.95]{zi4} %
\usepackage{mathtools} %
\usepackage{amsmath,amssymb}
\usepackage[bb=dsserif]{mathalpha}
\usepackage{microtype}
\usepackage{xcolor}
\usepackage{ninecolors}
\usepackage{graphicx}
\graphicspath{{./fig/}} %
\usepackage{physics2}
\usephysicsmodule{ab,ab.braket}
\usepackage{orcidlink}
\usepackage{hyperref}
\hypersetup{
  unicode,
  pdftitle={Simulating a chiral gauge theory},
  pdfauthor={Laurens Lootens, Hersh Singh},
  pdfprintscaling=None,
  breaklinks,
  colorlinks=true,
  linkcolor=azure5,
  citecolor=azure5,
  filecolor=magenta,
  urlcolor=blue5,
}
\usepackage[capitalise]{cleveref}
\crefname{section}{Sec.}{Secs.}
\crefname{appendix}{App.}{Apps.}
\crefname{subappendix}{App.}{Apps.}

\newcommand\NumC{1.002(1)}            %
\newcommand\NumDeltaOne{0.0270(3)}    %
\newcommand\NumDeltaUN{0.97}          %
\newcommand\NumDeltaW{0.262(2)}       %
\newcommand\NumDeltaOneBest{0.0269(2)}  %
\newcommand\NumDeltaUNBest{0.967(7)}    %
\newcommand\NumDeltaWBest{0.259(3)}     %
\newcommand\NumProduct{0.250(3)}        %
\newcommand\NumRRatio{1.017(3)}         %
\newcommand\NumChiMin{16}             %
\newcommand\NumChiMax{256}            %
\newcommand\NumChiTHMax{256}          %
\newcommand\NumTHSpread{0.08\%}       %

\newcommand\code[1]{\texttt{#1}} %
\newcommand\minisec[1]{{\emph{{#1}}}---}
\renewcommand\>{\rangle}
\newcommand\<{\langle}
\newcommand\del\partial
\newcommand\hc{\mathrm{h.c.}}
\DeclareMathOperator{\sgn}{sgn}
\newcommand\mc{\mathcal}

\newcommand\ZZ{\mathbb{Z}}
\newcommand\Zn{\ZZ_{N}}
\newcommand\Znn{\ZZ_{N^2}}
\newcommand\Znsq{\ZZ_{N^2}}
\newcommand\ZnW{\ZZ_{N}^{(w)}}

\newcommand\TFFZ{34-50}

\newcommand\Nf{N_{F}}
\newcommand\anom{{\cal{A}}}
\newcommand\Kinv{K^{-1}}
\newcommand\Vmat{\mathcal{V}}

\newcommand\qL[1]{q_{L,#1}}
\newcommand\qR[1]{q_{R,#1}}
\newcommand\psiL[1]{\psi_{L, #1}}
\newcommand\psiR[1]{\psi_{R, #1}}

\newcommand\phiboson{\varphi}
\newcommand\phidual{\tilde{\varphi}}
\newcommand\Vdual{\tilde{V}}
\newcommand\DeltaM{\Delta}
\newcommand\DeltaW{\tilde\Delta}

\newcommand\Hgauge{H_{\text{gauge}}}
\newcommand\Hfermions{H_{\text{matter}}}
\newcommand\textkin{\text{(kin)}}
\newcommand\texthop{\text{(hop)}}
\newcommand\hkin{h^{\textkin}}
\newcommand\hhop{h^{\texthop}}
\newcommand\helec{h^{\text{(elec)}}}

\newcommand\ClockOp{Z}
\newcommand\Qchi{\mathcal{Q}}

\NewDocumentCommand{\xm}{som}{%
  \IfBooleanTF{#1}{\hat{x}}{x}\IfNoValueTF{#2}{_{#3}}{_{#2,#3}}%
}
\NewDocumentCommand{\zm}{som}{%
  \IfBooleanTF{#1}{\hat{z}}{z}\IfNoValueTF{#2}{_{#3}}{_{#2,#3}}%
}
\NewDocumentCommand{\Xm}{som}{%
  \IfBooleanTF{#1}{\hat{X}}{X}\IfNoValueTF{#2}{_{#3}}{_{#2,#3}}%
}
\NewDocumentCommand{\Zm}{som}{%
  \IfBooleanTF{#1}{\hat{\ClockOp}}{\ClockOp}\IfNoValueTF{#2}{_{#3}}{_{#2,#3}}%
}
\NewDocumentCommand{\Xl}{som}{%
  \IfBooleanTF{#1}{\hat{\widetilde{X}}}{\widetilde{X}}\IfNoValueTF{#2}{_{#3}}{_{#2,#3}}%
}
\NewDocumentCommand{\Zl}{som}{%
  \IfBooleanTF{#1}{\hat{\widetilde{\ClockOp}}}{\widetilde{\ClockOp}}\IfNoValueTF{#2}{_{#3}}{_{#2,#3}}%
}
\NewDocumentCommand{\Ql}{om}{%
  \IfNoValueTF{#1}{\widetilde{Q}_{#2}}{\widetilde{Q}_{#1,#2}}%
}
\NewDocumentCommand{\Qm}{om}{%
  \IfNoValueTF{#1}{Q_{#2}}{Q_{#1,#2}}%
}
\NewCommandCopy{\Qw}{\Ql}
\NewDocumentCommand{\Qli}{om}{%
  \IfNoValueTF{#1}{\widetilde{q}_{#2}}{\widetilde{q}_{#1,#2}}%
}
\NewDocumentCommand{\Qmi}{om}{%
  \IfNoValueTF{#1}{q_{#2}}{q_{#1,#2}}%
}
\NewDocumentCommand{\nw}{o}{%
  \IfNoValueTF{#1}{\tilde{n}}{\tilde{n}_{#1}}%
}
\NewDocumentCommand{\nm}{o}{%
  \IfNoValueTF{#1}{n}{n_{#1}}%
}

\newcommand{\Vect}{\mathsf{Vec}}
\newcommand{\Rep}{\mathsf{Rep}}
\newcommand{\Mod}{\mathsf{Mod}}
\DeclareMathOperator{\End}{\mathsf{End}}

\begin{document}

\title{Simulating a chiral gauge theory}

\author{Laurens Lootens}
\email{ll708@cam.ac.uk}
\affiliation{Department of Applied Mathematics and Theoretical Physics, University of Cambridge, Cambridge CB3 0WA, UK}
\author{Hersh Singh\,\orcidlink{0000-0002-2002-6959}}
\email{hersh.singh@perimeterinstitute.ca}
\affiliation{Perimeter Institute for Theoretical Physics, Waterloo, ON N2L 2Y5, Canada}
\affiliation{Institute for Quantum Computing, University of Waterloo, Waterloo, ON N2L 3G1, Canada}

\begin{abstract}
  Chiral gauge theories underlie the Standard Model, yet none has been simulated nonperturbatively.
  Lattice constructions that evade the Nielsen--Ninomiya obstruction entail complex fermion measures or infinite-dimensional local Hilbert spaces. %
  Here we provide a gauge-invariant Hamiltonian lattice regularization of 1+1-dimensional abelian chiral gauge theories built from clock-and-shift qudits with $\ZZ_N$ gauge fields, in which the global chiral symmetry and its 't Hooft anomaly survive exactly mod $N$.
  Anomaly cancellation on the lattice lets the Gauss laws be imposed and solved exactly, reducing the theory to an unconstrained qudit chain.
  Using infinite matrix product states, we simulate the canonical 1+1-dimensional chiral gauge theory, the \TFFZ{} model, and find excellent agreement with analytic predictions, showing that a finite-dimensional local Hilbert space at $N=6$ reproduces the expected infrared physics of the continuum \TFFZ{} model.
  This provides an efficient and practical solution to the problem of simulating general abelian chiral gauge theories in 1+1 dimensions, enabling quantum simulations.
\end{abstract}

\maketitle
\acresetall

\section{Introduction}
\label{sec:intro}

Despite the enormous success of the Standard Model of particle physics, there is a fundamental problem: we do not know how to simulate it.
Lattice regularization remains the only known nonperturbative way to define \ac{QCD}
\cite{wilson_confinement_1974,kogut_hamiltonian_1975}.
Extending this to the full Standard Model is, however, obstructed by its nature as a \emph{chiral} gauge theory: the Nielsen--Ninomiya theorem \cite{nielsen_absence_1981,nielsen_absence_1981a,karsten_lattice_1981} shows that, under mild assumptions, lattice fermions come in vector-like pairs, so we cannot directly define and simulate chiral gauge theories.

The most successful approach to putting chiral fermions on the lattice has relied on the \ac{GW} relation \cite{ginsparg_remnant_1982} 
and its solutions \cite{neuberger_more_1998,neuberger_exactly_1998,hasenfratz_prospects_1998,kaplan_chiral_1993,kaplan_method_1992},
culminating in the realization that the \ac{GW} relation enforces an exact lattice chiral symmetry \cite{luscher_exact_1998} and enables a nonperturbative lattice definition of \emph{abelian} chiral gauge theories \cite{luscher_abelian_1999,luscher_chiral_2002a}. 
However, efforts to make L\"uscher's construction explicit
\cite{kadoh_another_2005,kadoh_numerical_2005,kadoh_simple_2008,kadoh_solving_2004} have not resulted in a numerical simulation.
Indeed, despite half a century of progress in lattice gauge theory, and a quarter century after L\"uscher's construction of abelian chiral gauge theories, we lack a \emph{single} lattice simulation of a chiral gauge theory with exact gauge invariance.%
\footnote{
  Kikukawa and collaborators have extended L\"uscher's construction to $SU(2) \times U(1)$ \cite{kadoh_construction_2008,kikukawa_gauge_2001,kikukawa_lattice_2002}.
  Recently, Craig has presented an \textsc{ai}-assisted proof extending it to nonabelian gauge groups, which include that of the Standard Model \cite{craig_gaugeinvariant_2026}.
  If validated, this would be a breakthrough in providing a nonperturbative definition of the Standard Model.
  Regardless, a \emph{practical} construction, amenable to numerical simulations, remains elusive. 
}%

The goal of this paper is to close that gap.
We propose a method that works for general 1+1-dimensional abelian chiral gauge theories.
We demonstrate the effectiveness of our proposal by performing the first nonperturbative simulation of the prototypical chiral gauge theory, the gauged \TFFZ{} model, in the thermodynamic limit using tensor networks and determining its \ac{IR} physics.

Advances in understanding bosonic anomalies on the lattice have led to lattice formulations of 1+1-dimensional abelian chiral gauge theories
\cite{berkowitz_exact_2023,seifnashri_exactly_2026,thorngren_chiral_2026}.
These first bosonize the continuum fermions and then put the bosonized theory on the lattice, such that the 't~Hooft anomalies are reproduced, allowing an exact gauging of the anomaly-free abelian chiral symmetry.
While well-defined, these do not immediately yield a numerical simulation, because chiral gauge theories so defined generically run into a sign problem with Monte Carlo methods.
On the other hand, tensor-network methods do not suffer from a sign problem but 
face difficulties when dealing with the infinite-dimensional local Hilbert spaces that the bosonic descriptions demand.
Indeed, finite-dimensional local Hilbert spaces are known to obstruct exact realizations of the chiral anomaly \cite{kapustin_anomalous_2025,fidkowski_nogo_2023,gioia_disentangling_2026}, so the constructions above use infinite-dimensional rotor degrees of freedom.

\begin{figure*}[t]
  \centering
  \includegraphics[width=\textwidth]{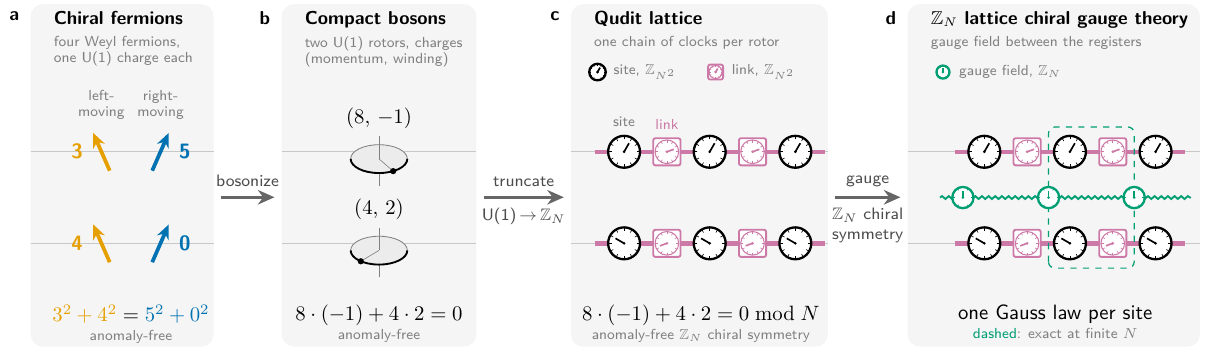}
  \caption{%
    The \TFFZ{} chiral gauge theory, from continuum Weyl fermions to a $\ZZ_N$ lattice gauge theory.
    Each row is one Dirac pair, followed across the four panels.
    \textbf{a}, Four Weyl fermions, a left-mover and a right-mover per row with $U(1)$ charges $(3,5)$ and $(4,0)$.
    \textbf{b}, Bosonized, each pair becomes one compact boson, drawn as a rotor, carrying momentum and winding charges $(8,-1)$ and $(4,2)$. %
    \textbf{c}, Put the compact bosons on a spatial lattice and truncate the local Hilbert space. Each rotor becomes one lattice chain of $\ZZ_{N^2}$ clocks, the momentum charge on the sites and the winding charge on the links.
    The chiral $U(1)$ becomes a $\ZZ_N$ symmetry, anomaly-free because $8\cdot(-1)+4\cdot 2 = 0 \bmod N$.
    \textbf{d}, Gauging the $\ZZ_N$ chiral symmetry adds a $\ZZ_N$ clock gauge field between each link register and the next site register.
    One Gauss law (dashed) ties the momentum charges on a site, the winding charges on the link to its right, and the two gauge clocks bounding it.
  }
  \label{fig:schematic}
\end{figure*}

To circumvent this, we present a formulation with finite $N^2$-dimensional local Hilbert spaces, which keeps only a $\ZZ_N$ subgroup of the chiral symmetry and its anomaly on a qudit chain.
This avoids the obstructions above, and is enough to gauge the $\ZZ_N$ subgroup exactly.
It also allows the use of tensor-network methods to perform numerical simulations of 1+1-dimensional abelian chiral gauge theories, which we find to be remarkably efficient.

In particular, our model consists of $\Zn$ gauge fields on the sites and $\Znn$ clock/shift degrees of freedom on sites and links.
\Cref{fig:schematic} shows how our model is related to the traditional fermionic description of a chiral gauge theory.
Our model reproduces the construction of Ref.~\cite{seifnashri_exactly_2026} in the $N\to \infty$ limit.
But, in fact, we find that the qudit model already reproduces the \ac{IR} physics of the continuum theory at $N = 6$ for the \TFFZ{} model, showing that infinite-dimensional local Hilbert spaces are not a necessary ingredient in a practical gauge-invariant lattice construction of chiral gauge theories.

The rest of this paper is organized as follows.
In \cref{sec:chiral}, we review the continuum theory and construct the qudit lattice model, which we then reduce to an unconstrained chain for the \TFFZ{} model. (An alternative category-theoretic construction is presented in \cref{sec:abstract-nonsense}.)
In \cref{sec:3450-numerics}, we validate the construction on the $\Zn$ Schwinger model and present our simulation of the \TFFZ{} model.
We conclude with an outlook in \cref{sec:conclusion}.

\section{The qudit chiral gauge theory}%
\label{sec:chiral}

\newcommand\qmv{\textsc{q}m\textsc{v}}

\subsection{Continuum: $U(1)$ chiral gauge theory and bosonization}

A 1+1-dimensional continuum $U(1)$ chiral gauge theory can be constructed with $\Nf$ left-handed Weyl fermions $\psiL{a}$
and $\Nf$ right-handed Weyl fermions $\psiR{a}$ for $a=1,\dotsc, \Nf$. 
Consider a $U(1)$ symmetry under which the left-handed fermions transform with charges $\qL{a}$ and the right-handed ones transform with charges $\qR{a}$.
\newcommand\Uchi{U(1)_\chi}
\begin{align}
	\Uchi:\quad \psiL{a} \mapsto e^{i \theta \qL{a} } \psiL{a}\, \quad
	\psiR{a} \mapsto e^{i \theta \qR{a} } \psiR{a}.
\end{align}
Generically, this symmetry has a 't Hooft anomaly which prevents its gauging.
However, when
\begin{align}
	\sum_{a=1}^{\Nf} \qL{a}^2 = \sum_{a=1}^{\Nf} \qR{a}^2  
  \label{eq:anomaly-cancellation}
\end{align}
the anomaly cancels and the symmetry can be gauged.
The theory is described by the continuum Hamiltonian
\begin{align}
	H = \int dx \Big[ \sum_{a=1}^{\Nf} \sum_{s = L,R} \chi_s\, \psi_{s,a}^{\dagger}\, i D_1 \psi_{s,a}
	  + \frac{e^2}{2} E^2 \Big] ,
  \label{eq:H-continuum}
\end{align}
where $\chi_L = -\chi_R = 1$, $D_1 = \del_1 - i q_{s,a} A_1$, and $[E(x), A_1(y)] = i \delta(x-y)$, subject to the Gauss law
\begin{align}
	\del_1 E = \sum_{a=1}^{\Nf} \sum_{s = L,R} q_{s,a}\, \psi_{s,a}^{\dagger} \psi_{s,a} .
\end{align}
A theory where all charges appear in pairs  $\qL{a} = \pm \qR{a}$ (up to relabelling of flavors) is said to be a vector-like gauge theory.
Otherwise, this is a chiral gauge theory.
The simplest nontrivial chiral gauge theory has two flavors, with the smallest charges $(\qL{},\qR{}) = (3,4 | 5,0)$, the so-called 34-50 model.
This theory has a long history of being the target of proposals for lattice chiral gauge theories \cite{bhattacharya_chiral_2006,giedt_chiral_2007,suzuki_perturbative_2007,chen_decoupling_2013,wang_nonperturbative_2023,wang_solution_2019,zeng_symmetric_2022,golterman_constraints_2025,morikawa_another_2024,morikawa_novel_2025}.

In 1+1 dimensions, bosonization provides a dictionary to map the continuum fermions to bosons \cite{coleman_quantum_1975,haldane_luttinger_1981,mandelstam_soliton_1975}, which has been used to study 1+1-dimensional chiral gauge theories both in the continuum \cite{halliday_quantization_1986,mouland_phases_2025}
and to motivate lattice formulations \cite{wang_solution_2019,wang_nonperturbative_2023,wang_symmetric_2022a}.
Our construction starts from this bosonized description.
Each flavor $(\psiL{a}, \psiR{a})$ bosonizes to a compact boson $\phi_a$ with dual $\tilde\phi_a$, so the $2\Nf$ Weyl fermions become $\Nf$ compact bosons.
Each boson carries a momentum symmetry $U(1)_m^{(a)}$, which shifts $\phi_a$, and a winding symmetry $U(1)_w^{(a)}$, which shifts $\tilde\phi_a$, with a mixed 't Hooft anomaly between the two.
This is the bosonized analog of the vector $U(1)$ and axial $U(1)$ symmetries of the fermionic theory.
The symmetry $\Uchi$ acts as a combination of these shifts,
\begin{align}
	\Uchi:\quad \phi_{a} \mapsto \phi_{a} + \theta\, \nm[a], \quad
	\tilde{\phi}_a \mapsto \tilde\phi_{a} + \theta\, \nw[a],
  \label{eq:Uchi-boson}
\end{align}
with momentum and winding charges fixed by the chiral charges (\cref{sec:bosonization}),
\begin{align}
	\nm[a] = \qL{a} + \qR{a}, \qquad \nw[a] = \tfrac{1}{2} \ab(\qL{a} - \qR{a}).
  \label{eq:qm-qw}
\end{align}
These are integers for all models in this work, where $\qL{a} = \qR{a} \bmod 2$ for each flavor.
Since $\Uchi$ acts only through shifts of $\phi_a$ and $\tilde\phi_a$, a $\Zn$ subgroup of it can be kept exact on the lattice by replacing each compact boson with a qudit chain, as we now describe.

\subsection{Lattice: Qudit $\Zn$ chiral gauge theory}

Our construction of a 1+1-dimensional lattice chiral gauge theory starts from $\Nf$ copies of a $\Znsq$ qudit clock model, which truncate the $U(1)$ rotors while keeping an exact chiral $\Zn$ symmetry and mod $N$ anomalies corresponding to the continuum 't Hooft anomalies.
With the right assignment of charges, anomaly cancellation is exact on the lattice, and the $\Zn$ chiral symmetry can be gauged.
\Cref{fig:schematic} shows how our model is related to the traditional fermionic description of a chiral gauge theory, via a continuum bosonization followed by a qudit truncation.
Here, we outline a construction parallel to Refs.~\cite{thorngren_chiral_2026,seifnashri_exactly_2026}, with details in \cref{sec:qudit-duality}.
We also present an alternative derivation in \cref{sec:abstract-nonsense}, using recently developed technology on gauging non-anomalous non-onsite symmetries. Besides verifying the more traditional computation, it offers a systematic construction for any 1+1-dimensional abelian gauge theory directly in a reduced (unconstrained) frame.

We take $\Nf$ copies of a $\Znsq$ clock-shift model labeled by $a = 1, \dotsc, \Nf$.
This is a $\Znsq$ analog of the modified Villain formulation of a compact boson \cite{sulejmanpasic_abelian_2019,cheng_liebschultzmattis_2023,fazza_lattice_2023,gorantla_modified_2021a}, and can be thought of as a qudit modified-Villain model.
For each flavor $a$, we have the clock-shift fields $(\Zm[a]{j}, \Xm[a]{j})$ on
sites $j$, and $(\Zl[a]{j + \frac{1}{2}},\Xl[a]{j + \frac{1}{2}})$ on the links 
$j+\frac12$.
Our convention is that the tuple $(Z,X)$ satisfies the algebra 
$	Z X {Z}^{\dagger} = \omega X $,
where $\omega = \exp(i 2 \pi / N^2 )$ is an $N^2$-root of unity.
The matter Hamiltonian is given by
\begin{align}
	\hkin_{a,j} &= \Zm[a]{j}  + \hc \\
  \hhop_{a,j} &= \Xm[a]{j}^{\dagger} \Zl[a]{j + \frac{1}{2}} \Xm[a]{j+1} + \hc
\end{align}
The local charges of the momentum and winding symmetries are
\begin{align}
	\Qm[a]{j} = \Xl[a]{j - \frac{1}{2}} \Zm[a]{j} \Xl[a]{j + \frac{1}{2}}^\dagger, \quad
  \Ql[a]{j + \frac{1}{2}} = \Zl[a]{j + \frac{1}{2}}
\end{align}
Before gauging the chiral symmetry, the local momentum and winding charges satisfy the constraints $\Qm{j}^N = 1$ and $\Ql{j+\frac12}^N = 1$, which imply that the global charges 
$Q = \prod_j Q_j$ and $\Ql{} = \prod_{j} \Ql{j + \frac{1}{2}}$
generate a $\Zn$ symmetry each.

The key feature of this formulation is that the mixed anomaly between the continuum $U(1)$ momentum and winding symmetries of the compact boson is preserved as an exact mod $N$ anomaly on the lattice, as seen in the anomalous algebra of the local charges
\begin{align}
	\Qm[a]{j} \Ql[a]{j + \frac{1}{2}} \Qm[a]{j}^{-1} &= \omega \Ql[a]{j + \frac{1}{2}}.
\end{align}
To get a chiral gauge theory, we need to gauge the charge 
\begin{align}
\Qchi_{j + \frac{1}{2}} &= \prod_{a}^{\Nf} \Qm[a]{j}^{\nm[a]} \Ql[a]{j + \frac{1}{2}}^{\nw[a]},
\end{align}
where $(\nm[a], \nw[a])$ are related to the Weyl fermion charges $(\qL{a}, \qR{a})$ as in \cref{eq:qm-qw}.
For these local charges to commute with each other, we need $[\Qchi_{j-\frac12}, \Qchi_{j+\frac12}] = 0$.
  Since $\Qchi_{j-\frac12} \Qchi_{j+\frac12} = \omega^{\sum_a \nm[a] \nw[a]} \Qchi_{j+\frac12} \Qchi_{j-\frac12}$, we only need mod $N$ anomaly cancellation,
\begin{align}
	\sum_{a=1}^{\Nf} \nm[a] \nw[a] = 0 \mod N,
  \label{eq:anomaly-cancellation-N2}
\end{align}
because the exponents $(\nm[a], \nw[a])$ are only defined mod $N$ before gauging, and can be shifted to make the sum vanish mod $N^2$.
For the models in this work, the sum vanishes exactly, which is the continuum condition \cref{eq:anomaly-cancellation}, since $\nm[a] \nw[a] = \frac{1}{2} (\qL{a}^2 - \qR{a}^2)$.
Now, we can gauge this symmetry by introducing $\Zn$ gauge fields $(U_j, V_j)$ on site $j$.%
\footnote{Note that for any finite $N$ this defines a larger family of chiral gauge theory models, in which the anomaly only vanishes for a particular value of $N$. In this work, we only focus on models which satisfy the above condition for any $N$. It is an interesting question whether the chiral gauge theory models where the anomaly only vanishes mod $N$ have a continuum limit at a fixed $N$. These are essentially chiral gauge theory analogues of quantum link models \cite{chandrasekharan_quantum_1997a,brower_qcd_1999,brower_dtheory_2004,Wiese:2021djl,Zache:2021ggw} and qubit regularizations \cite{Singh:2019uwd,Bhattacharya:2020gpm,Liu:2021tef,Caspar:2022llo,Nguyen:2022aaq,Liu:2023lsr,Chandrasekharan:2025pil}.
}
This gauge charge couples to the physical gauge fields via the Gauss constraint:
\begin{align}
	G_j \equiv \Qchi_{j + \frac{1}{2}} V^{-1}_j {V}_{j+1} = 1,
                        \label{eq:gauss-physical}
\end{align}
where $V_j$ is a $\ZZ_N$ electric field.
The other (Villain) Gauss constraints are
\begin{align}
	K_{j,a} \equiv \Qm[a]{j}^{N} U_j^{-\nw[a]} = 1, \quad 
	\tilde K_{j,a} \equiv \Ql[a]{j + \frac{1}{2}}^{N} U_{j+1}^{-\nm[a]} = 1, \quad 
  \label{eq:gauss-villain}
\end{align}
for $a = 1, \dotsc, \Nf$.
These deformations of the ungauged constraints ($\Qm[a]{j}^N= \Ql[a]{j+\frac12}^N = 1$) are fixed by invariance under the gauge transformation given by \cref{eq:gauss-physical}.
Note that all these constraints commute with each other.
The Villain constraints alongwih the anomaly cancellation condition imply that $\Qchi_{j+\frac12}^{N} = 1$. Therefore, $\Qchi_{j+\frac12}$ generates a $\Zn$ symmetry assuming that $\gcd(\nm[a],N) = 1$ or $\gcd(\nw[a], N) = 1$ for some $a = 1,\dots, \Nf$.
Finally, we can write a gauge-invariant electric field term,
\begin{align}
	\helec_j &= \omega^{\kappa} {V}^{\dagger}_j \prod_{a = 1 }^{\Nf}
             \Xm[a]{j}^{\nw[a]} \Xl[a]{j - \frac{1}{2}}^{\nm[a]} + \text{h.c.}
\end{align}
where $\kappa \in \ZZ_N$ is a discrete $\theta$ angle of the theory.
The full Hamiltonian of a general (abelian) qudit chiral gauge theory is the sum of the matter piece and a gauge field kinetic term, $H = \Hgauge  + \Hfermions$, where
\begin{align}
	\Hgauge &= - \sum_{j=1}^{L}\frac{J_e}{2} \helec_j \\
  \Hfermions &= - \sum_{j=1}^{L}\sum_{a=1}^{\Nf} \ab(\frac{J_p}{2}  \hkin_{a,j} + \frac{J_x}{2}  \hhop_{a,j})
\end{align}

\subsection{\TFFZ{} model}
\label{sec:3450-model}

Now, we can work out the \TFFZ{} model, whose charges $(\qL{}, \qR{}) = (3,4 | 5,0)$ come from a Pythagorean triple.
The momentum and winding charges given by \cref{eq:qm-qw} are then $(\nm[1], \nm[2]) = (8,4)$ and $(\nw[1], \nw[2]) = (-1,2)$, and the Gauss constraints of \cref{eq:gauss-villain,eq:gauss-physical} become
\begin{align}
	\Qm[1]{j}^{N} = U_j^{-1}, \quad \Ql[1]{j + \frac{1}{2}}^{N} &= U_{j+1}^{8} \\
	\Qm[2]{j}^{N} = U_j^{2}, \quad \Ql[2]{j + \frac{1}{2}}^{N} &= U_{j+1}^{4} \\
	\Qm[1]{j}^{8} \Qm[2]{j}^4 \Qw[1]{j + \frac{1}{2}}^{-1} \Qw[2]{j + \frac{1}{2}}^{2} &= V_{j} {V}^{\dagger}_{j+1} 
\end{align}
A qudit T-duality, given in \cref{sec:qudit-duality}, results in a dramatic simplification of this model.
It maps the chiral charge to the winding charge of flavor 1, so the physical Gauss law becomes $\Ql[1]{j+\frac12} = V_j V_{j+1}^{\dagger}$, and the other Gauss laws become trivial.
The gauge field can then be eliminated as in the charge-1 Schwinger model (\cref{sec:zn-schwinger}), and flavor 1 becomes an unconstrained $\Znsq$ register $(\Xm{j}, \Zm{j})$ on each site.

Flavor 2 is left as a $\ZZ_N$ clock chain in the modified Villain form, with the constraints $\Qm[2]{j}^N = \Ql[2]{j+\frac12}^N = 1$.
We eliminate these by gauging its $\ZZ_N$ winding symmetry $\ZnW$, which replaces the site and link registers of flavor 2 by a single unconstrained $\Znsq$ register $(\xm{j}, \zm{j})$ on each site (\cref{sec:3450-reduction}).
In terms of the original fields, this is the anomaly-free global symmetry with momentum-winding charge vector $(-4, 0 | 0, -1)$.
The result is a chain with two $\Znsq$ registers per site and no constraints,
\begin{align}
	H = -\sum_{j} \Big[ &\frac{J_p}{2} \ab( \Zm{j}^{-1} + \Zm{j}^{2} \zm{j}^{-1} )
	+ \frac{J_e}{2}\, \omega^{\kappa} \Xm{j} \notag\\
	&+ \frac{J_x}{2}\, \omega^{8}\, \Xm{j}^{-1} \zm{j}^{4} \ab(\xm{j}^{-1} \xm{j+1})^{2} \Zm{j+1}^{-8} \Xm{j+1} \notag\\
	&+ \frac{J_x}{2}\, \xm{j} \xm{j+1}^{-1} \Zm{j+1}^{4} + \hc \Big] .
	\label{eq:H-3450}
\end{align}
This unconstrained frame is what we use for all numerical simulations.

\begin{figure}[t]
  \centering
  \includegraphics[width=\columnwidth]{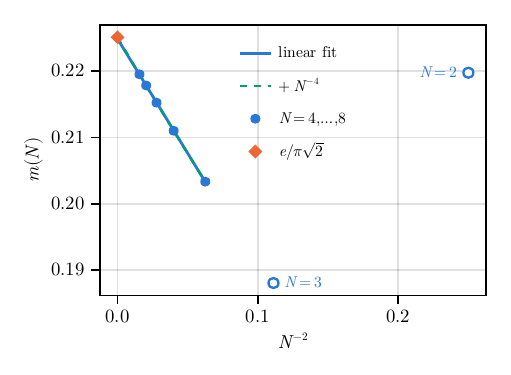}
  \caption{%
    Mass gap $m(N)$ of the $\ZZ_N$ Schwinger model against $N^{-2}$ at $e=1$,
    computed in the thermodynamic limit. Filled circles ($N = 4$--$8$) enter
    the linear fit (solid line) and the fit with an added $N^{-4}$ term
    (dashed); open circles ($N = 2,3$) are excluded. The diamond is the exact
    value $e/\pi\sqrt2$ of \cref{eq:gap}.%
  }
  \label{fig:gap_vs_N}
\end{figure}

\begin{figure*}[t]
  \centering
  \includegraphics[width=0.96\textwidth]{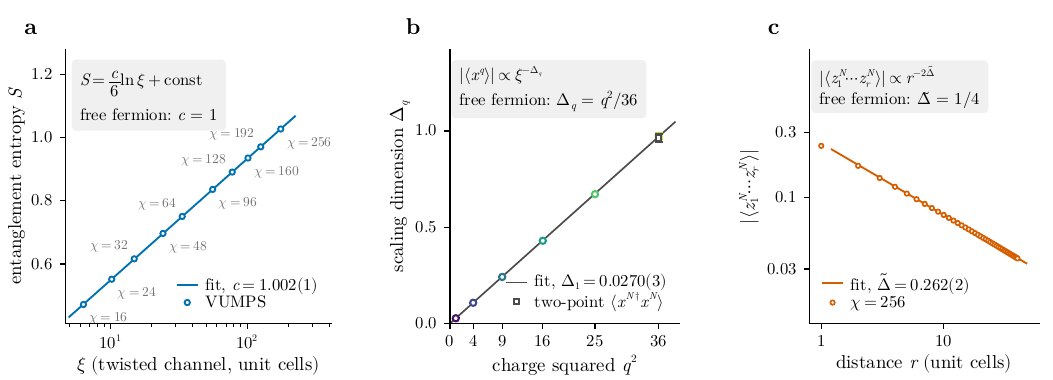}
  \caption{%
    Finite-entanglement scaling establishes that the \ac{IR} of the \TFFZ{} model at $N=6$ is massless Dirac fermion with a weak Thirring interactions, or equivalently, a compact boson slightly off the free-fermion radius.
    Panel (a) determines the central charge, and panels (b) and (c) determine scaling dimensions $\DeltaM, \DeltaW$ of the momentum and winding vertex operators, respectively.
    (\textbf{a}) Entanglement entropy $S$ against the correlation length $\xi$, for bond dimensions $\chi = \NumChiMin$ to $\NumChiMax$ (labels);
    solid line is a linear fit with $c = \NumC$.
    (\textbf{b}) Scaling dimensions $\Delta_q$ of the shift operators $x^q$ from their one-point functions (circles) against $q^2$; line, $\Delta_q = q^2\Delta_1$ with $\Delta_1 = \NumDeltaOne$; square, $\Delta(x^N)$ from the two-point function of $x^N$ at $\chi = \NumChiMax$.
    (\textbf{c}) Two-point function of the winding vertex, which in the reduced chain is the string $\prod_{i=1}^{r} z_i^N$ between the two endpoints, at $\chi = \NumChiMax$; line, fit with $\tilde\Delta = \NumDeltaW$.%
  }
  \label{fig:3450-c1}
\end{figure*}

\section{Simulation of the \TFFZ{} model}
\label{sec:3450-numerics}

To validate our construction, we first perform a simulation for a single-flavor $\ZZ_N$ vector-like gauge theory, which becomes the massless Schwinger model in the $N \to \infty$ limit. This is a translation-invariant chain with a finite number of states per site and no constraints, so it can be simulated directly in the thermodynamic limit using infinite \ac{MPS} methods. In particular, we use \ac{VUMPS} \cite{Zauner-Stauber:2017eqw} to compute the ground state, and the excitation ansatz \cite{haegeman_variational_2012,vanderstraeten_tangentspace_2019} to compute the mass gap.
\Cref{fig:gap_vs_N} shows this mass gap of the $\Zn$ Schwinger model as a function of $N$, which indeed extrapolates to the $m = \frac{e}{\pi\sqrt{2}}$ for the Schwinger boson.
The gap $m(N) = \min_k E_1(k)$ is the minimum of the lowest quasiparticle branch over a grid of $17$ momenta, and it lies at $k = 0$ for every $N$.
A linear fit in $N^{-2}$ to $N = 4$--$8$ already extrapolates to $0.1\%$ below the exact value $e/\pi\sqrt2 = 0.22508$, while adding an $N^{-4}$ term results in an agreement within $5 \times 10^{-5}$.
We exclude $N = 2, 3$ from the fits, since the $1/N$ matching of \cref{eq:couplings} is not yet reliable there.

We now move on to the \TFFZ{} model.
In the $N\to \infty$ limit, our model becomes identical to the one in Refs.~\cite{seifnashri_exactly_2026, lu_fermionic_2026}.
The question for us is: what is the \ac{IR} of the $\Zn$ chiral lattice gauge theory at finite $N$? 

We again use \ac{VUMPS} to obtain the ground state of the Hamiltonian in \cref{eq:H-3450} at $N=6$ and use finite entanglement scaling techniques
\cite{calabrese_entanglement_2009,pollmann_theory_2009,tagliacozzo_scaling_2008}
to extract conformal data at couplings $J_p = 1$, $J_x = 1/4$, $J_e = 1$ and $\kappa = 0$.
At a finite bond dimension, even a critical system develops a finite correlation length $\xi$.
Studying the scaling of correlation functions with respect to the length scale $\xi$ gives a precise method for extracting the critical behaviour such as scaling dimensions.
The correlation length $\xi$ is extracted from the symmetry-twisted transfer matrix: the unit-cell transfer matrix of the ground state with the on-site symmetry $z^N$ inserted on every site, whose leading eigenvalue $\lambda$ gives $\xi = -1/\ln|\lambda|$, the correlation length of the winding vertex.

\Cref{fig:3450-c1} shows the finite-entanglement scaling analysis.
The entanglement entropy grows with the correlation length $\xi$ of the transfer matrix as $S = (c/6)\ln\xi + \mathrm{const}$, with $c = \NumC$, in excellent agreement with the analytic prediction that the \ac{IR} is a free massless Dirac fermion or a compact boson \cite{mouland_phases_2025}.
The linear fit in \cref{fig:3450-c1}(a) works remarkably well even for bond dimensions as small as $\chi = 16$, suggesting that infinite-\ac{MPS} methods are well-suited for such models.

For a compact boson with radius $R$, the momentum and winding vertex scaling dimensions are $\Delta = {1}/{2 R^2}$ and $\tilde\Delta = {R^2}/{2}$, where the radius determines whether it sits at the free-fermion point ( $R_{\text{free}} = \frac{1}{\sqrt{2}}$) or away from it, which corresponds to an interacting Dirac fermion. %
We determine the radius from the next two panels.
As discussed in \cref{sec:zn-schwinger}, the qudit shift operator $x_j$ corresponds to the continuum compact boson as 
$x_j = e^{i \phi_j / N} $ to leading order in $N^{-1}$, with the scaling dimension of $x_j^q$ being $\Delta_{q} = (q^2 / N^2) \Delta$.
Though the $\Znsq$ symmetry generated by $g = \prod_{j} \zm{j} $ is unbroken, at finite bond dimension the one-point function $\braket< x^q >$ scales as $\xi^{- \Delta_{q}}$, which lets us extract $\Delta_q$ for each $q$. \Cref{fig:3450-c1}(b) shows that the fit $\Delta_q = q^2 \Delta_1$ yields a $\Delta_1$ consistent with a direct measurement of the two-point function $\braket< x^N_r x_1^N >$, giving a best-fit value of $\Delta_1 = \NumDeltaOneBest$.
Setting $q = N$, the momentum vertex $x^N$ then has dimension $\DeltaM = N^2 \Delta_1 = \NumDeltaUNBest$.

Since the winding symmetry is gauged completely in the frame of \cref{eq:H-3450}, the two-point function of the gauge-invariant flavor-2 winding vertex $\<\Xl[2]{i+\frac{1}{2}}^{-N} \Xl[2]{j+\frac12}^{N} \>$ becomes a string operator $\< z_{i+1}^N z_{i+2}^N \cdots z_j^N \>$, shown in \cref{fig:3450-c1}(c) as a function of the separation $r = |i-j|$ at $\chi = 256$.
The scaling dimension $\DeltaW$ is obtained by fitting to $r^{-2 \DeltaW}$ over $r = 4$--$40$ at each bond dimension.
A bond-dimension extrapolation using $\chi = 96,\dotsc,256$ then gives
$\DeltaW = \NumDeltaWBest$. 

The product $\DeltaM \DeltaW = \frac{1}{4}$ for a compact boson of any radius, while the ratio $R = (\DeltaW / \DeltaM)^{\frac{1}{4}}$ determines the radius. For the product, we find $\DeltaM \DeltaW = \NumProduct$, further validating the compact boson description.
However, for the radius, we measure $R / R_{\text{free}} = \NumRRatio$.
This answers the question above: at $N = 6$, the \ac{IR} of the $\Zn$ chiral lattice gauge theory is the predicted $c = 1$ theory, slightly off the free-fermion radius: a weakly-interacting massless Dirac fermion, as is expected due to the presence of an exactly marginal Thirring interaction.

Beyond the \ac{IR} spectrum, we can also test a hallmark of chiral gauge theories.
Just as $B+L$ is anomalously violated in the Standard Model, the \TFFZ{} model features an anomalous violation of fermion number in the presence of gauge fields: $\del_\mu j^\mu  = {\frac{\anom}{4\pi}} F^{\mu \nu} \epsilon_{\mu \nu} $, where $j^\mu$ is the vector $U(1)$ current and $\anom = \sum_{a} \qL{a} - \qR{a} = 2$  is the anomaly coefficient. Over a periodic ring, this means that fermion number changes by 2 units when the Chern-Simons number of the background gauge field changes by one,
\begin{align}
	\Delta F = 2 \Delta N_{\text{CS}} .
\end{align}

On the lattice, this anomaly is an exact operator statement.
Fermion number is generated by $Q_F = \prod_j \prod_a \Qm[a]{j} = e^{i\pi F/N}$, which commutes with the Gauss constraints and with every term of the Hamiltonian except the electric term, whose phase it shifts:
\begin{align}
	Q_F\, H(\kappa)\, Q_F^{-1} = H(\kappa + \anom/2) .
	\label{eq:F-kappa}
\end{align}
With $\theta = 2\pi\kappa/N$, this is the anomalous shift $\theta \to \theta + \anom\alpha$ for a rotation by $\alpha = \pi/N$.
The Villain constraints \cref{eq:gauss-villain} further give $e^{i\pi F} = \mathcal{W}^{\anom/2}$ with $\mathcal{W} = \prod_j U_j$, the lattice form of $\Delta F = \anom\, \Delta N_{\text{CS}}$.

\begin{figure}[t]
  \centering
  \includegraphics[width=\columnwidth]{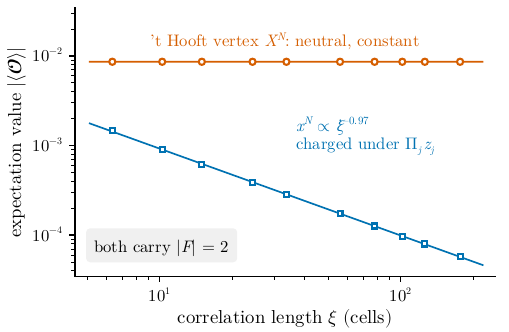}
  \caption{%
    Fermion number is not conserved in the \TFFZ{} model.
    Expectation values at $N=6$ of $X^N$ (circles) and $x^N$ (squares), with $X_j$ and $x_j$ the shift operators of \cref{eq:H-3450}, against the correlation length $\xi$, for bond dimensions $\chi = \NumChiMin$ to $\NumChiTHMax$, with power-law fits (lines).
    Both carry $|F| = 2$, but only $x^N$ is charged under the exact symmetry $\prod_j z_j$.%
  }
  \label{fig:3450-fviolation}
\end{figure}

\Cref{fig:3450-fviolation} shows the dynamical consequence.
The $N$-th power of the electric term $X_j$ of \cref{eq:H-3450} is the 't~Hooft vertex $e^{i\vec q\cdot\vec\phi} \sim \psiL{1}^3 \psiL{2}^4 (\psiR{1}^\dagger)^5$ of \cref{sec:bosonization}, with $F = \anom = 2$.
Its expectation value changes by \NumTHSpread{} over $\chi = \NumChiMin$ to $\NumChiTHMax$, while $x^N$, which also has $|F| = 2$ but is charged under the exact symmetry $\prod_j z_j$, falls as $\xi^{-\Delta}$ with $\Delta = \NumDeltaUN$.
A nonzero $\langle X^N\rangle$ therefore means that fermion number is not conserved.
This is a statement about the gauge dynamics, even though $X_j$ appears in the Hamiltonian: the electric term is the only term that violates $F$, and at $J_e = 0$, $Q_F$ is a symmetry that forbids $\langle X^N\rangle$.

\section{Conclusion and Outlook}%
\label{sec:conclusion}

Chiral gauge theories have long resisted simulation attempts. Even when a nonperturbative definition has been achieved, an efficient numerical approach has not been feasible due to sign problems or infinite-dimensional local Hilbert spaces.
In this work, we have presented a finite-dimensional Hamiltonian lattice regularization of 1+1-dimensional abelian chiral gauge theories.
This construction has yielded, to our knowledge, the first efficient numerical simulation in the thermodynamic limit of the \TFFZ{} model, the prototypical such theory.
At $N=6$, we find that the model flows to the $c=1$ theory predicted analytically, a compact boson slightly off the free-fermion radius.

Chiral gauge theories in 1+1 dimensions, and the \TFFZ{} model in particular, have long been the target of new proposals of lattice chiral gauge theories.
Ref.~\cite{kikukawa_monte_1998} presented a dynamical simulation in another 1+1-dimensional abelian chiral gauge theory, the 2-1111 model, using the overlap approach, even before L\"uscher's formalism. However, the approach could only reach modest volumes due to inherent sign problem with a Monte Carlo approach, and, more importantly, was not manifestly a gauge invariant regularization. Following the L\"uscher gauge-invariant formulation, { Kadoh and Kikukawa} made partial progress in implementing L\"uscher's solution \cite{kadoh_another_2005,kadoh_numerical_2005,kadoh_simple_2008,kadoh_solving_2004}, but this has also not resulted in a satisfactory numerical simulation.
In an alternate strand, Wang and Wen \cite{wang_nonperturbative_2023,wang_solution_2019} presented a thorough study of \ac{SMG} in 1+1 dimensions, upon which the later numerical work for the \TFFZ{} was based \cite{zeng_symmetric_2022,araki_symmetric_2026,zakharov_symmetric_2026}.
However, the numerical \ac{SMG} studies focus on the ungauged model, and did not include gauge fields, where the objections to decoupling the mirror sector lie \cite{golterman_constraints_2025,golterman_propagator_2024}.
It is also interesting to note that the tensor-network studies \ac{SMG} in the ungauged models required large bond dimensions, in contrast to the very modest bond dimensions required for our model.

Beyond static properties, our models open the possibility of investigating real-time dynamics of chiral gauge theories using quantum simulation techniques
\cite{banuls_simulating_2020,bauer_quantum_2023,Bauer:2023qgm}.
A qudit-based lattice model naturally lends itself to such simulations \cite{meth_simulating_2025,zache_fermionqudit_2023,popov_variational_2024,Gustafson:2022xdt,Kurkcuoglu:2024cfv}, complementing work on quantum simulation of Dirac fermions with an anomalous global chiral symmetry, as in \ac{QCD} \cite{lamm_exact_2026}.
These 1+1-dimensional theories could also serve as toy models for early-universe phenomena such as baryogenesis \cite{Grigoriev:1992nv,Aarts:1998td,Saffin:2011kc,carena_realtime_2025}, which involve nonequilibrium, nonperturbative dynamics of a chiral gauge theory.

While our work is most directly connected to bosonization-based formulations of 1+1-dimensional abelian chiral gauge theories \cite{thorngren_chiral_2026,seifnashri_exactly_2026,berkowitz_exact_2023}, it is worthwhile to explore connections to other proposals to understand how this approach might be generalized, or how one might address the issues in other approaches.
The connection to L\"uscher's definition \cite{luscher_abelian_1999,kadoh_construction_2008,craig_gaugeinvariant_2026} could be understood either by reverse-engineering bosonization \cite{baig_bosonization_2026a} or via anomaly inflow \cite{pedersen_reformulation_2023}.
Other promising proposals include recent domain-wall based constructions \cite{dang_gauge_2026,kaplan_regulating_2024,kaplan_weyl_2024,sen_hamiltonian_2026} and \ac{SMG} \cite{wang_nonperturbative_2023,eichten_chiral_1986,you_interacting_2014,you_interacting_2015,catterall_lattice_2024}.
Extension to higher dimensions or nonabelian gauge groups is not straightforward but certainly conceivable; see Refs.~\cite{onoda_lattice_2026,thorngren_chiral_2026,berkowitz_vectorvectoraxial_2026} for recent ideas in this vein.
A practical lattice definition of the Standard Model will require further new ideas, but 1+1-dimensional abelian chiral gauge theories now have an unambiguous, numerically efficient formulation against which new proposals can be tested and compared.

\begin{acknowledgments}
  We are grateful to Hank Lamm, Alessandro Mariani, Dan Hackett, David Tong and Frank Verstraete for helpful discussions along various stages of the manuscript.
  The tensor-network simulations were performed with \code{MPSKit.jl} \cite{devos_mpskit_2026} and \code{TensorKit.jl} \cite{Devos:2025yoj}, and in particular we thank Maarten Van Damme and Lukas Devos for useful comments.
  The research of HS at Perimeter Institute is supported in part by the Government of Canada through the Department of Innovation, Science and Economic Development Canada and by the Province of Ontario through the Ministry of Colleges and Universities.
  HS was also supported in part by the U.S.\ Department of Energy, Office of Science, Office of High Energy Physics Quantum Information Science Enabled Discovery (QuantISED) program ``Intersections of QIS and Theoretical Particle Physics''.
  This work was produced by Fermi Forward Discovery Group, LLC under Contract No.\ 89243024CSC000002 with the U.S.\ Department of Energy, Office of Science, Office of High Energy Physics.
  L.L. is supported by an EPSRC Postdoctoral Fellowship (grant No. EP/Y020456/1).

  \minisec{AI-use disclosure}
  We used Claude (Anthropic), an \textsc{ai} model, to assist with this work.
  After we conceived of the essential idea and carried out some analytic work, Claude greatly accelerated this research by helping with the simulation code and analysis.
  The code used to generate the figures in this work was almost entirely written by Claude; although we performed spot-checks against code written by us which convinced us of its correctness. 
  We take full responsibility for the content of this paper.
\end{acknowledgments}

\bibliography{refs.bib}

\begin{thebibliography}{117}%
\makeatletter
\providecommand \@ifxundefined [1]{%
 \@ifx{#1\undefined}
}%
\providecommand \@ifnum [1]{%
 \ifnum #1\expandafter \@firstoftwo
 \else \expandafter \@secondoftwo
 \fi
}%
\providecommand \@ifx [1]{%
 \ifx #1\expandafter \@firstoftwo
 \else \expandafter \@secondoftwo
 \fi
}%
\providecommand \natexlab [1]{#1}%
\providecommand \enquote  [1]{``#1''}%
\providecommand \bibnamefont  [1]{#1}%
\providecommand \bibfnamefont [1]{#1}%
\providecommand \citenamefont [1]{#1}%
\providecommand \href@noop [0]{\@secondoftwo}%
\providecommand \href [0]{\begingroup \@sanitize@url \@href}%
\providecommand \@href[1]{\@@startlink{#1}\@@href}%
\providecommand \@@href[1]{\endgroup#1\@@endlink}%
\providecommand \@sanitize@url [0]{\catcode `\\12\catcode `\$12\catcode
  `\&12\catcode `\#12\catcode `\^12\catcode `\_12\catcode `\%12\relax}%
\providecommand \@@startlink[1]{}%
\providecommand \@@endlink[0]{}%
\providecommand \url  [0]{\begingroup\@sanitize@url \@url }%
\providecommand \@url [1]{\endgroup\@href {#1}{\urlprefix }}%
\providecommand \urlprefix  [0]{URL }%
\providecommand \Eprint [0]{\href }%
\providecommand \doibase [0]{https://doi.org/}%
\providecommand \selectlanguage [0]{\@gobble}%
\providecommand \bibinfo  [0]{\@secondoftwo}%
\providecommand \bibfield  [0]{\@secondoftwo}%
\providecommand \translation [1]{[#1]}%
\providecommand \BibitemOpen [0]{}%
\providecommand \bibitemStop [0]{}%
\providecommand \bibitemNoStop [0]{.\EOS\space}%
\providecommand \EOS [0]{\spacefactor3000\relax}%
\providecommand \BibitemShut  [1]{\csname bibitem#1\endcsname}%
\let\auto@bib@innerbib\@empty
\bibitem [{\citenamefont {Wilson}(1974)}]{wilson_confinement_1974}%
  \BibitemOpen
  \bibfield  {author} {\bibinfo {author} {\bibfnamefont {K.~G.}\ \bibnamefont
  {Wilson}},\ }\bibfield  {title} {\bibinfo {title} {{Confinement of Quarks}},\
  }\href {https://doi.org/10.1103/PhysRevD.10.2445} {\bibfield  {journal}
  {\bibinfo  {journal} {Phys. Rev. D}\ }\textbf {\bibinfo {volume} {10}},\
  \bibinfo {pages} {2445} (\bibinfo {year} {1974})}\BibitemShut {NoStop}%
\bibitem [{\citenamefont {Kogut}\ and\ \citenamefont
  {Susskind}(1975)}]{kogut_hamiltonian_1975}%
  \BibitemOpen
  \bibfield  {author} {\bibinfo {author} {\bibfnamefont {J.~B.}\ \bibnamefont
  {Kogut}}\ and\ \bibinfo {author} {\bibfnamefont {L.}~\bibnamefont
  {Susskind}},\ }\bibfield  {title} {\bibinfo {title} {{Hamiltonian Formulation
  of Wilson's Lattice Gauge Theories}},\ }\href
  {https://doi.org/10.1103/PhysRevD.11.395} {\bibfield  {journal} {\bibinfo
  {journal} {Phys. Rev. D}\ }\textbf {\bibinfo {volume} {11}},\ \bibinfo
  {pages} {395} (\bibinfo {year} {1975})}\BibitemShut {NoStop}%
\bibitem [{\citenamefont {Nielsen}\ and\ \citenamefont
  {Ninomiya}(1981{\natexlab{a}})}]{nielsen_absence_1981}%
  \BibitemOpen
  \bibfield  {author} {\bibinfo {author} {\bibfnamefont {H.~B.}\ \bibnamefont
  {Nielsen}}\ and\ \bibinfo {author} {\bibfnamefont {M.}~\bibnamefont
  {Ninomiya}},\ }\bibfield  {title} {\bibinfo {title} {{Absence of Neutrinos on
  a Lattice. 1. Proof by Homotopy Theory}},\ }\href
  {https://doi.org/10.1016/0550-3213(81)90361-8} {\bibfield  {journal}
  {\bibinfo  {journal} {Nucl. Phys. B}\ }\textbf {\bibinfo {volume} {185}},\
  \bibinfo {pages} {20} (\bibinfo {year} {1981}{\natexlab{a}})},\ \bibinfo
  {note} {[Erratum: Nucl.Phys.B 195, 541 (1982)]}\BibitemShut {NoStop}%
\bibitem [{\citenamefont {Nielsen}\ and\ \citenamefont
  {Ninomiya}(1981{\natexlab{b}})}]{nielsen_absence_1981a}%
  \BibitemOpen
  \bibfield  {author} {\bibinfo {author} {\bibfnamefont {H.~B.}\ \bibnamefont
  {Nielsen}}\ and\ \bibinfo {author} {\bibfnamefont {M.}~\bibnamefont
  {Ninomiya}},\ }\bibfield  {title} {\bibinfo {title} {{Absence of Neutrinos on
  a Lattice. 2. Intuitive Topological Proof}},\ }\href
  {https://doi.org/10.1016/0550-3213(81)90524-1} {\bibfield  {journal}
  {\bibinfo  {journal} {Nucl. Phys. B}\ }\textbf {\bibinfo {volume} {193}},\
  \bibinfo {pages} {173} (\bibinfo {year} {1981}{\natexlab{b}})}\BibitemShut
  {NoStop}%
\bibitem [{\citenamefont {Karsten}(1981)}]{karsten_lattice_1981}%
  \BibitemOpen
  \bibfield  {author} {\bibinfo {author} {\bibfnamefont {L.~H.}\ \bibnamefont
  {Karsten}},\ }\bibfield  {title} {\bibinfo {title} {Lattice {{Fermions}} in
  {{Euclidean Space-time}}},\ }\href
  {https://doi.org/10.1016/0370-2693(81)90133-7} {\bibfield  {journal}
  {\bibinfo  {journal} {Phys. Lett. B}\ }\textbf {\bibinfo {volume} {104}},\
  \bibinfo {pages} {315} (\bibinfo {year} {1981})}\BibitemShut {NoStop}%
\bibitem [{\citenamefont {Ginsparg}\ and\ \citenamefont
  {Wilson}(1982)}]{ginsparg_remnant_1982}%
  \BibitemOpen
  \bibfield  {author} {\bibinfo {author} {\bibfnamefont {P.~H.}\ \bibnamefont
  {Ginsparg}}\ and\ \bibinfo {author} {\bibfnamefont {K.~G.}\ \bibnamefont
  {Wilson}},\ }\bibfield  {title} {\bibinfo {title} {{A Remnant of Chiral
  Symmetry on the Lattice}},\ }\href {https://doi.org/10.1103/PhysRevD.25.2649}
  {\bibfield  {journal} {\bibinfo  {journal} {Phys. Rev. D}\ }\textbf {\bibinfo
  {volume} {25}},\ \bibinfo {pages} {2649} (\bibinfo {year}
  {1982})}\BibitemShut {NoStop}%
\bibitem [{\citenamefont
  {Neuberger}(1998{\natexlab{a}})}]{neuberger_more_1998}%
  \BibitemOpen
  \bibfield  {author} {\bibinfo {author} {\bibfnamefont {H.}~\bibnamefont
  {Neuberger}},\ }\bibfield  {title} {\bibinfo {title} {{More about exactly
  massless quarks on the lattice}},\ }\href
  {https://doi.org/10.1016/S0370-2693(98)00355-4} {\bibfield  {journal}
  {\bibinfo  {journal} {Phys. Lett. B}\ }\textbf {\bibinfo {volume} {427}},\
  \bibinfo {pages} {353} (\bibinfo {year} {1998}{\natexlab{a}})},\ \Eprint
  {https://arxiv.org/abs/hep-lat/9801031} {arXiv:hep-lat/9801031} \BibitemShut
  {NoStop}%
\bibitem [{\citenamefont
  {Neuberger}(1998{\natexlab{b}})}]{neuberger_exactly_1998}%
  \BibitemOpen
  \bibfield  {author} {\bibinfo {author} {\bibfnamefont {H.}~\bibnamefont
  {Neuberger}},\ }\bibfield  {title} {\bibinfo {title} {{Exactly massless
  quarks on the lattice}},\ }\href
  {https://doi.org/10.1016/S0370-2693(97)01368-3} {\bibfield  {journal}
  {\bibinfo  {journal} {Phys. Lett. B}\ }\textbf {\bibinfo {volume} {417}},\
  \bibinfo {pages} {141} (\bibinfo {year} {1998}{\natexlab{b}})},\ \Eprint
  {https://arxiv.org/abs/hep-lat/9707022} {arXiv:hep-lat/9707022} \BibitemShut
  {NoStop}%
\bibitem [{\citenamefont {Hasenfratz}(1998)}]{hasenfratz_prospects_1998}%
  \BibitemOpen
  \bibfield  {author} {\bibinfo {author} {\bibfnamefont {P.}~\bibnamefont
  {Hasenfratz}},\ }\bibfield  {title} {\bibinfo {title} {{Prospects for perfect
  actions}},\ }\href {https://doi.org/10.1016/S0920-5632(97)00696-8} {\bibfield
   {journal} {\bibinfo  {journal} {Nucl. Phys. B Proc. Suppl.}\ }\textbf
  {\bibinfo {volume} {63}},\ \bibinfo {pages} {53} (\bibinfo {year} {1998})},\
  \Eprint {https://arxiv.org/abs/hep-lat/9709110} {arXiv:hep-lat/9709110}
  \BibitemShut {NoStop}%
\bibitem [{\citenamefont {Kaplan}(1993)}]{kaplan_chiral_1993}%
  \BibitemOpen
  \bibfield  {author} {\bibinfo {author} {\bibfnamefont {D.~B.}\ \bibnamefont
  {Kaplan}},\ }\bibfield  {title} {\bibinfo {title} {Chiral fermions on the
  lattice},\ }\href {https://doi.org/10.1016/0920-5632(93)90282-B} {\bibfield
  {journal} {\bibinfo  {journal} {Nuclear Physics B - Proceedings Supplements}\
  }\bibinfo {series} {Proceedings of the {{International Symposium}} On},\
  \textbf {\bibinfo {volume} {30}},\ \bibinfo {pages} {597} (\bibinfo {year}
  {1993})}\BibitemShut {NoStop}%
\bibitem [{\citenamefont {Kaplan}(1992)}]{kaplan_method_1992}%
  \BibitemOpen
  \bibfield  {author} {\bibinfo {author} {\bibfnamefont {D.~B.}\ \bibnamefont
  {Kaplan}},\ }\bibfield  {title} {\bibinfo {title} {{A Method for simulating
  chiral fermions on the lattice}},\ }\href
  {https://doi.org/10.1016/0370-2693(92)91112-M} {\bibfield  {journal}
  {\bibinfo  {journal} {Phys. Lett. B}\ }\textbf {\bibinfo {volume} {288}},\
  \bibinfo {pages} {342} (\bibinfo {year} {1992})},\ \Eprint
  {https://arxiv.org/abs/hep-lat/9206013} {arXiv:hep-lat/9206013} \BibitemShut
  {NoStop}%
\bibitem [{\citenamefont {Luscher}(1998)}]{luscher_exact_1998}%
  \BibitemOpen
  \bibfield  {author} {\bibinfo {author} {\bibfnamefont {M.}~\bibnamefont
  {Luscher}},\ }\bibfield  {title} {\bibinfo {title} {{Exact chiral symmetry on
  the lattice and the Ginsparg-Wilson relation}},\ }\href
  {https://doi.org/10.1016/S0370-2693(98)00423-7} {\bibfield  {journal}
  {\bibinfo  {journal} {Phys. Lett. B}\ }\textbf {\bibinfo {volume} {428}},\
  \bibinfo {pages} {342} (\bibinfo {year} {1998})},\ \Eprint
  {https://arxiv.org/abs/hep-lat/9802011} {arXiv:hep-lat/9802011} \BibitemShut
  {NoStop}%
\bibitem [{\citenamefont {Luscher}(1999)}]{luscher_abelian_1999}%
  \BibitemOpen
  \bibfield  {author} {\bibinfo {author} {\bibfnamefont {M.}~\bibnamefont
  {Luscher}},\ }\bibfield  {title} {\bibinfo {title} {{Abelian chiral gauge
  theories on the lattice with exact gauge invariance}},\ }\href
  {https://doi.org/10.1016/S0550-3213(99)00115-7} {\bibfield  {journal}
  {\bibinfo  {journal} {Nucl. Phys. B}\ }\textbf {\bibinfo {volume} {549}},\
  \bibinfo {pages} {295} (\bibinfo {year} {1999})},\ \Eprint
  {https://arxiv.org/abs/hep-lat/9811032} {arXiv:hep-lat/9811032} \BibitemShut
  {NoStop}%
\bibitem [{\citenamefont {L{\"u}scher}(2002)}]{luscher_chiral_2002a}%
  \BibitemOpen
  \bibfield  {author} {\bibinfo {author} {\bibfnamefont {M.}~\bibnamefont
  {L{\"u}scher}},\ }\bibfield  {title} {\bibinfo {title} {Chiral gauge theories
  revisited},\ }in\ \href {https://doi.org/10.1142/9789812778253_0002} {\emph
  {\bibinfo {booktitle} {Theory and {{Experiment Heading}} for {{New
  Physics}}}}},\ \bibinfo {series} {The {{Subnuclear Series}}}, Vol.\ \bibinfo
  {volume} {Volume 38}\ (\bibinfo  {publisher} {WORLD SCIENTIFIC},\ \bibinfo
  {year} {2002})\ pp.\ \bibinfo {pages} {41--89},\ \Eprint
  {https://arxiv.org/abs/hep-th/0102028} {arXiv:hep-th/0102028 [hep-th]}
  \BibitemShut {NoStop}%
\bibitem [{\citenamefont {Kadoh}\ and\ \citenamefont
  {Kikukawa}(2005{\natexlab{a}})}]{kadoh_another_2005}%
  \BibitemOpen
  \bibfield  {author} {\bibinfo {author} {\bibfnamefont {D.}~\bibnamefont
  {Kadoh}}\ and\ \bibinfo {author} {\bibfnamefont {Y.}~\bibnamefont
  {Kikukawa}},\ }\href {https://arxiv.org/abs/hep-lat/0504021} {\bibinfo
  {title} {Another approach to local cohomology problem in {{Abelian}} lattice
  gauge theories}} (\bibinfo {year} {2005}{\natexlab{a}}),\ \Eprint
  {https://arxiv.org/abs/hep-lat/0504021} {arXiv:hep-lat/0504021} \BibitemShut
  {NoStop}%
\bibitem [{\citenamefont {Kadoh}\ and\ \citenamefont
  {Kikukawa}(2005{\natexlab{b}})}]{kadoh_numerical_2005}%
  \BibitemOpen
  \bibfield  {author} {\bibinfo {author} {\bibfnamefont {D.}~\bibnamefont
  {Kadoh}}\ and\ \bibinfo {author} {\bibfnamefont {Y.}~\bibnamefont
  {Kikukawa}},\ }\bibfield  {title} {\bibinfo {title} {A {{Numerical}} solution
  to the local cohomology problem in {{U}}(1) chiral gauge theories},\ }\href
  {https://doi.org/10.1088/1126-6708/2005/01/024} {\bibfield  {journal}
  {\bibinfo  {journal} {JHEP}\ }\textbf {\bibinfo {volume} {2005}}\bibfield
  {number} {\bibinfo  {number} { (1)},\ \bibinfo {pages} {024}},\ }\Eprint
  {https://arxiv.org/abs/hep-lat/0401025} {arXiv:hep-lat/0401025} \BibitemShut
  {NoStop}%
\bibitem [{\citenamefont {Kadoh}\ and\ \citenamefont
  {Kikukawa}(2008{\natexlab{a}})}]{kadoh_simple_2008}%
  \BibitemOpen
  \bibfield  {author} {\bibinfo {author} {\bibfnamefont {D.}~\bibnamefont
  {Kadoh}}\ and\ \bibinfo {author} {\bibfnamefont {Y.}~\bibnamefont
  {Kikukawa}},\ }\bibfield  {title} {\bibinfo {title} {A {{Simple}}
  construction of fermion measure term in {{U}}(1) chiral lattice gauge
  theories with exact gauge invariance},\ }\href
  {https://doi.org/10.1088/1126-6708/2008/02/063} {\bibfield  {journal}
  {\bibinfo  {journal} {JHEP}\ }\textbf {\bibinfo {volume} {2008}}\bibfield
  {number} {\bibinfo  {number} { (2)},\ \bibinfo {pages} {063}},\ }\Eprint
  {https://arxiv.org/abs/0709.3656} {arXiv:0709.3656 [hep-lat]} \BibitemShut
  {NoStop}%
\bibitem [{\citenamefont {Kadoh}\ \emph {et~al.}(2004)\citenamefont {Kadoh},
  \citenamefont {Kikukawa},\ and\ \citenamefont
  {Nakayama}}]{kadoh_solving_2004}%
  \BibitemOpen
  \bibfield  {author} {\bibinfo {author} {\bibfnamefont {D.}~\bibnamefont
  {Kadoh}}, \bibinfo {author} {\bibfnamefont {Y.}~\bibnamefont {Kikukawa}},\
  and\ \bibinfo {author} {\bibfnamefont {Y.}~\bibnamefont {Nakayama}},\
  }\bibfield  {title} {\bibinfo {title} {Solving the local cohomology problem
  in {{U}}(1) chiral gauge theories within a finite lattice},\ }\href
  {https://doi.org/10.1088/1126-6708/2004/12/006} {\bibfield  {journal}
  {\bibinfo  {journal} {JHEP}\ }\textbf {\bibinfo {volume} {2004}}\bibfield
  {number} {\bibinfo  {number} { (12)},\ \bibinfo {pages} {006}},\ }\Eprint
  {https://arxiv.org/abs/hep-lat/0309022} {arXiv:hep-lat/0309022} \BibitemShut
  {NoStop}%
\bibitem [{\citenamefont {Kadoh}\ and\ \citenamefont
  {Kikukawa}(2008{\natexlab{b}})}]{kadoh_construction_2008}%
  \BibitemOpen
  \bibfield  {author} {\bibinfo {author} {\bibfnamefont {D.}~\bibnamefont
  {Kadoh}}\ and\ \bibinfo {author} {\bibfnamefont {Y.}~\bibnamefont
  {Kikukawa}},\ }\bibfield  {title} {\bibinfo {title} {A {{Construction}} of
  the {{Glashow-Weinberg-Salam}} model on the lattice with exact gauge
  invariance},\ }\href {https://doi.org/10.1088/1126-6708/2008/05/095}
  {\bibfield  {journal} {\bibinfo  {journal} {JHEP}\ }\textbf {\bibinfo
  {volume} {2008}}\bibfield  {number} {\bibinfo  {number} { (5)},\ \bibinfo
  {pages} {095}},\ }\bibinfo {note} {[Erratum: JHEP 03, 095 (2011)]},\ \Eprint
  {https://arxiv.org/abs/0709.3658} {arXiv:0709.3658 [hep-lat]} \BibitemShut
  {NoStop}%
\bibitem [{\citenamefont {Kikukawa}\ and\ \citenamefont
  {Nakayama}(2001)}]{kikukawa_gauge_2001}%
  \BibitemOpen
  \bibfield  {author} {\bibinfo {author} {\bibfnamefont {Y.}~\bibnamefont
  {Kikukawa}}\ and\ \bibinfo {author} {\bibfnamefont {Y.}~\bibnamefont
  {Nakayama}},\ }\bibfield  {title} {\bibinfo {title} {Gauge anomaly
  cancellations in {{SU}}(2)({{L}}) x {{U}}(1)({{Y}}) electroweak theory on the
  lattice},\ }\href {https://doi.org/10.1016/S0550-3213(00)00714-8} {\bibfield
  {journal} {\bibinfo  {journal} {Nucl. Phys. B}\ }\textbf {\bibinfo {volume}
  {597}},\ \bibinfo {pages} {519} (\bibinfo {year} {2001})},\ \Eprint
  {https://arxiv.org/abs/hep-lat/0005015} {arXiv:hep-lat/0005015} \BibitemShut
  {NoStop}%
\bibitem [{\citenamefont {Kikukawa}\ \emph {et~al.}(2002)\citenamefont
  {Kikukawa}, \citenamefont {Nakayama},\ and\ \citenamefont
  {Suzuki}}]{kikukawa_lattice_2002}%
  \BibitemOpen
  \bibfield  {author} {\bibinfo {author} {\bibfnamefont {Y.}~\bibnamefont
  {Kikukawa}}, \bibinfo {author} {\bibfnamefont {Y.}~\bibnamefont {Nakayama}},\
  and\ \bibinfo {author} {\bibfnamefont {H.}~\bibnamefont {Suzuki}},\
  }\bibfield  {title} {\bibinfo {title} {On the lattice construction of
  electroweak gauge theory},\ }\href
  {https://doi.org/10.1016/S0920-5632(01)01837-0} {\bibfield  {journal}
  {\bibinfo  {journal} {Nucl. Phys. B Proc. Suppl.}\ }\textbf {\bibinfo
  {volume} {106}},\ \bibinfo {pages} {763} (\bibinfo {year} {2002})},\ \Eprint
  {https://arxiv.org/abs/hep-lat/0111036} {arXiv:hep-lat/0111036} \BibitemShut
  {NoStop}%
\bibitem [{\citenamefont {Craig}(2026)}]{craig_gaugeinvariant_2026}%
  \BibitemOpen
  \bibfield  {author} {\bibinfo {author} {\bibfnamefont {N.}~\bibnamefont
  {Craig}},\ }\href@noop {} {\bibinfo {title} {{A gauge-invariant measure for
  lattice chiral fermions}}} (\bibinfo {year} {2026}),\ \Eprint
  {https://arxiv.org/abs/2610.09093} {arXiv:2610.09093 [hep-lat]} \BibitemShut
  {NoStop}%
\bibitem [{\citenamefont {Berkowitz}\ \emph {et~al.}(2024)\citenamefont
  {Berkowitz}, \citenamefont {Cherman},\ and\ \citenamefont
  {Jacobson}}]{berkowitz_exact_2023}%
  \BibitemOpen
  \bibfield  {author} {\bibinfo {author} {\bibfnamefont {E.}~\bibnamefont
  {Berkowitz}}, \bibinfo {author} {\bibfnamefont {A.}~\bibnamefont {Cherman}},\
  and\ \bibinfo {author} {\bibfnamefont {T.}~\bibnamefont {Jacobson}},\
  }\bibfield  {title} {\bibinfo {title} {{Exact lattice chiral symmetry in 2D
  gauge theory}},\ }\href {https://doi.org/10.1103/PhysRevD.110.014510}
  {\bibfield  {journal} {\bibinfo  {journal} {Phys. Rev. D}\ }\textbf {\bibinfo
  {volume} {110}},\ \bibinfo {pages} {014510} (\bibinfo {year} {2024})},\
  \Eprint {https://arxiv.org/abs/2310.17539} {arXiv:2310.17539 [hep-lat]}
  \BibitemShut {NoStop}%
\bibitem [{\citenamefont {Seifnashri}(2026)}]{seifnashri_exactly_2026}%
  \BibitemOpen
  \bibfield  {author} {\bibinfo {author} {\bibfnamefont {S.}~\bibnamefont
  {Seifnashri}},\ }\href@noop {} {\bibinfo {title} {Exactly {{Solvable}} 1+1d
  {{Chiral Lattice Gauge Theories}}}} (\bibinfo {year} {2026}),\ \Eprint
  {https://arxiv.org/abs/2601.14359} {arXiv:2601.14359} \BibitemShut {NoStop}%
\bibitem [{\citenamefont {Thorngren}\ \emph {et~al.}(2026)\citenamefont
  {Thorngren}, \citenamefont {Preskill},\ and\ \citenamefont
  {Fidkowski}}]{thorngren_chiral_2026}%
  \BibitemOpen
  \bibfield  {author} {\bibinfo {author} {\bibfnamefont {R.}~\bibnamefont
  {Thorngren}}, \bibinfo {author} {\bibfnamefont {J.}~\bibnamefont
  {Preskill}},\ and\ \bibinfo {author} {\bibfnamefont {L.}~\bibnamefont
  {Fidkowski}},\ }\bibfield  {title} {\bibinfo {title} {{Chiral lattice gauge
  theories from symmetry disentanglers}},\ }\href
  {https://doi.org/10.1007/JHEP07(2026)271} {\bibfield  {journal} {\bibinfo
  {journal} {JHEP}\ }\textbf {\bibinfo {volume} {2026}}\bibfield  {number}
  {\bibinfo  {number} { (7)},\ \bibinfo {pages} {271}},\ }\Eprint
  {https://arxiv.org/abs/2601.04304} {arXiv:2601.04304 [hep-th]} \BibitemShut
  {NoStop}%
\bibitem [{\citenamefont {Kapustin}\ and\ \citenamefont
  {Sopenko}(2025)}]{kapustin_anomalous_2025}%
  \BibitemOpen
  \bibfield  {author} {\bibinfo {author} {\bibfnamefont {A.}~\bibnamefont
  {Kapustin}}\ and\ \bibinfo {author} {\bibfnamefont {N.}~\bibnamefont
  {Sopenko}},\ }\bibfield  {title} {\bibinfo {title} {Anomalous symmetries of
  quantum spin chains and a generalization of the lieb--schultz--mattis
  theorem},\ }\href {https://doi.org/10.1007/s00220-025-05422-2} {\bibfield
  {journal} {\bibinfo  {journal} {Commun. Math. Phys.}\ }\textbf {\bibinfo
  {volume} {406}},\ \bibinfo {pages} {238} (\bibinfo {year} {2025})},\ \Eprint
  {https://arxiv.org/abs/2401.02533} {arXiv:2401.02533 [math-ph]} \BibitemShut
  {NoStop}%
\bibitem [{\citenamefont {Fidkowski}\ and\ \citenamefont
  {Xu}(2023)}]{fidkowski_nogo_2023}%
  \BibitemOpen
  \bibfield  {author} {\bibinfo {author} {\bibfnamefont {L.}~\bibnamefont
  {Fidkowski}}\ and\ \bibinfo {author} {\bibfnamefont {C.}~\bibnamefont {Xu}},\
  }\bibfield  {title} {\bibinfo {title} {{A No-Go Result for Implementing
  Chiral Symmetries by Locality-Preserving Unitaries in a Three-Dimensional
  Hamiltonian Lattice Model of Fermions}},\ }\href
  {https://doi.org/10.1103/PhysRevLett.131.196601} {\bibfield  {journal}
  {\bibinfo  {journal} {Phys. Rev. Lett.}\ }\textbf {\bibinfo {volume} {131}},\
  \bibinfo {pages} {196601} (\bibinfo {year} {2023})},\ \Eprint
  {https://arxiv.org/abs/2306.10105} {arXiv:2306.10105 [cond-mat.str-el]}
  \BibitemShut {NoStop}%
\bibitem [{\citenamefont {Gioia}\ \emph {et~al.}(2026)\citenamefont {Gioia},
  \citenamefont {Pace}, \citenamefont {Verresen}, \citenamefont {Shao},\ and\
  \citenamefont {Thorngren}}]{gioia_disentangling_2026}%
  \BibitemOpen
  \bibfield  {author} {\bibinfo {author} {\bibfnamefont {L.}~\bibnamefont
  {Gioia}}, \bibinfo {author} {\bibfnamefont {S.~D.}\ \bibnamefont {Pace}},
  \bibinfo {author} {\bibfnamefont {R.}~\bibnamefont {Verresen}}, \bibinfo
  {author} {\bibfnamefont {S.-H.}\ \bibnamefont {Shao}},\ and\ \bibinfo
  {author} {\bibfnamefont {R.}~\bibnamefont {Thorngren}},\ }\href
  {https://arxiv.org/abs/2609.30363} {\bibinfo {title} {Disentangling the toric
  code}} (\bibinfo {year} {2026}),\ \Eprint {https://arxiv.org/abs/2609.30363}
  {arXiv:2609.30363} \BibitemShut {NoStop}%
\bibitem [{\citenamefont {Bhattacharya}\ \emph {et~al.}(2006)\citenamefont
  {Bhattacharya}, \citenamefont {Martin},\ and\ \citenamefont
  {Poppitz}}]{bhattacharya_chiral_2006}%
  \BibitemOpen
  \bibfield  {author} {\bibinfo {author} {\bibfnamefont {T.}~\bibnamefont
  {Bhattacharya}}, \bibinfo {author} {\bibfnamefont {M.~R.}\ \bibnamefont
  {Martin}},\ and\ \bibinfo {author} {\bibfnamefont {E.}~\bibnamefont
  {Poppitz}},\ }\bibfield  {title} {\bibinfo {title} {Chiral {{Lattice Gauge
  Theories}} from {{Warped Domain Walls}} and {{Ginsparg-Wilson Fermions}}},\
  }\href {https://doi.org/10.1103/PhysRevD.74.085028} {\bibfield  {journal}
  {\bibinfo  {journal} {Phys. Rev. D}\ }\textbf {\bibinfo {volume} {74}},\
  \bibinfo {pages} {085028} (\bibinfo {year} {2006})},\ \Eprint
  {https://arxiv.org/abs/hep-lat/0605003} {arXiv:hep-lat/0605003 [hep-lat]}
  \BibitemShut {NoStop}%
\bibitem [{\citenamefont {Giedt}\ and\ \citenamefont
  {Poppitz}(2007)}]{giedt_chiral_2007}%
  \BibitemOpen
  \bibfield  {author} {\bibinfo {author} {\bibfnamefont {J.}~\bibnamefont
  {Giedt}}\ and\ \bibinfo {author} {\bibfnamefont {E.}~\bibnamefont
  {Poppitz}},\ }\bibfield  {title} {\bibinfo {title} {Chiral {{Lattice Gauge
  Theories}} and {{The Strong Coupling Dynamics}} of a {{Yukawa-Higgs Model}}
  with {{Ginsparg-Wilson Fermions}}},\ }\href
  {https://doi.org/10.1088/1126-6708/2007/10/076} {\bibfield  {journal}
  {\bibinfo  {journal} {J. High Energy Phys.}\ }\textbf {\bibinfo {volume}
  {2007}}\bibfield  {number} {\bibinfo  {number} { (10)},\ \bibinfo {pages}
  {076}},\ }\Eprint {https://arxiv.org/abs/hep-lat/0701004}
  {arXiv:hep-lat/0701004} \BibitemShut {NoStop}%
\bibitem [{\citenamefont {Suzuki}(2007)}]{suzuki_perturbative_2007}%
  \BibitemOpen
  \bibfield  {author} {\bibinfo {author} {\bibfnamefont {H.}~\bibnamefont
  {Suzuki}},\ }\href {https://doi.org/10.48550/arXiv.0706.3083} {\bibinfo
  {title} {Perturbative spectrum of a {{Yukawa-Higgs}} model with
  {{Ginsparg-Wilson}} fermions}} (\bibinfo {year} {2007}),\ \Eprint
  {https://arxiv.org/abs/0706.3083} {arXiv:0706.3083 [hep-lat]} \BibitemShut
  {NoStop}%
\bibitem [{\citenamefont {Chen}\ \emph {et~al.}(2013)\citenamefont {Chen},
  \citenamefont {Giedt},\ and\ \citenamefont {Poppitz}}]{chen_decoupling_2013}%
  \BibitemOpen
  \bibfield  {author} {\bibinfo {author} {\bibfnamefont {C.}~\bibnamefont
  {Chen}}, \bibinfo {author} {\bibfnamefont {J.}~\bibnamefont {Giedt}},\ and\
  \bibinfo {author} {\bibfnamefont {E.}~\bibnamefont {Poppitz}},\ }\bibfield
  {title} {\bibinfo {title} {On the decoupling of mirror fermions},\ }\href
  {https://doi.org/10.1007/JHEP04(2013)131} {\bibfield  {journal} {\bibinfo
  {journal} {J. High Energ. Phys.}\ }\textbf {\bibinfo {volume}
  {2013}}\bibfield  {number} {\bibinfo  {number} { (4)},\ \bibinfo {pages}
  {131}},\ }\Eprint {https://arxiv.org/abs/1211.6947} {arXiv:1211.6947
  [hep-lat]} \BibitemShut {NoStop}%
\bibitem [{\citenamefont {Wang}\ and\ \citenamefont
  {Wen}(2023)}]{wang_nonperturbative_2023}%
  \BibitemOpen
  \bibfield  {author} {\bibinfo {author} {\bibfnamefont {J.}~\bibnamefont
  {Wang}}\ and\ \bibinfo {author} {\bibfnamefont {X.-G.}\ \bibnamefont {Wen}},\
  }\bibfield  {title} {\bibinfo {title} {{Nonperturbative regularization of
  (1+1)-dimensional anomaly-free chiral fermions and bosons: On the equivalence
  of anomaly matching conditions and boundary gapping rules}},\ }\href
  {https://doi.org/10.1103/PhysRevB.107.014311} {\bibfield  {journal} {\bibinfo
   {journal} {Phys. Rev. B}\ }\textbf {\bibinfo {volume} {107}},\ \bibinfo
  {pages} {014311} (\bibinfo {year} {2023})},\ \Eprint
  {https://arxiv.org/abs/1307.7480} {arXiv:1307.7480 [hep-lat]} \BibitemShut
  {NoStop}%
\bibitem [{\citenamefont {Wang}\ and\ \citenamefont
  {Wen}(2019)}]{wang_solution_2019}%
  \BibitemOpen
  \bibfield  {author} {\bibinfo {author} {\bibfnamefont {J.}~\bibnamefont
  {Wang}}\ and\ \bibinfo {author} {\bibfnamefont {X.-G.}\ \bibnamefont {Wen}},\
  }\bibfield  {title} {\bibinfo {title} {{Solution to the 1+1 dimensional
  gauged chiral fermion problem}},\ }\href
  {https://doi.org/10.1103/PhysRevD.99.111501} {\bibfield  {journal} {\bibinfo
  {journal} {Phys. Rev. D}\ }\textbf {\bibinfo {volume} {99}},\ \bibinfo
  {pages} {111501} (\bibinfo {year} {2019})},\ \Eprint
  {https://arxiv.org/abs/1807.05998} {arXiv:1807.05998 [hep-lat]} \BibitemShut
  {NoStop}%
\bibitem [{\citenamefont {Zeng}\ \emph {et~al.}(2022)\citenamefont {Zeng},
  \citenamefont {Zhu}, \citenamefont {Wang},\ and\ \citenamefont
  {You}}]{zeng_symmetric_2022}%
  \BibitemOpen
  \bibfield  {author} {\bibinfo {author} {\bibfnamefont {M.}~\bibnamefont
  {Zeng}}, \bibinfo {author} {\bibfnamefont {Z.}~\bibnamefont {Zhu}}, \bibinfo
  {author} {\bibfnamefont {J.}~\bibnamefont {Wang}},\ and\ \bibinfo {author}
  {\bibfnamefont {Y.-Z.}\ \bibnamefont {You}},\ }\bibfield  {title} {\bibinfo
  {title} {{Symmetric Mass Generation in the 1+1 Dimensional Chiral Fermion
  3-4-5-0 Model}},\ }\href {https://doi.org/10.1103/PhysRevLett.128.185301}
  {\bibfield  {journal} {\bibinfo  {journal} {Phys. Rev. Lett.}\ }\textbf
  {\bibinfo {volume} {128}},\ \bibinfo {pages} {185301} (\bibinfo {year}
  {2022})},\ \Eprint {https://arxiv.org/abs/2202.12355} {arXiv:2202.12355
  [cond-mat.str-el]} \BibitemShut {NoStop}%
\bibitem [{\citenamefont {Golterman}\ and\ \citenamefont
  {Shamir}(2026)}]{golterman_constraints_2025}%
  \BibitemOpen
  \bibfield  {author} {\bibinfo {author} {\bibfnamefont {M.}~\bibnamefont
  {Golterman}}\ and\ \bibinfo {author} {\bibfnamefont {Y.}~\bibnamefont
  {Shamir}},\ }\bibfield  {title} {\bibinfo {title} {Constraints on the
  symmetric mass generation paradigm for lattice chiral gauge theories},\
  }\href {https://doi.org/10.1103/qd1t-wzy1} {\bibfield  {journal} {\bibinfo
  {journal} {Phys. Rev. D}\ }\textbf {\bibinfo {volume} {113}},\ \bibinfo
  {pages} {014503} (\bibinfo {year} {2026})},\ \Eprint
  {https://arxiv.org/abs/2505.20436} {arXiv:2505.20436 [hep-lat]} \BibitemShut
  {NoStop}%
\bibitem [{\citenamefont {Morikawa}\ \emph {et~al.}(2024)\citenamefont
  {Morikawa}, \citenamefont {Onoda},\ and\ \citenamefont
  {Suzuki}}]{morikawa_another_2024}%
  \BibitemOpen
  \bibfield  {author} {\bibinfo {author} {\bibfnamefont {O.}~\bibnamefont
  {Morikawa}}, \bibinfo {author} {\bibfnamefont {S.}~\bibnamefont {Onoda}},\
  and\ \bibinfo {author} {\bibfnamefont {H.}~\bibnamefont {Suzuki}},\
  }\bibfield  {title} {\bibinfo {title} {Yet another lattice formulation of
  {{2D}} \${{U}}(1)\$ chiral gauge theory via bosonization},\ }\href
  {https://doi.org/10.1093/ptep/ptae075} {\bibfield  {journal} {\bibinfo
  {journal} {PTEP}\ }\textbf {\bibinfo {volume} {2024}},\ \bibinfo {pages}
  {063B01} (\bibinfo {year} {2024})},\ \Eprint
  {https://arxiv.org/abs/2403.03420} {arXiv:2403.03420 [hep-lat]} \BibitemShut
  {NoStop}%
\bibitem [{\citenamefont {Onoda}\ \emph {et~al.}(2025)\citenamefont {Onoda},
  \citenamefont {Morikawa},\ and\ \citenamefont
  {Suzuki}}]{morikawa_novel_2025}%
  \BibitemOpen
  \bibfield  {author} {\bibinfo {author} {\bibfnamefont {S.}~\bibnamefont
  {Onoda}}, \bibinfo {author} {\bibfnamefont {O.}~\bibnamefont {Morikawa}},\
  and\ \bibinfo {author} {\bibfnamefont {H.}~\bibnamefont {Suzuki}},\
  }\bibfield  {title} {\bibinfo {title} {Novel {{Lattice Formulation}} of {{2D
  Chiral Gauge Theory}} via {{Bosonization}}},\ }\href
  {https://doi.org/10.22323/1.466.0363} {\bibfield  {journal} {\bibinfo
  {journal} {PoS}\ }\textbf {\bibinfo {volume} {LATTICE2024}},\ \bibinfo
  {pages} {363} (\bibinfo {year} {2025})},\ \Eprint
  {https://arxiv.org/abs/2501.18949} {arXiv:2501.18949 [hep-lat]} \BibitemShut
  {NoStop}%
\bibitem [{\citenamefont {Coleman}(1975)}]{coleman_quantum_1975}%
  \BibitemOpen
  \bibfield  {author} {\bibinfo {author} {\bibfnamefont {S.~R.}\ \bibnamefont
  {Coleman}},\ }\bibfield  {title} {\bibinfo {title} {{The Quantum Sine-Gordon
  Equation as the Massive Thirring Model}},\ }\href
  {https://doi.org/10.1103/PhysRevD.11.2088} {\bibfield  {journal} {\bibinfo
  {journal} {Phys. Rev. D}\ }\textbf {\bibinfo {volume} {11}},\ \bibinfo
  {pages} {2088} (\bibinfo {year} {1975})}\BibitemShut {NoStop}%
\bibitem [{\citenamefont {Haldane}(1981)}]{haldane_luttinger_1981}%
  \BibitemOpen
  \bibfield  {author} {\bibinfo {author} {\bibfnamefont {F.~D.~M.}\
  \bibnamefont {Haldane}},\ }\bibfield  {title} {\bibinfo {title} {{Luttinger
  liquid theory of one-dimensional quantum fluids. I. Properties of the
  Luttinger model and their extension to the general 1D interacting spinless
  Fermi gas}},\ }\href {https://doi.org/10.1088/0022-3719/14/19/010} {\bibfield
   {journal} {\bibinfo  {journal} {J. Phys. C}\ }\textbf {\bibinfo {volume}
  {14}},\ \bibinfo {pages} {2585} (\bibinfo {year} {1981})}\BibitemShut
  {NoStop}%
\bibitem [{\citenamefont {Mandelstam}(1975)}]{mandelstam_soliton_1975}%
  \BibitemOpen
  \bibfield  {author} {\bibinfo {author} {\bibfnamefont {S.}~\bibnamefont
  {Mandelstam}},\ }\bibfield  {title} {\bibinfo {title} {{Soliton Operators for
  the Quantized Sine-Gordon Equation}},\ }\href
  {https://doi.org/10.1103/PhysRevD.11.3026} {\bibfield  {journal} {\bibinfo
  {journal} {Phys. Rev. D}\ }\textbf {\bibinfo {volume} {11}},\ \bibinfo
  {pages} {3026} (\bibinfo {year} {1975})}\BibitemShut {NoStop}%
\bibitem [{\citenamefont {Halliday}\ \emph {et~al.}(1986)\citenamefont
  {Halliday}, \citenamefont {Rabinovici}, \citenamefont {Schwimmer},\ and\
  \citenamefont {Chanowitz}}]{halliday_quantization_1986}%
  \BibitemOpen
  \bibfield  {author} {\bibinfo {author} {\bibfnamefont {I.~G.}\ \bibnamefont
  {Halliday}}, \bibinfo {author} {\bibfnamefont {E.}~\bibnamefont
  {Rabinovici}}, \bibinfo {author} {\bibfnamefont {A.}~\bibnamefont
  {Schwimmer}},\ and\ \bibinfo {author} {\bibfnamefont {M.~S.}\ \bibnamefont
  {Chanowitz}},\ }\bibfield  {title} {\bibinfo {title} {{Quantization of
  Anomalous Two-dimensional Models}},\ }\href
  {https://doi.org/10.1016/0550-3213(86)90163-X} {\bibfield  {journal}
  {\bibinfo  {journal} {Nucl. Phys. B}\ }\textbf {\bibinfo {volume} {268}},\
  \bibinfo {pages} {413} (\bibinfo {year} {1986})}\BibitemShut {NoStop}%
\bibitem [{\citenamefont {Mouland}\ \emph {et~al.}(2026)\citenamefont
  {Mouland}, \citenamefont {Tong},\ and\ \citenamefont
  {Zan}}]{mouland_phases_2025}%
  \BibitemOpen
  \bibfield  {author} {\bibinfo {author} {\bibfnamefont {R.}~\bibnamefont
  {Mouland}}, \bibinfo {author} {\bibfnamefont {D.}~\bibnamefont {Tong}},\ and\
  \bibinfo {author} {\bibfnamefont {B.}~\bibnamefont {Zan}},\ }\bibfield
  {title} {\bibinfo {title} {{Phases of 2d gauge theories and symmetric mass
  generation}},\ }\href {https://doi.org/10.1007/JHEP04(2026)154} {\bibfield
  {journal} {\bibinfo  {journal} {JHEP}\ }\textbf {\bibinfo {volume}
  {2026}}\bibfield  {number} {\bibinfo  {number} { (4)},\ \bibinfo {pages}
  {154}},\ }\Eprint {https://arxiv.org/abs/2509.12305} {arXiv:2509.12305
  [hep-th]} \BibitemShut {NoStop}%
\bibitem [{\citenamefont {Wang}\ and\ \citenamefont
  {You}(2022)}]{wang_symmetric_2022a}%
  \BibitemOpen
  \bibfield  {author} {\bibinfo {author} {\bibfnamefont {J.}~\bibnamefont
  {Wang}}\ and\ \bibinfo {author} {\bibfnamefont {Y.-Z.}\ \bibnamefont {You}},\
  }\bibfield  {title} {\bibinfo {title} {{Symmetric Mass Generation}},\ }\href
  {https://doi.org/10.3390/sym14071475} {\bibfield  {journal} {\bibinfo
  {journal} {Symmetry}\ }\textbf {\bibinfo {volume} {14}},\ \bibinfo {pages}
  {1475} (\bibinfo {year} {2022})},\ \Eprint {https://arxiv.org/abs/2204.14271}
  {arXiv:2204.14271 [cond-mat.str-el]} \BibitemShut {NoStop}%
\bibitem [{\citenamefont {Sulejmanpasic}\ and\ \citenamefont
  {Gattringer}(2019)}]{sulejmanpasic_abelian_2019}%
  \BibitemOpen
  \bibfield  {author} {\bibinfo {author} {\bibfnamefont {T.}~\bibnamefont
  {Sulejmanpasic}}\ and\ \bibinfo {author} {\bibfnamefont {C.}~\bibnamefont
  {Gattringer}},\ }\bibfield  {title} {\bibinfo {title} {{Abelian gauge
  theories on the lattice: $\theta$-Terms and compact gauge theory with(out)
  monopoles}},\ }\href {https://doi.org/10.1016/j.nuclphysb.2019.114616}
  {\bibfield  {journal} {\bibinfo  {journal} {Nucl. Phys. B}\ }\textbf
  {\bibinfo {volume} {943}},\ \bibinfo {pages} {114616} (\bibinfo {year}
  {2019})},\ \Eprint {https://arxiv.org/abs/1901.02637} {arXiv:1901.02637
  [hep-lat]} \BibitemShut {NoStop}%
\bibitem [{\citenamefont {Cheng}\ and\ \citenamefont
  {Seiberg}(2023)}]{cheng_liebschultzmattis_2023}%
  \BibitemOpen
  \bibfield  {author} {\bibinfo {author} {\bibfnamefont {M.}~\bibnamefont
  {Cheng}}\ and\ \bibinfo {author} {\bibfnamefont {N.}~\bibnamefont
  {Seiberg}},\ }\bibfield  {title} {\bibinfo {title} {{Lieb-Schultz-Mattis,
  Luttinger, and 't Hooft - anomaly matching in lattice systems}},\ }\href
  {https://doi.org/10.21468/SciPostPhys.15.2.051} {\bibfield  {journal}
  {\bibinfo  {journal} {SciPost Phys.}\ }\textbf {\bibinfo {volume} {15}},\
  \bibinfo {pages} {051} (\bibinfo {year} {2023})},\ \Eprint
  {https://arxiv.org/abs/2211.12543} {arXiv:2211.12543 [cond-mat.str-el]}
  \BibitemShut {NoStop}%
\bibitem [{\citenamefont {Fazza}\ and\ \citenamefont
  {Sulejmanpasic}(2023)}]{fazza_lattice_2023}%
  \BibitemOpen
  \bibfield  {author} {\bibinfo {author} {\bibfnamefont {L.}~\bibnamefont
  {Fazza}}\ and\ \bibinfo {author} {\bibfnamefont {T.}~\bibnamefont
  {Sulejmanpasic}},\ }\bibfield  {title} {\bibinfo {title} {Lattice quantum
  {{Villain Hamiltonians}}: Compact scalars, {{U}}(1) gauge theories, fracton
  models and quantum {{Ising}} model dualities},\ }\href
  {https://doi.org/10.1007/JHEP05(2023)017} {\bibfield  {journal} {\bibinfo
  {journal} {J. High Energ. Phys.}\ }\textbf {\bibinfo {volume}
  {2023}}\bibfield  {number} {\bibinfo  {number} { (5)},\ \bibinfo {pages}
  {17}},\ }\Eprint {https://arxiv.org/abs/2211.13047} {arXiv:2211.13047
  [hep-th]} \BibitemShut {NoStop}%
\bibitem [{\citenamefont {Gorantla}\ \emph {et~al.}(2021)\citenamefont
  {Gorantla}, \citenamefont {Lam}, \citenamefont {Seiberg},\ and\ \citenamefont
  {Shao}}]{gorantla_modified_2021a}%
  \BibitemOpen
  \bibfield  {author} {\bibinfo {author} {\bibfnamefont {P.}~\bibnamefont
  {Gorantla}}, \bibinfo {author} {\bibfnamefont {H.~T.}\ \bibnamefont {Lam}},
  \bibinfo {author} {\bibfnamefont {N.}~\bibnamefont {Seiberg}},\ and\ \bibinfo
  {author} {\bibfnamefont {S.-H.}\ \bibnamefont {Shao}},\ }\bibfield  {title}
  {\bibinfo {title} {{A modified Villain formulation of fractons and other
  exotic theories}},\ }\href {https://doi.org/10.1063/5.0060808} {\bibfield
  {journal} {\bibinfo  {journal} {J. Math. Phys.}\ }\textbf {\bibinfo {volume}
  {62}},\ \bibinfo {pages} {102301} (\bibinfo {year} {2021})},\ \Eprint
  {https://arxiv.org/abs/2103.01257} {arXiv:2103.01257 [cond-mat.str-el]}
  \BibitemShut {NoStop}%
\bibitem [{\citenamefont {Chandrasekharan}\ and\ \citenamefont
  {Wiese}(1997)}]{chandrasekharan_quantum_1997a}%
  \BibitemOpen
  \bibfield  {author} {\bibinfo {author} {\bibfnamefont {S.}~\bibnamefont
  {Chandrasekharan}}\ and\ \bibinfo {author} {\bibfnamefont {U.~J.}\
  \bibnamefont {Wiese}},\ }\bibfield  {title} {\bibinfo {title} {Quantum link
  models: {{A}} discrete approach to gauge theories},\ }\href
  {https://doi.org/10.1016/S0550-3213(97)80041-7} {\bibfield  {journal}
  {\bibinfo  {journal} {Nuclear Physics B}\ }\textbf {\bibinfo {volume}
  {492}},\ \bibinfo {pages} {455} (\bibinfo {year} {1997})},\ \Eprint
  {https://arxiv.org/abs/hep-lat/9609042} {arXiv:hep-lat/9609042 [hep-lat]}
  \BibitemShut {NoStop}%
\bibitem [{\citenamefont {Brower}\ \emph {et~al.}(1999)\citenamefont {Brower},
  \citenamefont {Chandrasekharan},\ and\ \citenamefont
  {Wiese}}]{brower_qcd_1999}%
  \BibitemOpen
  \bibfield  {author} {\bibinfo {author} {\bibfnamefont {R.}~\bibnamefont
  {Brower}}, \bibinfo {author} {\bibfnamefont {S.}~\bibnamefont
  {Chandrasekharan}},\ and\ \bibinfo {author} {\bibfnamefont {U.~J.}\
  \bibnamefont {Wiese}},\ }\bibfield  {title} {\bibinfo {title} {{{QCD}} as a
  quantum link model},\ }\href {https://doi.org/10.1103/PhysRevD.60.094502}
  {\bibfield  {journal} {\bibinfo  {journal} {Phys.Rev.}\ }\textbf {\bibinfo
  {volume} {D60}},\ \bibinfo {pages} {094502} (\bibinfo {year} {1999})},\
  \Eprint {https://arxiv.org/abs/hep-th/9704106} {arXiv:hep-th/9704106
  [hep-th]} \BibitemShut {NoStop}%
\bibitem [{\citenamefont {Brower}\ \emph {et~al.}(2004)\citenamefont {Brower},
  \citenamefont {Chandrasekharan}, \citenamefont {Riederer},\ and\
  \citenamefont {Wiese}}]{brower_dtheory_2004}%
  \BibitemOpen
  \bibfield  {author} {\bibinfo {author} {\bibfnamefont {R.}~\bibnamefont
  {Brower}}, \bibinfo {author} {\bibfnamefont {S.}~\bibnamefont
  {Chandrasekharan}}, \bibinfo {author} {\bibfnamefont {S.}~\bibnamefont
  {Riederer}},\ and\ \bibinfo {author} {\bibfnamefont {U.~J.}\ \bibnamefont
  {Wiese}},\ }\bibfield  {title} {\bibinfo {title} {D-theory: Field
  quantization by dimensional reduction of discrete variables},\ }\href
  {https://doi.org/10.1016/j.nuclphysb.2004.06.007} {\bibfield  {journal}
  {\bibinfo  {journal} {Nuclear Physics B}\ }\textbf {\bibinfo {volume}
  {693}},\ \bibinfo {pages} {149} (\bibinfo {year} {2004})},\ \Eprint
  {https://arxiv.org/abs/hep-lat/0309182} {arXiv:hep-lat/0309182 [hep-lat]}
  \BibitemShut {NoStop}%
\bibitem [{\citenamefont {Wiese}(2021)}]{Wiese:2021djl}%
  \BibitemOpen
  \bibfield  {author} {\bibinfo {author} {\bibfnamefont {U.-J.}\ \bibnamefont
  {Wiese}},\ }\bibfield  {title} {\bibinfo {title} {{From quantum link models
  to D-theory: a resource efficient framework for the quantum simulation and
  computation of gauge theories}},\ }\href
  {https://doi.org/10.1098/rsta.2021.0068} {\bibfield  {journal} {\bibinfo
  {journal} {Phil. Trans. A. Math. Phys. Eng. Sci.}\ }\textbf {\bibinfo
  {volume} {380}},\ \bibinfo {pages} {20210068} (\bibinfo {year} {2021})},\
  \Eprint {https://arxiv.org/abs/2107.09335} {arXiv:2107.09335 [hep-lat]}
  \BibitemShut {NoStop}%
\bibitem [{\citenamefont {Zache}\ \emph {et~al.}(2022)\citenamefont {Zache},
  \citenamefont {Van~Damme}, \citenamefont {Halimeh}, \citenamefont {Hauke},\
  and\ \citenamefont {Banerjee}}]{Zache:2021ggw}%
  \BibitemOpen
  \bibfield  {author} {\bibinfo {author} {\bibfnamefont {T.~V.}\ \bibnamefont
  {Zache}}, \bibinfo {author} {\bibfnamefont {M.}~\bibnamefont {Van~Damme}},
  \bibinfo {author} {\bibfnamefont {J.~C.}\ \bibnamefont {Halimeh}}, \bibinfo
  {author} {\bibfnamefont {P.}~\bibnamefont {Hauke}},\ and\ \bibinfo {author}
  {\bibfnamefont {D.}~\bibnamefont {Banerjee}},\ }\bibfield  {title} {\bibinfo
  {title} {{Toward the continuum limit of a (1+1)D quantum link Schwinger
  model}},\ }\href {https://doi.org/10.1103/PhysRevD.106.L091502} {\bibfield
  {journal} {\bibinfo  {journal} {Phys. Rev. D}\ }\textbf {\bibinfo {volume}
  {106}},\ \bibinfo {pages} {L091502} (\bibinfo {year} {2022})},\ \Eprint
  {https://arxiv.org/abs/2104.00025} {arXiv:2104.00025 [hep-lat]} \BibitemShut
  {NoStop}%
\bibitem [{\citenamefont {Singh}\ and\ \citenamefont
  {Chandrasekharan}(2019)}]{Singh:2019uwd}%
  \BibitemOpen
  \bibfield  {author} {\bibinfo {author} {\bibfnamefont {H.}~\bibnamefont
  {Singh}}\ and\ \bibinfo {author} {\bibfnamefont {S.}~\bibnamefont
  {Chandrasekharan}},\ }\bibfield  {title} {\bibinfo {title} {{Qubit
  regularization of the $O(3)$ sigma model}},\ }\href
  {https://doi.org/10.1103/PhysRevD.100.054505} {\bibfield  {journal} {\bibinfo
   {journal} {Phys. Rev. D}\ }\textbf {\bibinfo {volume} {100}},\ \bibinfo
  {pages} {054505} (\bibinfo {year} {2019})},\ \Eprint
  {https://arxiv.org/abs/1905.13204} {arXiv:1905.13204 [hep-lat]} \BibitemShut
  {NoStop}%
\bibitem [{\citenamefont {Bhattacharya}\ \emph {et~al.}(2021)\citenamefont
  {Bhattacharya}, \citenamefont {Buser}, \citenamefont {Chandrasekharan},
  \citenamefont {Gupta},\ and\ \citenamefont {Singh}}]{Bhattacharya:2020gpm}%
  \BibitemOpen
  \bibfield  {author} {\bibinfo {author} {\bibfnamefont {T.}~\bibnamefont
  {Bhattacharya}}, \bibinfo {author} {\bibfnamefont {A.~J.}\ \bibnamefont
  {Buser}}, \bibinfo {author} {\bibfnamefont {S.}~\bibnamefont
  {Chandrasekharan}}, \bibinfo {author} {\bibfnamefont {R.}~\bibnamefont
  {Gupta}},\ and\ \bibinfo {author} {\bibfnamefont {H.}~\bibnamefont {Singh}},\
  }\bibfield  {title} {\bibinfo {title} {{Qubit regularization of asymptotic
  freedom}},\ }\href {https://doi.org/10.1103/PhysRevLett.126.172001}
  {\bibfield  {journal} {\bibinfo  {journal} {Phys. Rev. Lett.}\ }\textbf
  {\bibinfo {volume} {126}},\ \bibinfo {pages} {172001} (\bibinfo {year}
  {2021})},\ \Eprint {https://arxiv.org/abs/2012.02153} {arXiv:2012.02153
  [hep-lat]} \BibitemShut {NoStop}%
\bibitem [{\citenamefont {Liu}\ and\ \citenamefont
  {Chandrasekharan}(2022)}]{Liu:2021tef}%
  \BibitemOpen
  \bibfield  {author} {\bibinfo {author} {\bibfnamefont {H.}~\bibnamefont
  {Liu}}\ and\ \bibinfo {author} {\bibfnamefont {S.}~\bibnamefont
  {Chandrasekharan}},\ }\bibfield  {title} {\bibinfo {title} {{Qubit
  Regularization and Qubit Embedding Algebras}},\ }\href
  {https://doi.org/10.3390/sym14020305} {\bibfield  {journal} {\bibinfo
  {journal} {Symmetry}\ }\textbf {\bibinfo {volume} {14}},\ \bibinfo {pages}
  {305} (\bibinfo {year} {2022})},\ \Eprint {https://arxiv.org/abs/2112.02090}
  {arXiv:2112.02090 [hep-lat]} \BibitemShut {NoStop}%
\bibitem [{\citenamefont {Caspar}\ and\ \citenamefont
  {Singh}(2022)}]{Caspar:2022llo}%
  \BibitemOpen
  \bibfield  {author} {\bibinfo {author} {\bibfnamefont {S.}~\bibnamefont
  {Caspar}}\ and\ \bibinfo {author} {\bibfnamefont {H.}~\bibnamefont {Singh}},\
  }\bibfield  {title} {\bibinfo {title} {{From Asymptotic Freedom to
  {\ensuremath{\theta}} Vacua: Qubit Embeddings of the O(3) Nonlinear
  {\ensuremath{\sigma}} Model}},\ }\href
  {https://doi.org/10.1103/PhysRevLett.129.022003} {\bibfield  {journal}
  {\bibinfo  {journal} {Phys. Rev. Lett.}\ }\textbf {\bibinfo {volume} {129}},\
  \bibinfo {pages} {022003} (\bibinfo {year} {2022})},\ \Eprint
  {https://arxiv.org/abs/2203.15766} {arXiv:2203.15766 [hep-lat]} \BibitemShut
  {NoStop}%
\bibitem [{\citenamefont {Nguyen}\ and\ \citenamefont
  {Singh}(2023)}]{Nguyen:2022aaq}%
  \BibitemOpen
  \bibfield  {author} {\bibinfo {author} {\bibfnamefont {M.}~\bibnamefont
  {Nguyen}}\ and\ \bibinfo {author} {\bibfnamefont {H.}~\bibnamefont {Singh}},\
  }\bibfield  {title} {\bibinfo {title} {{Lattice regularizations of
  {\ensuremath{\theta}} vacua: Anomalies and qubit models}},\ }\href
  {https://doi.org/10.1103/PhysRevD.107.014507} {\bibfield  {journal} {\bibinfo
   {journal} {Phys. Rev. D}\ }\textbf {\bibinfo {volume} {107}},\ \bibinfo
  {pages} {014507} (\bibinfo {year} {2023})},\ \Eprint
  {https://arxiv.org/abs/2209.12630} {arXiv:2209.12630 [hep-lat]} \BibitemShut
  {NoStop}%
\bibitem [{\citenamefont {Liu}\ \emph {et~al.}(2025)\citenamefont {Liu},
  \citenamefont {Bhattacharya}, \citenamefont {Chandrasekharan},\ and\
  \citenamefont {Gupta}}]{Liu:2023lsr}%
  \BibitemOpen
  \bibfield  {author} {\bibinfo {author} {\bibfnamefont {H.}~\bibnamefont
  {Liu}}, \bibinfo {author} {\bibfnamefont {T.}~\bibnamefont {Bhattacharya}},
  \bibinfo {author} {\bibfnamefont {S.}~\bibnamefont {Chandrasekharan}},\ and\
  \bibinfo {author} {\bibfnamefont {R.}~\bibnamefont {Gupta}},\ }\bibfield
  {title} {\bibinfo {title} {{Phases of 2D massless QCD with qubit
  regularization}},\ }\href {https://doi.org/10.1103/PhysRevD.111.094511}
  {\bibfield  {journal} {\bibinfo  {journal} {Phys. Rev. D}\ }\textbf {\bibinfo
  {volume} {111}},\ \bibinfo {pages} {094511} (\bibinfo {year} {2025})},\
  \Eprint {https://arxiv.org/abs/2312.17734} {arXiv:2312.17734 [hep-lat]}
  \BibitemShut {NoStop}%
\bibitem [{\citenamefont {Chandrasekharan}(2025)}]{Chandrasekharan:2025pil}%
  \BibitemOpen
  \bibfield  {author} {\bibinfo {author} {\bibfnamefont {S.}~\bibnamefont
  {Chandrasekharan}},\ }\bibfield  {title} {\bibinfo {title} {{Qubit
  Regularization of Quantum Field Theories}},\ }\href
  {https://doi.org/10.22323/1.466.0001} {\bibfield  {journal} {\bibinfo
  {journal} {PoS}\ }\textbf {\bibinfo {volume} {LATTICE2024}},\ \bibinfo
  {pages} {001} (\bibinfo {year} {2025})},\ \Eprint
  {https://arxiv.org/abs/2502.05716} {arXiv:2502.05716 [hep-lat]} \BibitemShut
  {NoStop}%
\bibitem [{\citenamefont {Zauner-Stauber}\ \emph {et~al.}(2018)\citenamefont
  {Zauner-Stauber}, \citenamefont {Vanderstraeten}, \citenamefont {Fishman},
  \citenamefont {Verstraete},\ and\ \citenamefont
  {Haegeman}}]{Zauner-Stauber:2017eqw}%
  \BibitemOpen
  \bibfield  {author} {\bibinfo {author} {\bibfnamefont {V.}~\bibnamefont
  {Zauner-Stauber}}, \bibinfo {author} {\bibfnamefont {L.}~\bibnamefont
  {Vanderstraeten}}, \bibinfo {author} {\bibfnamefont {M.~T.}\ \bibnamefont
  {Fishman}}, \bibinfo {author} {\bibfnamefont {F.}~\bibnamefont
  {Verstraete}},\ and\ \bibinfo {author} {\bibfnamefont {J.}~\bibnamefont
  {Haegeman}},\ }\bibfield  {title} {\bibinfo {title} {{Variational
  optimization algorithms for uniform matrix product states}},\ }\href
  {https://doi.org/10.1103/PhysRevB.97.045145} {\bibfield  {journal} {\bibinfo
  {journal} {Phys. Rev. B}\ }\textbf {\bibinfo {volume} {97}},\ \bibinfo
  {pages} {045145} (\bibinfo {year} {2018})},\ \Eprint
  {https://arxiv.org/abs/1701.07035} {arXiv:1701.07035 [quant-ph]} \BibitemShut
  {NoStop}%
\bibitem [{\citenamefont {Haegeman}\ \emph {et~al.}(2012)\citenamefont
  {Haegeman}, \citenamefont {Pirvu}, \citenamefont {Weir}, \citenamefont
  {Cirac}, \citenamefont {Osborne}, \citenamefont {Verschelde},\ and\
  \citenamefont {Verstraete}}]{haegeman_variational_2012}%
  \BibitemOpen
  \bibfield  {author} {\bibinfo {author} {\bibfnamefont {J.}~\bibnamefont
  {Haegeman}}, \bibinfo {author} {\bibfnamefont {B.}~\bibnamefont {Pirvu}},
  \bibinfo {author} {\bibfnamefont {D.~J.}\ \bibnamefont {Weir}}, \bibinfo
  {author} {\bibfnamefont {J.~I.}\ \bibnamefont {Cirac}}, \bibinfo {author}
  {\bibfnamefont {T.~J.}\ \bibnamefont {Osborne}}, \bibinfo {author}
  {\bibfnamefont {H.}~\bibnamefont {Verschelde}},\ and\ \bibinfo {author}
  {\bibfnamefont {F.}~\bibnamefont {Verstraete}},\ }\bibfield  {title}
  {\bibinfo {title} {{Variational matrix product ansatz for dispersion
  relations}},\ }\href {https://doi.org/10.1103/PhysRevB.85.100408} {\bibfield
  {journal} {\bibinfo  {journal} {Phys. Rev. B}\ }\textbf {\bibinfo {volume}
  {85}},\ \bibinfo {pages} {100408} (\bibinfo {year} {2012})},\ \Eprint
  {https://arxiv.org/abs/1103.2286} {arXiv:1103.2286 [quant-ph]} \BibitemShut
  {NoStop}%
\bibitem [{\citenamefont {Vanderstraeten}\ \emph {et~al.}(2019)\citenamefont
  {Vanderstraeten}, \citenamefont {Haegeman},\ and\ \citenamefont
  {Verstraete}}]{vanderstraeten_tangentspace_2019}%
  \BibitemOpen
  \bibfield  {author} {\bibinfo {author} {\bibfnamefont {L.}~\bibnamefont
  {Vanderstraeten}}, \bibinfo {author} {\bibfnamefont {J.}~\bibnamefont
  {Haegeman}},\ and\ \bibinfo {author} {\bibfnamefont {F.}~\bibnamefont
  {Verstraete}},\ }\bibfield  {title} {\bibinfo {title} {{Tangent-space methods
  for uniform matrix product states}},\ }\href
  {https://doi.org/10.21468/SciPostPhysLectNotes.7} {\bibfield  {journal}
  {\bibinfo  {journal} {SciPost Phys. Lect. Notes}\ }\textbf {\bibinfo {volume}
  {7}},\ \bibinfo {pages} {1} (\bibinfo {year} {2019})},\ \Eprint
  {https://arxiv.org/abs/1810.07006} {arXiv:1810.07006 [cond-mat.str-el]}
  \BibitemShut {NoStop}%
\bibitem [{\citenamefont {Lu}\ and\ \citenamefont
  {Shao}(2026)}]{lu_fermionic_2026}%
  \BibitemOpen
  \bibfield  {author} {\bibinfo {author} {\bibfnamefont {Z.}~\bibnamefont
  {Lu}}\ and\ \bibinfo {author} {\bibfnamefont {S.-H.}\ \bibnamefont {Shao}},\
  }\href {https://arxiv.org/abs/2608.02728} {\bibinfo {title} {Fermionic
  {{Villain}} model with exact lattice chiral symmetries}} (\bibinfo {year}
  {2026}),\ \Eprint {https://arxiv.org/abs/2608.02728} {arXiv:2608.02728
  [hep-th]} \BibitemShut {NoStop}%
\bibitem [{\citenamefont {Calabrese}\ and\ \citenamefont
  {Cardy}(2009)}]{calabrese_entanglement_2009}%
  \BibitemOpen
  \bibfield  {author} {\bibinfo {author} {\bibfnamefont {P.}~\bibnamefont
  {Calabrese}}\ and\ \bibinfo {author} {\bibfnamefont {J.}~\bibnamefont
  {Cardy}},\ }\bibfield  {title} {\bibinfo {title} {{Entanglement entropy and
  conformal field theory}},\ }\href
  {https://doi.org/10.1088/1751-8113/42/50/504005} {\bibfield  {journal}
  {\bibinfo  {journal} {J. Phys. A}\ }\textbf {\bibinfo {volume} {42}},\
  \bibinfo {pages} {504005} (\bibinfo {year} {2009})},\ \Eprint
  {https://arxiv.org/abs/0905.4013} {arXiv:0905.4013 [cond-mat.stat-mech]}
  \BibitemShut {NoStop}%
\bibitem [{\citenamefont {Pollmann}\ \emph {et~al.}(2009)\citenamefont
  {Pollmann}, \citenamefont {Mukerjee}, \citenamefont {Turner},\ and\
  \citenamefont {Moore}}]{pollmann_theory_2009}%
  \BibitemOpen
  \bibfield  {author} {\bibinfo {author} {\bibfnamefont {F.}~\bibnamefont
  {Pollmann}}, \bibinfo {author} {\bibfnamefont {S.}~\bibnamefont {Mukerjee}},
  \bibinfo {author} {\bibfnamefont {A.~M.}\ \bibnamefont {Turner}},\ and\
  \bibinfo {author} {\bibfnamefont {J.~E.}\ \bibnamefont {Moore}},\ }\bibfield
  {title} {\bibinfo {title} {{Theory of Finite-Entanglement Scaling at
  One-Dimensional Quantum Critical Points}},\ }\href
  {https://doi.org/10.1103/PhysRevLett.102.255701} {\bibfield  {journal}
  {\bibinfo  {journal} {Phys. Rev. Lett.}\ }\textbf {\bibinfo {volume} {102}},\
  \bibinfo {pages} {255701} (\bibinfo {year} {2009})},\ \Eprint
  {https://arxiv.org/abs/0812.2903} {arXiv:0812.2903 [cond-mat.str-el]}
  \BibitemShut {NoStop}%
\bibitem [{\citenamefont {Tagliacozzo}\ \emph {et~al.}(2008)\citenamefont
  {Tagliacozzo}, \citenamefont {de~Oliveira}, \citenamefont {Iblisdir},\ and\
  \citenamefont {Latorre}}]{tagliacozzo_scaling_2008}%
  \BibitemOpen
  \bibfield  {author} {\bibinfo {author} {\bibfnamefont {L.}~\bibnamefont
  {Tagliacozzo}}, \bibinfo {author} {\bibfnamefont {T.~R.}\ \bibnamefont
  {de~Oliveira}}, \bibinfo {author} {\bibfnamefont {S.}~\bibnamefont
  {Iblisdir}},\ and\ \bibinfo {author} {\bibfnamefont {J.~I.}\ \bibnamefont
  {Latorre}},\ }\bibfield  {title} {\bibinfo {title} {{Scaling of entanglement
  support for Matrix Product States}},\ }\href
  {https://doi.org/10.1103/PhysRevB.78.024410} {\bibfield  {journal} {\bibinfo
  {journal} {Phys. Rev. B}\ }\textbf {\bibinfo {volume} {78}},\ \bibinfo
  {pages} {024410} (\bibinfo {year} {2008})},\ \Eprint
  {https://arxiv.org/abs/0712.1976} {arXiv:0712.1976 [cond-mat.stat-mech]}
  \BibitemShut {NoStop}%
\bibitem [{\citenamefont {Kikukawa}\ \emph {et~al.}(1998)\citenamefont
  {Kikukawa}, \citenamefont {Narayanan},\ and\ \citenamefont
  {Neuberger}}]{kikukawa_monte_1998}%
  \BibitemOpen
  \bibfield  {author} {\bibinfo {author} {\bibfnamefont {Y.}~\bibnamefont
  {Kikukawa}}, \bibinfo {author} {\bibfnamefont {R.}~\bibnamefont
  {Narayanan}},\ and\ \bibinfo {author} {\bibfnamefont {H.}~\bibnamefont
  {Neuberger}},\ }\bibfield  {title} {\bibinfo {title} {Monte {{Carlo}}
  evaluation of a fermion number violating observable in {{2D}}},\ }\href
  {https://doi.org/10.1103/PhysRevD.57.1233} {\bibfield  {journal} {\bibinfo
  {journal} {Phys. Rev. D}\ }\textbf {\bibinfo {volume} {57}},\ \bibinfo
  {pages} {1233} (\bibinfo {year} {1998})},\ \Eprint
  {https://arxiv.org/abs/hep-lat/9705006} {arXiv:hep-lat/9705006} \BibitemShut
  {NoStop}%
\bibitem [{\citenamefont {Araki}\ \emph {et~al.}(2026)\citenamefont {Araki},
  \citenamefont {Fukaya}, \citenamefont {Onogi},\ and\ \citenamefont
  {Yamaguchi}}]{araki_symmetric_2026}%
  \BibitemOpen
  \bibfield  {author} {\bibinfo {author} {\bibfnamefont {S.}~\bibnamefont
  {Araki}}, \bibinfo {author} {\bibfnamefont {H.}~\bibnamefont {Fukaya}},
  \bibinfo {author} {\bibfnamefont {T.}~\bibnamefont {Onogi}},\ and\ \bibinfo
  {author} {\bibfnamefont {S.}~\bibnamefont {Yamaguchi}},\ }\href
  {https://doi.org/10.48550/arXiv.2608.29963} {\bibinfo {title} {Symmetric
  {{Mass Generation}} for {{Domain-Wall Fermions}}}} (\bibinfo {year} {2026}),\
  \Eprint {https://arxiv.org/abs/2608.29963} {arXiv:2608.29963 [hep-lat]}
  \BibitemShut {NoStop}%
\bibitem [{\citenamefont {Zakharov}\ \emph {et~al.}(2026)\citenamefont
  {Zakharov}, \citenamefont {Ueda}, \citenamefont {Verstraete},\ and\
  \citenamefont {Beenakker}}]{zakharov_symmetric_2026}%
  \BibitemOpen
  \bibfield  {author} {\bibinfo {author} {\bibfnamefont {V.~A.}\ \bibnamefont
  {Zakharov}}, \bibinfo {author} {\bibfnamefont {A.}~\bibnamefont {Ueda}},
  \bibinfo {author} {\bibfnamefont {F.}~\bibnamefont {Verstraete}},\ and\
  \bibinfo {author} {\bibfnamefont {C.~W.~J.}\ \bibnamefont {Beenakker}},\
  }\bibfield  {title} {\bibinfo {title} {Symmetric mass generation of
  interacting chiral fermions on a one-dimensional lattice without fermion
  doubling},\ }\href {https://doi.org/10.1103/2v9h-25qs} {\bibfield  {journal}
  {\bibinfo  {journal} {Phys. Rev. Res.}\ }\textbf {\bibinfo {volume} {8}},\
  \bibinfo {pages} {043009} (\bibinfo {year} {2026})},\ \Eprint
  {https://arxiv.org/abs/2606.24713} {arXiv:2606.24713 [hep-th]} \BibitemShut
  {NoStop}%
\bibitem [{\citenamefont {Golterman}\ and\ \citenamefont
  {Shamir}(2024)}]{golterman_propagator_2024}%
  \BibitemOpen
  \bibfield  {author} {\bibinfo {author} {\bibfnamefont {M.}~\bibnamefont
  {Golterman}}\ and\ \bibinfo {author} {\bibfnamefont {Y.}~\bibnamefont
  {Shamir}},\ }\bibfield  {title} {\bibinfo {title} {{Propagator Zeros and
  Lattice Chiral Gauge Theories}},\ }\href
  {https://doi.org/10.1103/PhysRevLett.132.081903} {\bibfield  {journal}
  {\bibinfo  {journal} {Phys. Rev. Lett.}\ }\textbf {\bibinfo {volume} {132}},\
  \bibinfo {pages} {081903} (\bibinfo {year} {2024})},\ \Eprint
  {https://arxiv.org/abs/2311.12790} {arXiv:2311.12790 [hep-lat]} \BibitemShut
  {NoStop}%
\bibitem [{\citenamefont {Ba{\~n}uls}\ \emph {et~al.}(2020)\citenamefont
  {Ba{\~n}uls} \emph {et~al.}}]{banuls_simulating_2020}%
  \BibitemOpen
  \bibfield  {author} {\bibinfo {author} {\bibfnamefont {M.~C.}\ \bibnamefont
  {Ba{\~n}uls}} \emph {et~al.},\ }\bibfield  {title} {\bibinfo {title}
  {{Simulating Lattice Gauge Theories within Quantum Technologies}},\ }\href
  {https://doi.org/10.1140/epjd/e2020-100571-8} {\bibfield  {journal} {\bibinfo
   {journal} {Eur. Phys. J. D}\ }\textbf {\bibinfo {volume} {74}},\ \bibinfo
  {pages} {165} (\bibinfo {year} {2020})},\ \Eprint
  {https://arxiv.org/abs/1911.00003} {arXiv:1911.00003 [quant-ph]} \BibitemShut
  {NoStop}%
\bibitem [{\citenamefont {Bauer}\ \emph
  {et~al.}(2023{\natexlab{a}})\citenamefont {Bauer} \emph
  {et~al.}}]{bauer_quantum_2023}%
  \BibitemOpen
  \bibfield  {author} {\bibinfo {author} {\bibfnamefont {C.~W.}\ \bibnamefont
  {Bauer}} \emph {et~al.},\ }\bibfield  {title} {\bibinfo {title} {{Quantum
  Simulation for High-Energy Physics}},\ }\href
  {https://doi.org/10.1103/PRXQuantum.4.027001} {\bibfield  {journal} {\bibinfo
   {journal} {PRX Quantum}\ }\textbf {\bibinfo {volume} {4}},\ \bibinfo {pages}
  {027001} (\bibinfo {year} {2023}{\natexlab{a}})},\ \Eprint
  {https://arxiv.org/abs/2204.03381} {arXiv:2204.03381 [quant-ph]} \BibitemShut
  {NoStop}%
\bibitem [{\citenamefont {Bauer}\ \emph
  {et~al.}(2023{\natexlab{b}})\citenamefont {Bauer}, \citenamefont {Davoudi},
  \citenamefont {Klco},\ and\ \citenamefont {Savage}}]{Bauer:2023qgm}%
  \BibitemOpen
  \bibfield  {author} {\bibinfo {author} {\bibfnamefont {C.~W.}\ \bibnamefont
  {Bauer}}, \bibinfo {author} {\bibfnamefont {Z.}~\bibnamefont {Davoudi}},
  \bibinfo {author} {\bibfnamefont {N.}~\bibnamefont {Klco}},\ and\ \bibinfo
  {author} {\bibfnamefont {M.~J.}\ \bibnamefont {Savage}},\ }\bibfield  {title}
  {\bibinfo {title} {{Quantum simulation of fundamental particles and
  forces}},\ }\href {https://doi.org/10.1038/s42254-023-00599-8} {\bibfield
  {journal} {\bibinfo  {journal} {Nature Rev. Phys.}\ }\textbf {\bibinfo
  {volume} {5}},\ \bibinfo {pages} {420} (\bibinfo {year}
  {2023}{\natexlab{b}})},\ \Eprint {https://arxiv.org/abs/2404.06298}
  {arXiv:2404.06298 [hep-ph]} \BibitemShut {NoStop}%
\bibitem [{\citenamefont {Meth}\ \emph {et~al.}(2025)\citenamefont {Meth} \emph
  {et~al.}}]{meth_simulating_2025}%
  \BibitemOpen
  \bibfield  {author} {\bibinfo {author} {\bibfnamefont {M.}~\bibnamefont
  {Meth}} \emph {et~al.},\ }\bibfield  {title} {\bibinfo {title} {{Simulating
  two-dimensional lattice gauge theories on a qudit quantum computer}},\ }\href
  {https://doi.org/10.1038/s41567-025-02797-w} {\bibfield  {journal} {\bibinfo
  {journal} {Nature Phys.}\ }\textbf {\bibinfo {volume} {21}},\ \bibinfo
  {pages} {570} (\bibinfo {year} {2025})},\ \Eprint
  {https://arxiv.org/abs/2310.12110} {arXiv:2310.12110 [quant-ph]} \BibitemShut
  {NoStop}%
\bibitem [{\citenamefont {Zache}\ \emph {et~al.}(2023)\citenamefont {Zache},
  \citenamefont {Gonz{\'a}lez-Cuadra},\ and\ \citenamefont
  {Zoller}}]{zache_fermionqudit_2023}%
  \BibitemOpen
  \bibfield  {author} {\bibinfo {author} {\bibfnamefont {T.~V.}\ \bibnamefont
  {Zache}}, \bibinfo {author} {\bibfnamefont {D.}~\bibnamefont
  {Gonz{\'a}lez-Cuadra}},\ and\ \bibinfo {author} {\bibfnamefont
  {P.}~\bibnamefont {Zoller}},\ }\bibfield  {title} {\bibinfo {title}
  {{Fermion-qudit quantum processors for simulating lattice gauge theories with
  matter}},\ }\href {https://doi.org/10.22331/q-2023-10-16-1140} {\bibfield
  {journal} {\bibinfo  {journal} {Quantum}\ }\textbf {\bibinfo {volume} {7}},\
  \bibinfo {pages} {1140} (\bibinfo {year} {2023})},\ \Eprint
  {https://arxiv.org/abs/2303.08683} {arXiv:2303.08683 [quant-ph]} \BibitemShut
  {NoStop}%
\bibitem [{\citenamefont {Popov}\ \emph {et~al.}(2024)\citenamefont {Popov},
  \citenamefont {Meth}, \citenamefont {Lewenstein}, \citenamefont {Hauke},
  \citenamefont {Ringbauer}, \citenamefont {Zohar},\ and\ \citenamefont
  {Kasper}}]{popov_variational_2024}%
  \BibitemOpen
  \bibfield  {author} {\bibinfo {author} {\bibfnamefont {P.~P.}\ \bibnamefont
  {Popov}}, \bibinfo {author} {\bibfnamefont {M.}~\bibnamefont {Meth}},
  \bibinfo {author} {\bibfnamefont {M.}~\bibnamefont {Lewenstein}}, \bibinfo
  {author} {\bibfnamefont {P.}~\bibnamefont {Hauke}}, \bibinfo {author}
  {\bibfnamefont {M.}~\bibnamefont {Ringbauer}}, \bibinfo {author}
  {\bibfnamefont {E.}~\bibnamefont {Zohar}},\ and\ \bibinfo {author}
  {\bibfnamefont {V.}~\bibnamefont {Kasper}},\ }\bibfield  {title} {\bibinfo
  {title} {{Variational quantum simulation of U(1) lattice gauge theories with
  qudit systems}},\ }\href {https://doi.org/10.1103/PhysRevResearch.6.013202}
  {\bibfield  {journal} {\bibinfo  {journal} {Phys. Rev. Res.}\ }\textbf
  {\bibinfo {volume} {6}},\ \bibinfo {pages} {013202} (\bibinfo {year}
  {2024})},\ \Eprint {https://arxiv.org/abs/2307.15173} {arXiv:2307.15173
  [quant-ph]} \BibitemShut {NoStop}%
\bibitem [{\citenamefont {Gustafson}\ \emph {et~al.}(2022)\citenamefont
  {Gustafson}, \citenamefont {Lamm}, \citenamefont {Lovelace},\ and\
  \citenamefont {Musk}}]{Gustafson:2022xdt}%
  \BibitemOpen
  \bibfield  {author} {\bibinfo {author} {\bibfnamefont {E.~J.}\ \bibnamefont
  {Gustafson}}, \bibinfo {author} {\bibfnamefont {H.}~\bibnamefont {Lamm}},
  \bibinfo {author} {\bibfnamefont {F.}~\bibnamefont {Lovelace}},\ and\
  \bibinfo {author} {\bibfnamefont {D.}~\bibnamefont {Musk}},\ }\bibfield
  {title} {\bibinfo {title} {{Primitive quantum gates for an SU(2) discrete
  subgroup: Binary tetrahedral}},\ }\href
  {https://doi.org/10.1103/PhysRevD.106.114501} {\bibfield  {journal} {\bibinfo
   {journal} {Phys. Rev. D}\ }\textbf {\bibinfo {volume} {106}},\ \bibinfo
  {pages} {114501} (\bibinfo {year} {2022})},\ \Eprint
  {https://arxiv.org/abs/2208.12309} {arXiv:2208.12309 [quant-ph]} \BibitemShut
  {NoStop}%
\bibitem [{\citenamefont {K{\"u}rk{\c{c}}{\"u}o{\u{g}}lu}\ \emph
  {et~al.}(2026)\citenamefont {K{\"u}rk{\c{c}}{\"u}o{\u{g}}lu}, \citenamefont
  {Lamm},\ and\ \citenamefont {Maestri}}]{Kurkcuoglu:2024cfv}%
  \BibitemOpen
  \bibfield  {author} {\bibinfo {author} {\bibfnamefont {D.~M.}\ \bibnamefont
  {K{\"u}rk{\c{c}}{\"u}o{\u{g}}lu}}, \bibinfo {author} {\bibfnamefont
  {H.}~\bibnamefont {Lamm}},\ and\ \bibinfo {author} {\bibfnamefont
  {A.}~\bibnamefont {Maestri}},\ }\bibfield  {title} {\bibinfo {title} {{Qudit
  gate decomposition for lattice gauge theories on superconducting
  radio-frequency cavities}},\ }\href {https://doi.org/10.1103/bd87-g26f}
  {\bibfield  {journal} {\bibinfo  {journal} {Phys. Rev. A}\ }\textbf {\bibinfo
  {volume} {114}},\ \bibinfo {pages} {022460} (\bibinfo {year} {2026})},\
  \Eprint {https://arxiv.org/abs/2410.16414} {arXiv:2410.16414 [quant-ph]}
  \BibitemShut {NoStop}%
\bibitem [{\citenamefont {Lamm}\ \emph {et~al.}(2026)\citenamefont {Lamm},
  \citenamefont {Roggero}, \citenamefont {Singh},\ and\ \citenamefont
  {Spagnoli}}]{lamm_exact_2026}%
  \BibitemOpen
  \bibfield  {author} {\bibinfo {author} {\bibfnamefont {H.}~\bibnamefont
  {Lamm}}, \bibinfo {author} {\bibfnamefont {A.}~\bibnamefont {Roggero}},
  \bibinfo {author} {\bibfnamefont {H.}~\bibnamefont {Singh}},\ and\ \bibinfo
  {author} {\bibfnamefont {L.}~\bibnamefont {Spagnoli}},\ }\href@noop {}
  {\bibinfo {title} {{Exact chiral symmetry with quantum signal processing}}}
  (\bibinfo {year} {2026}),\ \Eprint {https://arxiv.org/abs/2607.28524}
  {arXiv:2607.28524 [hep-lat]} \BibitemShut {NoStop}%
\bibitem [{\citenamefont {Grigoriev}\ \emph {et~al.}(1992)\citenamefont
  {Grigoriev}, \citenamefont {Shaposhnikov},\ and\ \citenamefont
  {Turok}}]{Grigoriev:1992nv}%
  \BibitemOpen
  \bibfield  {author} {\bibinfo {author} {\bibfnamefont {D.~Y.}\ \bibnamefont
  {Grigoriev}}, \bibinfo {author} {\bibfnamefont {M.~E.}\ \bibnamefont
  {Shaposhnikov}},\ and\ \bibinfo {author} {\bibfnamefont {N.}~\bibnamefont
  {Turok}},\ }\bibfield  {title} {\bibinfo {title} {{Electroweak baryogenesis:
  A Numerical study in (1+1)-dimensions}},\ }\href
  {https://doi.org/10.1016/0370-2693(92)91608-C} {\bibfield  {journal}
  {\bibinfo  {journal} {Phys. Lett. B}\ }\textbf {\bibinfo {volume} {275}},\
  \bibinfo {pages} {395} (\bibinfo {year} {1992})}\BibitemShut {NoStop}%
\bibitem [{\citenamefont {Aarts}\ and\ \citenamefont
  {Smit}(1999)}]{Aarts:1998td}%
  \BibitemOpen
  \bibfield  {author} {\bibinfo {author} {\bibfnamefont {G.}~\bibnamefont
  {Aarts}}\ and\ \bibinfo {author} {\bibfnamefont {J.}~\bibnamefont {Smit}},\
  }\bibfield  {title} {\bibinfo {title} {{Real time dynamics with fermions on a
  lattice}},\ }\href {https://doi.org/10.1016/S0550-3213(99)00320-X} {\bibfield
   {journal} {\bibinfo  {journal} {Nucl. Phys. B}\ }\textbf {\bibinfo {volume}
  {555}},\ \bibinfo {pages} {355} (\bibinfo {year} {1999})},\ \Eprint
  {https://arxiv.org/abs/hep-ph/9812413} {arXiv:hep-ph/9812413} \BibitemShut
  {NoStop}%
\bibitem [{\citenamefont {Saffin}\ and\ \citenamefont
  {Tranberg}(2011)}]{Saffin:2011kc}%
  \BibitemOpen
  \bibfield  {author} {\bibinfo {author} {\bibfnamefont {P.~M.}\ \bibnamefont
  {Saffin}}\ and\ \bibinfo {author} {\bibfnamefont {A.}~\bibnamefont
  {Tranberg}},\ }\bibfield  {title} {\bibinfo {title} {{Real-time Fermions for
  Baryogenesis Simulations}},\ }\href {https://doi.org/10.1007/JHEP07(2011)066}
  {\bibfield  {journal} {\bibinfo  {journal} {JHEP}\ }\textbf {\bibinfo
  {volume} {07}},\ \bibinfo {pages} {066}},\ \Eprint
  {https://arxiv.org/abs/1105.5546} {arXiv:1105.5546 [hep-ph]} \BibitemShut
  {NoStop}%
\bibitem [{\citenamefont {Carena}\ \emph {et~al.}(2026)\citenamefont {Carena},
  \citenamefont {Li}, \citenamefont {Ou},\ and\ \citenamefont
  {Singh}}]{carena_realtime_2025}%
  \BibitemOpen
  \bibfield  {author} {\bibinfo {author} {\bibfnamefont {M.}~\bibnamefont
  {Carena}}, \bibinfo {author} {\bibfnamefont {Y.-Y.}\ \bibnamefont {Li}},
  \bibinfo {author} {\bibfnamefont {T.}~\bibnamefont {Ou}},\ and\ \bibinfo
  {author} {\bibfnamefont {H.}~\bibnamefont {Singh}},\ }\bibfield  {title}
  {\bibinfo {title} {Real-{{Time Simulation}} of {{Asymmetry Generation}} in
  {{Fermion-Bubble Collisions}}},\ }\href {https://doi.org/10.1103/nphy-2y8q}
  {\bibfield  {journal} {\bibinfo  {journal} {Phys. Rev. D}\ }\textbf {\bibinfo
  {volume} {113}},\ \bibinfo {pages} {014502} (\bibinfo {year} {2026})},\
  \Eprint {https://arxiv.org/abs/2412.10365} {arXiv:2412.10365 [hep-ph]}
  \BibitemShut {NoStop}%
\bibitem [{\citenamefont {Baig}\ \emph {et~al.}(2026)\citenamefont {Baig},
  \citenamefont {Chen}, \citenamefont {Cherman},\ and\ \citenamefont
  {Neuzil}}]{baig_bosonization_2026a}%
  \BibitemOpen
  \bibfield  {author} {\bibinfo {author} {\bibfnamefont {S.~U.}\ \bibnamefont
  {Baig}}, \bibinfo {author} {\bibfnamefont {S.}~\bibnamefont {Chen}}, \bibinfo
  {author} {\bibfnamefont {A.}~\bibnamefont {Cherman}},\ and\ \bibinfo {author}
  {\bibfnamefont {M.}~\bibnamefont {Neuzil}},\ }\href
  {https://doi.org/10.48550/arXiv.2607.09935} {\bibinfo {title} {Bosonization
  versus the {{Nielsen-Ninomiya}} theorem}} (\bibinfo {year} {2026}),\ \Eprint
  {https://arxiv.org/abs/2607.09935} {arXiv:2607.09935 [hep-th]} \BibitemShut
  {NoStop}%
\bibitem [{\citenamefont {Pedersen}\ and\ \citenamefont
  {Kikukawa}(2023)}]{pedersen_reformulation_2023}%
  \BibitemOpen
  \bibfield  {author} {\bibinfo {author} {\bibfnamefont {J.~W.}\ \bibnamefont
  {Pedersen}}\ and\ \bibinfo {author} {\bibfnamefont {Y.}~\bibnamefont
  {Kikukawa}},\ }\bibfield  {title} {\bibinfo {title} {{Reformulation of
  anomaly inflow on the lattice and construction of lattice chiral gauge
  theories}},\ }\href {https://doi.org/10.22323/1.430.0381} {\bibfield
  {journal} {\bibinfo  {journal} {PoS}\ }\textbf {\bibinfo {volume}
  {LATTICE2022}},\ \bibinfo {pages} {381} (\bibinfo {year} {2023})}\BibitemShut
  {NoStop}%
\bibitem [{\citenamefont {Dang}\ \emph {et~al.}(2026)\citenamefont {Dang},
  \citenamefont {Karur},\ and\ \citenamefont {Sen}}]{dang_gauge_2026}%
  \BibitemOpen
  \bibfield  {author} {\bibinfo {author} {\bibfnamefont {J.}~\bibnamefont
  {Dang}}, \bibinfo {author} {\bibfnamefont {R.}~\bibnamefont {Karur}},\ and\
  \bibinfo {author} {\bibfnamefont {S.}~\bibnamefont {Sen}},\ }\href@noop {}
  {\bibinfo {title} {{Gauge field flow for chiral gauge theories on a disk
  boundary}}} (\bibinfo {year} {2026}),\ \Eprint
  {https://arxiv.org/abs/2606.05305} {arXiv:2606.05305 [hep-lat]} \BibitemShut
  {NoStop}%
\bibitem [{\citenamefont {Kaplan}\ and\ \citenamefont
  {Sen}(2026)}]{kaplan_regulating_2024}%
  \BibitemOpen
  \bibfield  {author} {\bibinfo {author} {\bibfnamefont {D.~B.}\ \bibnamefont
  {Kaplan}}\ and\ \bibinfo {author} {\bibfnamefont {S.}~\bibnamefont {Sen}},\
  }\bibfield  {title} {\bibinfo {title} {{Regulated chiral gauge theory and the
  strong $CP$ problem}},\ }\href {https://doi.org/10.1103/wld4-f12c} {\bibfield
   {journal} {\bibinfo  {journal} {Phys. Rev. D}\ }\textbf {\bibinfo {volume}
  {114}},\ \bibinfo {pages} {014508} (\bibinfo {year} {2026})},\ \Eprint
  {https://arxiv.org/abs/2412.02024} {arXiv:2412.02024 [hep-lat]} \BibitemShut
  {NoStop}%
\bibitem [{\citenamefont {Kaplan}\ and\ \citenamefont
  {Sen}(2024)}]{kaplan_weyl_2024}%
  \BibitemOpen
  \bibfield  {author} {\bibinfo {author} {\bibfnamefont {D.~B.}\ \bibnamefont
  {Kaplan}}\ and\ \bibinfo {author} {\bibfnamefont {S.}~\bibnamefont {Sen}},\
  }\bibfield  {title} {\bibinfo {title} {{Weyl Fermions on a Finite Lattice}},\
  }\href {https://doi.org/10.1103/PhysRevLett.132.141604} {\bibfield  {journal}
  {\bibinfo  {journal} {Phys. Rev. Lett.}\ }\textbf {\bibinfo {volume} {132}},\
  \bibinfo {pages} {141604} (\bibinfo {year} {2024})},\ \Eprint
  {https://arxiv.org/abs/2312.04012} {arXiv:2312.04012 [hep-lat]} \BibitemShut
  {NoStop}%
\bibitem [{\citenamefont {Sen}(2026)}]{sen_hamiltonian_2026}%
  \BibitemOpen
  \bibfield  {author} {\bibinfo {author} {\bibfnamefont {S.}~\bibnamefont
  {Sen}},\ }\href@noop {} {\bibinfo {title} {{Hamiltonian framework for Chiral
  Gauge Theories on a Disk Boundary}}} (\bibinfo {year} {2026}),\ \Eprint
  {https://arxiv.org/abs/2610.11035} {arXiv:2610.11035 [hep-lat]} \BibitemShut
  {NoStop}%
\bibitem [{\citenamefont {Eichten}\ and\ \citenamefont
  {Preskill}(1986)}]{eichten_chiral_1986}%
  \BibitemOpen
  \bibfield  {author} {\bibinfo {author} {\bibfnamefont {E.}~\bibnamefont
  {Eichten}}\ and\ \bibinfo {author} {\bibfnamefont {J.}~\bibnamefont
  {Preskill}},\ }\bibfield  {title} {\bibinfo {title} {{Chiral Gauge Theories
  on the Lattice}},\ }\href {https://doi.org/10.1016/0550-3213(86)90207-5}
  {\bibfield  {journal} {\bibinfo  {journal} {Nucl. Phys. B}\ }\textbf
  {\bibinfo {volume} {268}},\ \bibinfo {pages} {179} (\bibinfo {year}
  {1986})}\BibitemShut {NoStop}%
\bibitem [{\citenamefont {You}\ \emph {et~al.}(2014)\citenamefont {You},
  \citenamefont {BenTov},\ and\ \citenamefont {Xu}}]{you_interacting_2014}%
  \BibitemOpen
  \bibfield  {author} {\bibinfo {author} {\bibfnamefont {Y.}~\bibnamefont
  {You}}, \bibinfo {author} {\bibfnamefont {Y.}~\bibnamefont {BenTov}},\ and\
  \bibinfo {author} {\bibfnamefont {C.}~\bibnamefont {Xu}},\ }\href@noop {}
  {\bibinfo {title} {{Interacting Topological Superconductors and possible
  Origin of $16n$ Chiral Fermions in the Standard Model}}} (\bibinfo {year}
  {2014}),\ \Eprint {https://arxiv.org/abs/1402.4151} {arXiv:1402.4151
  [cond-mat.str-el]} \BibitemShut {NoStop}%
\bibitem [{\citenamefont {You}\ and\ \citenamefont
  {Xu}(2015)}]{you_interacting_2015}%
  \BibitemOpen
  \bibfield  {author} {\bibinfo {author} {\bibfnamefont {Y.-Z.}\ \bibnamefont
  {You}}\ and\ \bibinfo {author} {\bibfnamefont {C.}~\bibnamefont {Xu}},\
  }\bibfield  {title} {\bibinfo {title} {{Interacting Topological Insulator and
  Emergent Grand Unified Theory}},\ }\href
  {https://doi.org/10.1103/PhysRevB.91.125147} {\bibfield  {journal} {\bibinfo
  {journal} {Phys. Rev. B}\ }\textbf {\bibinfo {volume} {91}},\ \bibinfo
  {pages} {125147} (\bibinfo {year} {2015})},\ \Eprint
  {https://arxiv.org/abs/1412.4784} {arXiv:1412.4784 [cond-mat.str-el]}
  \BibitemShut {NoStop}%
\bibitem [{\citenamefont {Catterall}(2024)}]{catterall_lattice_2024}%
  \BibitemOpen
  \bibfield  {author} {\bibinfo {author} {\bibfnamefont {S.}~\bibnamefont
  {Catterall}},\ }\bibfield  {title} {\bibinfo {title} {{Lattice regularization
  of reduced K\"ahler-Dirac fermions and connections to chiral fermions}},\
  }\href {https://doi.org/10.21468/SciPostPhys.16.4.108} {\bibfield  {journal}
  {\bibinfo  {journal} {SciPost Phys.}\ }\textbf {\bibinfo {volume} {16}},\
  \bibinfo {pages} {108} (\bibinfo {year} {2024})},\ \Eprint
  {https://arxiv.org/abs/2311.02487} {arXiv:2311.02487 [hep-lat]} \BibitemShut
  {NoStop}%
\bibitem [{\citenamefont {Onoda}(2026)}]{onoda_lattice_2026}%
  \BibitemOpen
  \bibfield  {author} {\bibinfo {author} {\bibfnamefont {S.}~\bibnamefont
  {Onoda}},\ }\href {https://doi.org/10.48550/arXiv.2606.12358} {\bibinfo
  {title} {Lattice chiral non-{{Abelian}} gauge symmetry via bosonization}}
  (\bibinfo {year} {2026}),\ \Eprint {https://arxiv.org/abs/2606.12358}
  {arXiv:2606.12358 [hep-lat]} \BibitemShut {NoStop}%
\bibitem [{\citenamefont {Berkowitz}\ \emph {et~al.}(2026)\citenamefont
  {Berkowitz}, \citenamefont {Chen},\ and\ \citenamefont
  {Cherman}}]{berkowitz_vectorvectoraxial_2026}%
  \BibitemOpen
  \bibfield  {author} {\bibinfo {author} {\bibfnamefont {E.}~\bibnamefont
  {Berkowitz}}, \bibinfo {author} {\bibfnamefont {S.}~\bibnamefont {Chen}},\
  and\ \bibinfo {author} {\bibfnamefont {A.}~\bibnamefont {Cherman}},\
  }\bibfield  {title} {\bibinfo {title} {{A Vector-Vector-Axial Anomaly in
  4D}},\ }in\ \href@noop {} {\emph {\bibinfo {booktitle} {{43rd International
  Symposium on Lattice Field Theory}}}}\ (\bibinfo {year} {2026})\ \Eprint
  {https://arxiv.org/abs/2609.17369} {arXiv:2609.17369 [hep-lat]} \BibitemShut
  {NoStop}%
\bibitem [{\citenamefont {Devos}\ \emph {et~al.}(2026)\citenamefont {Devos},
  \citenamefont {Van~Damme},\ and\ \citenamefont
  {Haegeman}}]{devos_mpskit_2026}%
  \BibitemOpen
  \bibfield  {author} {\bibinfo {author} {\bibfnamefont {L.}~\bibnamefont
  {Devos}}, \bibinfo {author} {\bibfnamefont {M.}~\bibnamefont {Van~Damme}},\
  and\ \bibinfo {author} {\bibfnamefont {J.}~\bibnamefont {Haegeman}},\ }\href
  {https://doi.org/10.5281/zenodo.10654900} {\bibinfo {title}
  {{{MPSKit}}.jl}},\ \bibinfo {howpublished}
  {\url{https://github.com/QuantumKitHub/MPSKit.jl}} (\bibinfo {year}
  {2026})\BibitemShut {NoStop}%
\bibitem [{\citenamefont {Devos}\ and\ \citenamefont
  {Haegeman}(2025)}]{Devos:2025yoj}%
  \BibitemOpen
  \bibfield  {author} {\bibinfo {author} {\bibfnamefont {L.}~\bibnamefont
  {Devos}}\ and\ \bibinfo {author} {\bibfnamefont {J.}~\bibnamefont
  {Haegeman}},\ }\href@noop {} {\bibinfo {title} {{TensorKit.jl: A Julia
  package for large-scale tensor computations, with a hint of category
  theory}}} (\bibinfo {year} {2025}),\ \Eprint
  {https://arxiv.org/abs/2508.10076} {arXiv:2508.10076 [cs.MS]} \BibitemShut
  {NoStop}%
\bibitem [{\citenamefont {Giveon}\ \emph {et~al.}(1994)\citenamefont {Giveon},
  \citenamefont {Porrati},\ and\ \citenamefont
  {Rabinovici}}]{giveon_target_1994}%
  \BibitemOpen
  \bibfield  {author} {\bibinfo {author} {\bibfnamefont {A.}~\bibnamefont
  {Giveon}}, \bibinfo {author} {\bibfnamefont {M.}~\bibnamefont {Porrati}},\
  and\ \bibinfo {author} {\bibfnamefont {E.}~\bibnamefont {Rabinovici}},\
  }\bibfield  {title} {\bibinfo {title} {{Target space duality in string
  theory}},\ }\href {https://doi.org/10.1016/0370-1573(94)90070-1} {\bibfield
  {journal} {\bibinfo  {journal} {Phys. Rept.}\ }\textbf {\bibinfo {volume}
  {244}},\ \bibinfo {pages} {77} (\bibinfo {year} {1994})},\ \Eprint
  {https://arxiv.org/abs/hep-th/9401139} {arXiv:hep-th/9401139} \BibitemShut
  {NoStop}%
\bibitem [{\citenamefont {Vancraeynest-De~Cuiper}\ \emph
  {et~al.}(2026)\citenamefont {Vancraeynest-De~Cuiper}, \citenamefont
  {Garre-Rubio}, \citenamefont {Verstraete}, \citenamefont {Vervoort},
  \citenamefont {Williamson},\ and\ \citenamefont
  {Lootens}}]{cuiper2025gauging}%
  \BibitemOpen
  \bibfield  {author} {\bibinfo {author} {\bibfnamefont {B.}~\bibnamefont
  {Vancraeynest-De~Cuiper}}, \bibinfo {author} {\bibfnamefont {J.}~\bibnamefont
  {Garre-Rubio}}, \bibinfo {author} {\bibfnamefont {F.}~\bibnamefont
  {Verstraete}}, \bibinfo {author} {\bibfnamefont {K.}~\bibnamefont
  {Vervoort}}, \bibinfo {author} {\bibfnamefont {D.~J.}\ \bibnamefont
  {Williamson}},\ and\ \bibinfo {author} {\bibfnamefont {L.}~\bibnamefont
  {Lootens}},\ }\bibfield  {title} {\bibinfo {title} {From gauging to duality
  in one-dimensional quantum lattice models},\ }\href
  {https://doi.org/10.21468/SciPostPhys.20.4.113} {\bibfield  {journal}
  {\bibinfo  {journal} {SciPost Phys.}\ }\textbf {\bibinfo {volume} {20}},\
  \bibinfo {pages} {113} (\bibinfo {year} {2026})},\ \Eprint
  {https://arxiv.org/abs/2509.22051} {arXiv:2509.22051 [cond-mat.str-el]}
  \BibitemShut {NoStop}%
\bibitem [{\citenamefont {Lootens}\ \emph {et~al.}(2023)\citenamefont
  {Lootens}, \citenamefont {Delcamp}, \citenamefont {Ortiz},\ and\
  \citenamefont {Verstraete}}]{lootens2023dualities}%
  \BibitemOpen
  \bibfield  {author} {\bibinfo {author} {\bibfnamefont {L.}~\bibnamefont
  {Lootens}}, \bibinfo {author} {\bibfnamefont {C.}~\bibnamefont {Delcamp}},
  \bibinfo {author} {\bibfnamefont {G.}~\bibnamefont {Ortiz}},\ and\ \bibinfo
  {author} {\bibfnamefont {F.}~\bibnamefont {Verstraete}},\ }\bibfield  {title}
  {\bibinfo {title} {Dualities in one-dimensional quantum lattice models:
  {Symmetric} hamiltonians and matrix product operator intertwiners},\ }\href
  {https://doi.org/10.1103/PRXQuantum.4.020357} {\bibfield  {journal} {\bibinfo
   {journal} {PRX Quantum}\ }\textbf {\bibinfo {volume} {4}},\ \bibinfo {pages}
  {020357} (\bibinfo {year} {2023})},\ \Eprint
  {https://arxiv.org/abs/2112.09091} {arXiv:2112.09091 [quant-ph]} \BibitemShut
  {NoStop}%
\bibitem [{\citenamefont {Lootens}\ \emph {et~al.}(2024)\citenamefont
  {Lootens}, \citenamefont {Delcamp},\ and\ \citenamefont
  {Verstraete}}]{lootens_dualities_2024}%
  \BibitemOpen
  \bibfield  {author} {\bibinfo {author} {\bibfnamefont {L.}~\bibnamefont
  {Lootens}}, \bibinfo {author} {\bibfnamefont {C.}~\bibnamefont {Delcamp}},\
  and\ \bibinfo {author} {\bibfnamefont {F.}~\bibnamefont {Verstraete}},\
  }\bibfield  {title} {\bibinfo {title} {{Dualities in One-Dimensional Quantum
  Lattice Models: Topological Sectors}},\ }\href
  {https://doi.org/10.1103/PRXQuantum.5.010338} {\bibfield  {journal} {\bibinfo
   {journal} {PRX Quantum}\ }\textbf {\bibinfo {volume} {5}},\ \bibinfo {pages}
  {010338} (\bibinfo {year} {2024})},\ \Eprint
  {https://arxiv.org/abs/2211.03777} {arXiv:2211.03777 [quant-ph]} \BibitemShut
  {NoStop}%
\bibitem [{\citenamefont {Ji}\ and\ \citenamefont {Wen}(2020)}]{Ji:2019jhk}%
  \BibitemOpen
  \bibfield  {author} {\bibinfo {author} {\bibfnamefont {W.}~\bibnamefont
  {Ji}}\ and\ \bibinfo {author} {\bibfnamefont {X.-G.}\ \bibnamefont {Wen}},\
  }\bibfield  {title} {\bibinfo {title} {{Categorical symmetry and
  noninvertible anomaly in symmetry-breaking and topological phase
  transitions}},\ }\href {https://doi.org/10.1103/PhysRevResearch.2.033417}
  {\bibfield  {journal} {\bibinfo  {journal} {Phys. Rev. Res.}\ }\textbf
  {\bibinfo {volume} {2}},\ \bibinfo {pages} {033417} (\bibinfo {year}
  {2020})},\ \Eprint {https://arxiv.org/abs/1912.13492} {arXiv:1912.13492
  [cond-mat.str-el]} \BibitemShut {NoStop}%
\bibitem [{\citenamefont {Kong}\ \emph {et~al.}(2020)\citenamefont {Kong},
  \citenamefont {Lan}, \citenamefont {Wen}, \citenamefont {Zhang},\ and\
  \citenamefont {Zheng}}]{Kong:2020cie}%
  \BibitemOpen
  \bibfield  {author} {\bibinfo {author} {\bibfnamefont {L.}~\bibnamefont
  {Kong}}, \bibinfo {author} {\bibfnamefont {T.}~\bibnamefont {Lan}}, \bibinfo
  {author} {\bibfnamefont {X.-G.}\ \bibnamefont {Wen}}, \bibinfo {author}
  {\bibfnamefont {Z.-H.}\ \bibnamefont {Zhang}},\ and\ \bibinfo {author}
  {\bibfnamefont {H.}~\bibnamefont {Zheng}},\ }\bibfield  {title} {\bibinfo
  {title} {{Algebraic higher symmetry and categorical symmetry -- a holographic
  and entanglement view of symmetry}},\ }\href
  {https://doi.org/10.1103/PhysRevResearch.2.043086} {\bibfield  {journal}
  {\bibinfo  {journal} {Phys. Rev. Res.}\ }\textbf {\bibinfo {volume} {2}},\
  \bibinfo {pages} {043086} (\bibinfo {year} {2020})},\ \Eprint
  {https://arxiv.org/abs/2005.14178} {arXiv:2005.14178 [cond-mat.str-el]}
  \BibitemShut {NoStop}%
\bibitem [{\citenamefont {Freed}\ \emph {et~al.}(2022)\citenamefont {Freed},
  \citenamefont {Moore},\ and\ \citenamefont {Teleman}}]{Freed:2022qnc}%
  \BibitemOpen
  \bibfield  {author} {\bibinfo {author} {\bibfnamefont {D.~S.}\ \bibnamefont
  {Freed}}, \bibinfo {author} {\bibfnamefont {G.~W.}\ \bibnamefont {Moore}},\
  and\ \bibinfo {author} {\bibfnamefont {C.}~\bibnamefont {Teleman}},\
  }\href@noop {} {\bibinfo {title} {{Topological symmetry in quantum field
  theory}}} (\bibinfo {year} {2022}),\ \Eprint
  {https://arxiv.org/abs/2209.07471} {arXiv:2209.07471 [hep-th]} \BibitemShut
  {NoStop}%
\bibitem [{\citenamefont {Kitaev}\ and\ \citenamefont
  {Kong}(2012)}]{kitaev_models_2012}%
  \BibitemOpen
  \bibfield  {author} {\bibinfo {author} {\bibfnamefont {A.}~\bibnamefont
  {Kitaev}}\ and\ \bibinfo {author} {\bibfnamefont {L.}~\bibnamefont {Kong}},\
  }\bibfield  {title} {\bibinfo {title} {Models for gapped boundaries and
  domain walls},\ }\href {https://doi.org/10.1007/s00220-012-1500-5} {\bibfield
   {journal} {\bibinfo  {journal} {Communications in Mathematical Physics}\
  }\textbf {\bibinfo {volume} {313}},\ \bibinfo {pages} {351} (\bibinfo {year}
  {2012})},\ \Eprint {https://arxiv.org/abs/1104.5047} {arXiv:1104.5047
  [cond-mat.str-el]} \BibitemShut {NoStop}%
\bibitem [{\citenamefont {Davydov}\ \emph {et~al.}(2010)\citenamefont
  {Davydov}, \citenamefont {Mueger}, \citenamefont {Nikshych},\ and\
  \citenamefont {Ostrik}}]{Davydov:2010kfz}%
  \BibitemOpen
  \bibfield  {author} {\bibinfo {author} {\bibfnamefont {A.}~\bibnamefont
  {Davydov}}, \bibinfo {author} {\bibfnamefont {M.}~\bibnamefont {Mueger}},
  \bibinfo {author} {\bibfnamefont {D.}~\bibnamefont {Nikshych}},\ and\
  \bibinfo {author} {\bibfnamefont {V.}~\bibnamefont {Ostrik}},\ }\href@noop {}
  {\bibinfo {title} {{The Witt group of non-degenerate braided fusion
  categories}}} (\bibinfo {year} {2010}),\ \Eprint
  {https://arxiv.org/abs/1009.2117} {arXiv:1009.2117 [math.QA]} \BibitemShut
  {NoStop}%
\bibitem [{\citenamefont {Kong}(2014)}]{Kong:2013aya}%
  \BibitemOpen
  \bibfield  {author} {\bibinfo {author} {\bibfnamefont {L.}~\bibnamefont
  {Kong}},\ }\bibfield  {title} {\bibinfo {title} {{Anyon condensation and
  tensor categories}},\ }\href
  {https://doi.org/10.1016/j.nuclphysb.2014.07.003} {\bibfield  {journal}
  {\bibinfo  {journal} {Nucl. Phys. B}\ }\textbf {\bibinfo {volume} {886}},\
  \bibinfo {pages} {436} (\bibinfo {year} {2014})},\ \Eprint
  {https://arxiv.org/abs/1307.8244} {arXiv:1307.8244 [cond-mat.str-el]}
  \BibitemShut {NoStop}%
\bibitem [{\citenamefont {Dijkgraaf}\ and\ \citenamefont
  {Witten}(1990)}]{Dijkgraaf:1989pz}%
  \BibitemOpen
  \bibfield  {author} {\bibinfo {author} {\bibfnamefont {R.}~\bibnamefont
  {Dijkgraaf}}\ and\ \bibinfo {author} {\bibfnamefont {E.}~\bibnamefont
  {Witten}},\ }\bibfield  {title} {\bibinfo {title} {{Topological Gauge
  Theories and Group Cohomology}},\ }\href {https://doi.org/10.1007/BF02096988}
  {\bibfield  {journal} {\bibinfo  {journal} {Commun. Math. Phys.}\ }\textbf
  {\bibinfo {volume} {129}},\ \bibinfo {pages} {393} (\bibinfo {year}
  {1990})}\BibitemShut {NoStop}%
\bibitem [{\citenamefont {de~Wild~Propitius}(1995)}]{deWildPropitius:1995cf}%
  \BibitemOpen
  \bibfield  {author} {\bibinfo {author} {\bibfnamefont {M.~D.~F.}\
  \bibnamefont {de~Wild~Propitius}},\ }\emph {\bibinfo {title} {{Topological
  interactions in broken gauge theories}}},\ \href@noop {} {Ph.D. thesis},\
  \bibinfo  {school} {Amsterdam U.} (\bibinfo {year} {1995}),\ \Eprint
  {https://arxiv.org/abs/hep-th/9511195} {arXiv:hep-th/9511195} \BibitemShut
  {NoStop}%
\bibitem [{\citenamefont {Bais}\ and\ \citenamefont
  {Slingerland}(2009)}]{Bais:2008ni}%
  \BibitemOpen
  \bibfield  {author} {\bibinfo {author} {\bibfnamefont {F.~A.}\ \bibnamefont
  {Bais}}\ and\ \bibinfo {author} {\bibfnamefont {J.~K.}\ \bibnamefont
  {Slingerland}},\ }\bibfield  {title} {\bibinfo {title} {{Condensate induced
  transitions between topologically ordered phases}},\ }\href
  {https://doi.org/10.1103/PhysRevB.79.045316} {\bibfield  {journal} {\bibinfo
  {journal} {Phys. Rev. B}\ }\textbf {\bibinfo {volume} {79}},\ \bibinfo
  {pages} {045316} (\bibinfo {year} {2009})},\ \Eprint
  {https://arxiv.org/abs/0808.0627} {arXiv:0808.0627 [cond-mat.mes-hall]}
  \BibitemShut {NoStop}%
\bibitem [{\citenamefont {Ostrik}(2002)}]{Ostrik:2002ohv}%
  \BibitemOpen
  \bibfield  {author} {\bibinfo {author} {\bibfnamefont {V.}~\bibnamefont
  {Ostrik}},\ }\href@noop {} {\bibinfo {title} {{Module categories over the
  Drinfeld double of a finite group}}} (\bibinfo {year} {2002}),\ \Eprint
  {https://arxiv.org/abs/math/0202130} {arXiv:math/0202130} \BibitemShut
  {NoStop}%
\bibitem [{\citenamefont {Naidu}(2007)}]{naidu2007morita}%
  \BibitemOpen
  \bibfield  {author} {\bibinfo {author} {\bibfnamefont {D.}~\bibnamefont
  {Naidu}},\ }\bibfield  {title} {\bibinfo {title} {{Categorical Morita
  equivalence for group-theoretical categories}},\ }\href@noop {} {\bibfield
  {journal} {\bibinfo  {journal} {Comm. Algebra}\ }\textbf {\bibinfo {volume}
  {35}},\ \bibinfo {pages} {3544} (\bibinfo {year} {2007})},\ \Eprint
  {https://arxiv.org/abs/math/0605530} {arXiv:math/0605530 [math.QA]}
  \BibitemShut {NoStop}%
\bibitem [{\citenamefont {Tachikawa}(2020)}]{Tachikawa:2017gyf}%
  \BibitemOpen
  \bibfield  {author} {\bibinfo {author} {\bibfnamefont {Y.}~\bibnamefont
  {Tachikawa}},\ }\bibfield  {title} {\bibinfo {title} {{On gauging finite
  subgroups}},\ }\href {https://doi.org/10.21468/SciPostPhys.8.1.015}
  {\bibfield  {journal} {\bibinfo  {journal} {SciPost Phys.}\ }\textbf
  {\bibinfo {volume} {8}},\ \bibinfo {pages} {015} (\bibinfo {year} {2020})},\
  \Eprint {https://arxiv.org/abs/1712.09542} {arXiv:1712.09542 [hep-th]}
  \BibitemShut {NoStop}%
\bibitem [{\citenamefont {Naidu}\ and\ \citenamefont
  {Nikshych}(2008)}]{naidu2008lagrangian}%
  \BibitemOpen
  \bibfield  {author} {\bibinfo {author} {\bibfnamefont {D.}~\bibnamefont
  {Naidu}}\ and\ \bibinfo {author} {\bibfnamefont {D.}~\bibnamefont
  {Nikshych}},\ }\bibfield  {title} {\bibinfo {title} {{Lagrangian
  subcategories and braided tensor equivalences of twisted quantum doubles of
  finite groups}},\ }\href {https://doi.org/10.1007/s00220-008-0441-5}
  {\bibfield  {journal} {\bibinfo  {journal} {Commun. Math. Phys.}\ }\textbf
  {\bibinfo {volume} {279}},\ \bibinfo {pages} {845} (\bibinfo {year}
  {2008})},\ \Eprint {https://arxiv.org/abs/0705.0665} {arXiv:0705.0665
  [math.QA]} \BibitemShut {NoStop}%
\bibitem [{\citenamefont {Davydov}(2010)}]{davydov2010modular}%
  \BibitemOpen
  \bibfield  {author} {\bibinfo {author} {\bibfnamefont {A.}~\bibnamefont
  {Davydov}},\ }\bibfield  {title} {\bibinfo {title} {{Modular invariants for
  group-theoretical modular data. I}},\ }\href@noop {} {\bibfield  {journal}
  {\bibinfo  {journal} {J. Algebra}\ }\textbf {\bibinfo {volume} {323}},\
  \bibinfo {pages} {1321} (\bibinfo {year} {2010})},\ \Eprint
  {https://arxiv.org/abs/0908.1044} {arXiv:0908.1044 [math.QA]} \BibitemShut
  {NoStop}%
\bibitem [{\citenamefont {Haegeman}\ \emph {et~al.}(2015)\citenamefont
  {Haegeman}, \citenamefont {Van~Acoleyen}, \citenamefont {Schuch},
  \citenamefont {Cirac},\ and\ \citenamefont {Verstraete}}]{Haegeman:2014maa}%
  \BibitemOpen
  \bibfield  {author} {\bibinfo {author} {\bibfnamefont {J.}~\bibnamefont
  {Haegeman}}, \bibinfo {author} {\bibfnamefont {K.}~\bibnamefont
  {Van~Acoleyen}}, \bibinfo {author} {\bibfnamefont {N.}~\bibnamefont
  {Schuch}}, \bibinfo {author} {\bibfnamefont {J.~I.}\ \bibnamefont {Cirac}},\
  and\ \bibinfo {author} {\bibfnamefont {F.}~\bibnamefont {Verstraete}},\
  }\bibfield  {title} {\bibinfo {title} {{Gauging quantum states: from global
  to local symmetries in many-body systems}},\ }\href
  {https://doi.org/10.1103/PhysRevX.5.011024} {\bibfield  {journal} {\bibinfo
  {journal} {Phys. Rev. X}\ }\textbf {\bibinfo {volume} {5}},\ \bibinfo {pages}
  {011024} (\bibinfo {year} {2015})},\ \Eprint
  {https://arxiv.org/abs/1407.1025} {arXiv:1407.1025 [quant-ph]} \BibitemShut
  {NoStop}%
\end{thebibliography}%
\appendix
\crefalias{section}{appendix}
\crefalias{subsection}{subappendix}

\section{$\Zn$ Schwinger Model with \texorpdfstring{charge $\tilde{n}$}{Charge-ñ} }
\label{sec:zn-schwinger}

In this appendix, we show our construction for the charge-$\nw$ $\Zn$ Schwinger model.  This illustrates the model in a simpler setting and serves as an nontrivial numerical check of our construction.

We start with $\ZZ_{N^2}$ valued fields on sites and links.
On sites we have the unitary matrices $X_j, \Zm{j}$, and on links we have $\Xl{j + \frac{1}{2}},\Zl{j + \frac{1}{2}}$ with the clock--shift algebra:
\begin{align}
	\Zm{j} X_j \Zm{j}^{\dagger} = \omega X_j, \quad
	\Zl{j + \frac{1}{2}} \Xl{j + \frac{1}{2}} \Zl{j + \frac{1}{2}}^{\dagger} = \omega \Xl{j + \frac{1}{2}} 
\end{align}
where $\omega = \exp(i 2 \pi / N^2 )$ is the $N^2$-root of unity.
We also have $\ZZ_{N}$ valued fields on sites $U_j, V_j$ satisfying the clock shift algebra
\begin{align}
	V_j U_j {V}^{\dagger}_j &= \hat\omega U_j
\end{align}
where $\hat \omega = \exp(i 2 \pi / N) $.
Define the local charges
\begin{align}
	Q_{j} = \Xl{j - \frac{1}{2}} \Zm{j} \Xl{j + \frac{1}{2}}^\dagger, \quad
  \Ql{j + \frac{1}{2}} = \Zl{j + \frac{1}{2}}.
\end{align}
The Gauss constraints are
\begin{align}
  \hat K_j &\equiv Q_j^N U_{j}^{-\nw} = 1\\
  \hat K_{j + \frac{1}{2}} &\equiv \Ql{j + \frac{1}{2}}^{N} = 1 \\
  \hat G_{j + \frac{1}{2}} &\equiv \Zl{j + \frac{1}{2}}^{\nw} {V}_{j+1} {V}^{\dagger}_{j} = 1 
\end{align}
Note that the global charges  for the momentum and shift symmetries
\begin{align}
	Q = \prod_j Q_j, \quad \Ql{} = \prod_{j} \Ql{j + \frac{1}{2}}
\end{align}
satisfy 
\begin{align}
	Q^N =  \prod_{j} U_j^{\nw}, \quad \Ql{}^{N} = 1,
\end{align}
due to the Gauss constraints.
This implies that when $U_j = 1$ (no gauge fields), both the global momentum charge $Q$ and winding charge $\Ql{}$ are $\ZZ_{N}$ quantized. However, when we add a background gauge field for the winding symmetry, the momentum charge fractionalizes.
This is a manifestation of the mixed 't Hooft anomaly between winding and momentum symmetries.
The anomaly is most clearly visible in the anomalous algebra of momentum and winding charges
\begin{align}
	\Qm{j} \Ql{j + \frac{1}{2}} \Qm{j}^{-1} &= \omega \Ql{j + \frac{1}{2}}.
\end{align}
The $\Zn$ Schwinger model Hamiltonian is
\begin{align}
	H &= - \sum_j 
  \frac{J_p}{2} \Zm{j} 
      + \frac{J_x}{2} \ X_{j+1} \Zl{j + \frac{1}{2}}  {X}^{\dagger}_{j} 
+ \frac{J_e}{2}\  {V}_j ({X}^{\dagger}_j)^{\nw}
      +  \hc
\label{eq:zn-schwinger-app}
\end{align}

For $\nw=1$, one of Gauss constraints is
\begin{align}
\Zl{j + \frac{1}{2}} &= V_j {V}^{\dagger}_{j+1}.
\end{align}
Using this and defining new matter fields, dressed with the electric field,
\begin{align}
	\hat X_j = {V}^{\dagger}_j X_j, \quad \Zm*{j} = \Zm{j}.
\end{align}
the Hamiltonian becomes
\begin{align}
	H &= - \sum_j \ab (
      \frac{J_p}{2} \Zm*{j} 
      + \frac{J_x}{2} \ \hat X_{j+1} {\hat X}^{\dagger}_{j} 
      + \frac{J_e}{2}\  \hat X_j
      ) +  \hc
\end{align}
in terms of unconstrained variables $\hat X_j, \Zm*{j}$ with local Hilbert space dimension $N^2$.

\subsection{Matching with the $U(1)$ theory}

In order to determine the couplings of our model, we can match with Hamiltonian for the  charge-$\nw$ Schwinger model obtained in  Ref.~\cite{seifnashri_exactly_2026},
\begin{align}
	H &= \frac{e^2}{2} \sum_{j} \ab( E_j - \frac{\nw}{2\pi} \phi_j)^2 
      + \frac{1}{2 R^2} \sum_{j}  p_j^2 \notag\\
      &\quad\quad+ \frac{R^2}{2} \sum_j \ab(\widetilde p_{j + \frac{1}{2}} + \frac{1}{2\pi} (\delta \phi)_{j + \frac{1}{2}}  )^2
\end{align}
with the Gauss constraints
\begin{align}
	e^{2\pi i q_j } = e^{i \nw a_j}, \quad
                  e^{2 \pi i \tilde q_{j + \frac{1}{2}}} = 1, \quad
  \nw \tilde q_{j + \frac{1}{2}} + (\delta E)_{j + \frac{1}{2}} = 0,
\end{align}
where the (quantized) local momentum and winding charges are
\begin{align}
	q_j &= p_j - \frac{1}{2\pi} (\delta \tilde \phi)_j, \quad \tilde q_{j + \frac{1}{2}} = \tilde p_{j + \frac{1}{2}}.
\end{align}
and the lattice difference operator acts as $(\delta E)_{j+\frac12} = E_{j+1} - E_j$ and $(\delta \tilde \phi)_{j} = \tilde \phi_{j+\frac12} - \tilde \phi_{j-\frac12} $ for site and link fields, respectively.
The dictionary between the clock fields in $\Znn$ model and the $U(1)$ rotor fields is therefore
\begin{align}
	\Xm{j} &= e^{i \phi_j / N}, \quad
	\Zm{j} = e^{i 2\pi p_j / N}, \\ 
	\Xl{j + \frac{1}{2}} &= e^{i \tilde \phi_{j+\frac12} / N}, \quad
	\Zl{j + \frac{1}{2}} = e^{i 2 \pi \tilde p_{j+\frac12} / N}, \quad \\
  U_j &= e^{i a_j }, \quad V_{j} = e^{i 2 \pi E_j / N},
\end{align}
at leading order in $N^{-1}$,
resulting in the correspondence between the couplings
\begin{align}
	J_p &= \ab(\frac{N}{2 \pi R} )^2, \quad
	J_x = \ab( \frac{R N}{2\pi})^2, \quad
  J_e =  \ab(\frac{e N}{2 \pi})^2
\end{align}
The Schwinger model is obtained at the free-fermion radius of $R = \frac{1}{\sqrt{2}}$, resulting in the couplings
\begin{align}
	J_p = \frac{N^2}{2\pi^2},\quad
  J_x = \frac{N^2}{8\pi^2} = \frac{1}{4} J_p, \quad
  J_e = \frac{e^2 N^2}{4\pi^2} = \frac{e^2}{2} J_p.
  \label{eq:couplings}
\end{align}
With these choice of couplings, the mass gap in the limit $N \to \infty$ is \cite{seifnashri_exactly_2026,lu_fermionic_2026}
\begin{align}
	\Delta = \frac{e}{\pi \sqrt{2}}.
  \label{eq:gap}
\end{align}
We check this numerically in \cref{fig:gap_vs_N}. The gap of the finite-$N$ chain approaches $e/\pi\sqrt2$ with a leading correction consistent with $O(N^{-2})$.

\section{Exact reduction of $\ZZ_N$ \TFFZ{} lattice gauge theory to an unconstrained qudit chain}
\label{sec:qudit-duality}

  In this appendix, we construct the qudit T-duality used in \cref{sec:3450-model}.
It maps the chiral charge of the \TFFZ{} model to the winding charge of flavor 1, which simplifies all the Gauss constraints.
On the charge vector $(\nm[1], \nm[2] \,|\, \nw[1], \nw[2])$, it acts as
\begin{align}
  (8, 4 \,|\, -1, 2) \mapsto (0, 0 \,|\, 1, 0) .
\end{align}
This is achieved by the local Clifford map
\begin{align}
  \Xm[1]{j} &\mapsto \Xm[1]{j}^{-1} \Xm[2]{j}^{-2} \Xl[2]{j - \frac{1}{2}}^{-4} \Xl[1]{j + \frac{1}{2}}^{8}, \notag\\
  \Xm[2]{j} &\mapsto \Xm[2]{j}^{-1} \Xl[1]{j+\frac{1}{2}}^{4}, \notag\\
  \Xl[1]{j + \frac{1}{2}} &\mapsto \Xl[1]{j + \frac{1}{2}}^{-1}, \notag\\
  \Xl[2]{j + \frac{1}{2}} &\mapsto \Xl[1]{j + \frac{1}{2}}^{2} \Xl[2]{j + \frac{1}{2}}^{-1}, \notag\\
  \Zm[1]{j} &\mapsto \Zm[1]{j}^{-1}, \notag\\
  \Zm[2]{j} &\mapsto \Zm[1]{j}^{2} \Zm[2]{j}^{-1}, \notag\\
  \Zl[1]{j+\frac12} &\mapsto \omega^{8} \Zl[1]{j+\frac12}^{-1} \Zl[2]{j+\frac12}^{-2} \Qm[2]{j}^{-4} \Qm[1]{j+1}^{8}, \notag\\
  \Zl[2]{j+\frac12} &\mapsto \Zl[2]{j+\frac12}^{-1} \Qm[1]{j+1}^{4} .
  \label{eq:tduality}
\end{align}
This preserves the relation $Z X Z^\dagger = \omega X$, and all other commutators, for every $N$.
  This is a qudit version of the T-duality $T_M$ of Ref.~\cite{seifnashri_exactly_2026}, which realizes the same $O(2,2;\ZZ)$ element
\begin{align}
  M = \begin{bmatrix} -1 & 2 & 8 & 4 \\ 0 & -1 & -4 & 0 \\ 0 & 0 & -1 & 0 \\ 0 & 0 & -2 & -1 \end{bmatrix}
\end{align}
on the charge vectors of the modified Villain model.
With $X = e^{i\phi/N}$ and $Z = e^{2\pi i p/N}$, \cref{eq:tduality} reproduces its action on the Villain fields, up to the phase $\omega^{8}$, which arises from operator ordering and tends to $1$ as $N \to \infty$.
Under \cref{eq:tduality}, the charges transform as
\begin{align}
	\Qm[1]{j} &= \Xl[1]{j-\frac12} \Zm[1]{j} \Xl[1]{j+\frac12}^\dagger &&\mapsto \Qm[1]{j}^{-1} \notag\\
  \Qm[2]{j} &= \Xl[2]{j-\frac12} \Zm[2]{j} \Xl[2]{j+\frac12}^\dagger &&\mapsto \Qm[1]{j}^{2} \Qm[2]{j}^{-1} \notag\\
  \Ql[1]{j+\frac12} &= \Zl[1]{j+\frac12} &&\mapsto  {\omega^{8}} \Ql[1]{j+\frac12}^{-1} \Ql[2]{j+\frac12}^{-2} \Qm[2]{j}^{-4} \Qm[1]{j+1}^{8} \notag\\
  \Ql[2]{j+\frac12} &= \Zl[2]{j+\frac12} &&\mapsto  \Ql[2]{j+\frac12}^{-1} \Qm[1]{j+1}^{4}
\end{align}
Importantly, the local chiral charge transforms as
\begin{align}
  \Qchi_{j + \frac{1}{2}} = \Qm[1]{j}^{8} \Qm[2]{j}^4 \Qw[1]{j + \frac{1}{2}}^{-1} \Qw[2]{j + \frac{1}{2}}^{2}
  \mapsto  \Ql[1]{j+\frac12}.
\end{align}
In this frame, all Gauss constraints simplify to
\begin{align}
	\Qm[1]{j}^{N} &\equiv U_j, &\qquad   \Ql[1]{j+\frac12}^{N} &\equiv 1 \\
	\Qm[2]{j}^{N} &\equiv 1, &\qquad   \Ql[2]{j+\frac12}^{N} &\equiv 1 . \\
\Ql[1]{j+\frac12} &= V_j V_{j+1}^{\dagger}
\end{align}
The transformed Hamiltonian has kinetic terms and hopping terms
\begin{align}
	\hkin_{1,j} &= \Zm[1]{j}^{-1}  + \hc \notag \\
	\hkin_{2,j} &= \Zm[1]{j}^{2} \Zm[2]{j}^{-1}  + \hc \notag\\
  \hhop_{1,j}  &=
                 \omega^{8}\,
                 \Xm[1]{j}^{-1} \Zm[2]{j}^{4} \Xm[2]{j}^{-2}\,
                 \Zl[1]{j + \frac{1}{2}} \Zl[2]{j + \frac{1}{2}}^{2}\,
                 \Zm[1]{j+1}^{-8} \Xm[1]{j+1} \Xm[2]{j+1}^{2}
                 + \hc \notag\\
	\hhop_{2,j}
              &= \Xm[2]{j}^{1} \Zl[2]{j+\frac12}^{-1}  \Zm[1]{j+1}^{4} \Xm[2]{j+1}^{-1}  + \hc
\end{align}
and the electric term 
\begin{align}
	  \helec_j &= \Xm[1]{j} V^{\dagger}_{j} + \hc
\end{align}

Note now that the electric field can be eliminated by the same argument as for the charge-1 Schwinger model, which sends
\begin{align}
	\Xm[1]{j} {V}^{\dagger}_j \mapsto \Xm[1]{j}
\end{align}
so that electric field term becomes simply
\begin{align}
	\helec_j &= \Xm[1]{j} + \hc.
\end{align}
Using the physical Gauss law for $\Ql[1]{j+\frac12}$, the hopping terms $\hhop_{a,j}$ simplifies to
\begin{align}
  \hhop_{1,j} &=
                \omega^{8}\,
                \Xm[1]{j}^{-1}
                \Zm[2]{j}^{4}
                \Zm[1]{j+1}^{-8}
                \Xm[1]{j+1}
                \Xm[2]{j}^{-2}\,
                \Zl[2]{j + \frac{1}{2}}^{2}\,
                \Xm[2]{j+1}^{2}
                 + \hc \notag\\
	\hhop_{2,j}
  &=
  \Xm[2]{j}^{1}
  \Zl[2]{j+\frac12}^{-1}
  \Xm[2]{j+1}^{-1}
  \Zm[1]{j+1}^{4}
  + \hc.
\end{align}
As written, this now has two $\Znn$ degrees of freedom per site, and one $\Znn$ degree of freedom $\Zl[2]{j+\frac12}$ per link.
However, there are two Gauss constraints left:
\begin{align}
	\Qm[2]{j}^N = 1 &\implies \Xl[2]{j-\frac12}^N \Zm[2]{j}^N \Xl[2]{j+\frac12}^{-N} = 1 \\
  \Ql[2]{j-\frac12}^{N} = 1
  &\implies
  \Zl[2]{j-\frac12}^{N}  = 1
\end{align}
This means that $(\Xm[1]{j}, \Zm[1]{j})$ is an unconstrained $\Znn$ register, while $(\Xm[2]{j}, \Zm[2]{j})$ and $( \Xl[2]{j+\frac12}, \Zl[2]{j+\frac12} )$ are two $\Znsq$ registers coupled via Gauss constraints, together forming a $\ZZ_N$ clock chain in the modified Villain representation.

\subsection{Solving the flavor-2 constraints}
\label{sec:3450-reduction}

These constraints can be further solved to reduce the model to two unconstrained $\Znsq$ chains.
It is clear that this model has an unbroken global $\Zn$ symmetry which is the global winding symmetry of flavor 2: 
\begin{align}
	\Ql{2} = \prod_{j} \Zl[2]{j + \frac12}
\end{align}
To solve this further, we define invariant densities
\begin{align}
  \Qli[2]{j+\frac12} &= \Xm[2]{j}^{-1}\, \Zl[2]{j + \frac{1}{2}}\, \Xm[2]{j+1} \\
  \Qmi[2]{j} &= \Zm[2]{j} 
\end{align}
which commute with the Gauss constraints.
Note that the kinetic and hopping terms can now be written in terms of these as
\begin{align}
	\hkin_{1,j} &= \Zm[1]{j}^{-1}  + \hc,\\
	\hkin_{2,j} &= \Zm[1]{j}^{2} \Qmi[2]{j}^{-1}  + \hc, \\
  \hhop_{1,j} &= \omega^{8}\,
                \Xm[1]{j}^{-1} \Zm[1]{j+1}^{-8} \Xm[1]{j+1}\ 
                \Qmi[2]{j}^{4}
                \Qli[2]{j+\frac12}^2
                 + \hc, \notag\\
	\hhop_{2,j} &= \Qli[2]{j+\frac12}^{-1} \Zm[1]{j+1}^{4} + \hc.
\end{align}
Gauss constraints imply that
\begin{align}
	\Qli[2]{j+\frac12}^N &= (\Xm[2]{j}^{-1} \Xm[2]{j+1})^{N} ,\quad
  \Qmi[2]{j}^N &= (\Xl[2]{j-\frac12}^{-1} \Xl[2]{j+\frac12})^{N} .
\end{align}
We can now eliminate the remaining constraints by gauging the $\ZnW$ symmetry completely. To do so, we introduce site fields $\xm{j}, \zm{j}$ such that
\begin{align}
	\Qli[2]{j+\frac12} &= \xm{j}^{-1} \xm{j+1}, \\
  \Qmi[2]{j} &= \zm{j}.
\end{align}
(Note that doing this implies we restrict to the zero charge sector for the $\ZnW$ symmetry. Charged sectors correspond to defects.)
The Gauss constraints imply that
\begin{align}
  \xm{j}^{-N} \Xm[2]{j}^N &= \xm{j+1}^{-N} \Xm[2]{j+1}^N .
\end{align}
Now, the constraints have been eliminated since we can replace a pair of $\Znsq$ registers and the constraints by a single $\Znsq$ \emph{unconstrained} register given by 
\begin{equation}
  \underbrace{(\Xm[2]{j},\Zm[2]{j}),\ (\Xl[2]{j+\frac12},\Zl[2]{j+\frac12})}_{\Znsq \otimes \Znsq \text{ + Gauss law}}
  \ \longrightarrow\
  \underbrace{(\xm{j},\zm{j})}_{\Znsq}
\end{equation}
Substituting into the terms above gives the Hamiltonian of \cref{eq:H-3450}.

In terms of the original fields, what does gauging $\ZnW$ symmetry of flavor 2 mean? We can pull back the generator $\Qli[2]{j}$ via the qudit duality and show that 
\begin{align}
	\Ql[2]{j+\frac12}^{-1} \Qm[1]{j+1}^{-4} \mapsto \Ql[2]{j+\frac12}.
\end{align}
Therefore, this frame corresponds to gauging the (anomaly-free) global symmetry given by the momentum-winding charge vector $q_A = (-4, 0 | 0, -1) $.

\section{Relevant facts about bosonization}%
\label{sec:bosonization}

In this appendix, we collect some known facts about bosonization as they pertain to our construction.
Consider a 1+1-dimensional $U(1)$ gauge theory of $\Nf$ left-handed Weyl fermions $\psiL{a}$ with charges $\qL{a}$ and $\Nf$ right-handed Weyl fermions $\psiR{a}$ with charges $\qR{a}$, for $a = 1, \dotsc, \Nf$.
The gauge anomaly cancels when the charges satisfy \cref{eq:anomaly-cancellation},
\begin{align}
	\sum_{a=1}^{\Nf} \qL{a}^2 = \sum_{a=1}^{\Nf} \qR{a}^2 .
\end{align}
We bosonize the $2n$ Weyl fermions, $n \equiv \Nf$, into $2n$ compact chiral bosons $\phi_I \sim \phi_I + 2\pi$, $I = 1, \dotsc, 2n$, ordered with the left-movers first.
Their chiralities are encoded in the matrix
\begin{align}
	K_{IJ} = \chi_I \delta_{IJ}, \qquad
	\chi_I = \begin{cases} +1 & \text{left-moving}, \\ -1 & \text{right-moving}, \end{cases}
\end{align}
and the bosons satisfy
\begin{align}
	[\phi_I(x), \phi_J(y)] = -i\pi \Kinv_{IJ} \sgn(x-y).
\end{align}
With this normalization $e^{i\phi_I(x)} e^{i\phi_I(y)} = - e^{i\phi_I(y)} e^{i\phi_I(x)}$ for $x \neq y$, so the vertex operator $e^{i\phi_I}$ is the bosonized Weyl fermion, up to Klein factors.
The densities $\rho_I = \frac{1}{2\pi} \del_x \phi_I$ and integer charges $Q_I = \int dx\, \rho_I(x)$ satisfy
\begin{align}
	[\rho_I(x), \rho_J(y)] = \frac{i}{2\pi} \Kinv_{IJ}\, \delta'(x-y).
\end{align}

The $U(1)$ symmetry of the fermions is generated by $\Qchi = \vec q \cdot \vec Q$, with the charge vector
\begin{align}
  \vec q = (\qL{1}, \dotsc, \qL{n} \,|\, \qR{1}, \dotsc, \qR{n}) \in \ZZ^{2n}.
\end{align}
It shifts the chiral bosons as
\begin{align}
	e^{i\theta \Qchi}\, \vec \phi\, e^{-i\theta \Qchi} = \vec \phi + \theta\, \Kinv \vec q ,
\end{align}
so $e^{i\phi_I}$ carries charge $(\Kinv \vec q)_I = \chi_I q_I$, that is $\qL{a}$ for left-movers and $-\qR{a}$ for right-movers. By the remark above, this sign is immaterial.
The symmetry can be gauged only if the local charge density commutes with itself, so that the Gauss law constraints at different points are compatible.
The density commutator is
\begin{align}
	[\vec q \cdot \vec \rho(x), \vec q \cdot \vec \rho(y)] &= \frac{i}{2\pi} \left(\vec q^{\,T} \Kinv \vec q\right) \delta'(x-y), \\
	\vec q^{\,T} \Kinv \vec q &= \sum_{a=1}^{n} \qL{a}^2 - \sum_{a=1}^{n} \qR{a}^2 ,
\end{align}
and this Schwinger term vanishes exactly when the anomaly cancellation condition \cref{eq:anomaly-cancellation} holds.
Gauging the $U(1)$ symmetry then gives the chiral gauge theory.

We gauge it by promoting $\theta$ to a function $\theta(x)$ and introducing a gauge field $A_\mu$, with gauge transformations
\begin{align}
	\vec \phi(x) &\mapsto \vec \phi(x) + \theta(x)\, \Kinv \vec q , \\
	A_\mu(x) &\mapsto A_\mu(x) - \del_\mu \theta(x) .
\end{align}
The combination $D_x \vec \phi = \del_x \vec \phi + A_x \Kinv \vec q$ is gauge invariant.
In the temporal gauge $A_t = 0$, the electric field $E = \frac{1}{e^2} \del_t A_x$ is conjugate to $A_x$, $[A_x(x), E(y)] = i \delta(x-y)$.
The Hamiltonian for the gauged model is
\begin{align}
	H = \int dx \left[ \frac{e^2}{2} E^2 + \frac{1}{4\pi} (D_x \vec \phi)^T \Vmat \, D_x \vec \phi \right] .
\end{align}
where $e$ is the gauge coupling and $\Vmat$ is a symmetric positive-definite velocity matrix, with $\Vmat_{IJ} = \delta_{IJ}$ for free Weyl fermions of unit velocity.
The Gauss law operator
\begin{align}
	G(x) = \del_x E(x) + \frac{1}{2\pi} \vec q \cdot \del_x \vec \phi(x)
\end{align}
generates the gauge transformations above, and $[G(x), G(y)] = [\vec q \cdot \vec \rho(x), \vec q \cdot \vec \rho(y)]$, so the constraints $G(x) = 0$ are compatible exactly when the anomaly cancels.

\subsection{Vertex operators}

The local operators of the fermionic theory are the vertex operators
\begin{align}
	V_{\vec l} = e^{i \vec l \cdot \vec \phi}, \qquad \vec l \in \Lambda = \ZZ^{2n},
\end{align}
where integrality of $\vec l$ makes $V_{\vec l}$ single-valued under $\phi_I \mapsto \phi_I + 2\pi$.
With the bilinear form $\<\vec l, \vec l'\> = \vec l^{\,T} \Kinv \vec l'$, the shift of $\vec \phi$ under the $U(1)$ symmetry gives
\begin{align}
	e^{i\theta \Qchi}\, V_{\vec l}\, e^{-i\theta \Qchi} = e^{i\theta \<\vec l, \vec q\>}\, V_{\vec l} ,
\end{align}
so $V_{\vec l}$ carries charge $\<\vec l, \vec q\>$, and its spin is $h - \bar h = \frac{1}{2} \<\vec l, \vec l\>$.
The gauge-invariant vertex operators therefore form the perpendicular lattice
\begin{align}
	\Lambda_\perp = \{ \vec l \in \Lambda : \<\vec l, \vec q\> = 0 \} .
\end{align}

For a single flavor, write $\vec \phi = (\phi_L, \phi_R)$ and $\vec q = (\qL{} \,|\, \qR{})$.
The compact boson and its dual are
\begin{align}
	\phiboson = \vec l_- \cdot \vec \phi = \phi_L - \phi_R, \qquad
	\phidual = \vec l_+ \cdot \vec \phi = \tfrac{1}{2} (\phi_L + \phi_R),
\end{align}
with $\vec l_- = (1, -1)$ and $\vec l_+ = \frac{1}{2}(1, 1)$, so that $\phi_{L,R} = \phidual \pm \frac{1}{2} \phiboson$.
The vertex operators $V = e^{i\phiboson}$ and $\Vdual = e^{i\phidual}$ carry the momentum and winding charges of \cref{eq:qm-qw},
\begin{align}
	\nm = \<\vec l_-, \vec q \> = \qL{} + \qR{}, \qquad
	\nw = \<\vec l_+, \vec q \> = \tfrac{1}{2} (\qL{} - \qR{}) ,
\end{align}
and $e^{i(m \phiboson + w \phidual)}$ has charge $m \nm + w \nw$ and spin $m w$.
For the \TFFZ{} model, the flavors $\vec q = (3 \,|\, 5)$ and $(4 \,|\, 0)$ give $(\nm, \nw) = (8, -1)$ and $(4, 2)$.
Since $\nm[a] \nw[a] = \frac{1}{2} (\qL{a}^2 - \qR{a}^2)$, the anomaly cancellation condition becomes $\sum_a \nm[a] \nw[a] = 0$, here $-8 + 8 = 0$.
The map inverts as $\qL{}, \qR{} = \frac{1}{2} \nm \pm \nw$, so integer charges with $\qL{} = \qR{} \bmod 2$ correspond exactly to integer $\nw$ and even $\nm$.

\subsection{IR spectrum}

A T-duality is a change of basis of the chiral bosons that preserves their commutation relations,
\begin{align}
	\vec \phi \mapsto \tilde{\vec \phi} = M \vec \phi,  \qquad M^T K M = K ,
\end{align}
with integer $M$; cf.\ Ref.~\cite{giveon_target_1994}.
The condition $M^T K M = K$ is equivalent to $M \Kinv M^T = \Kinv$, so $[\tilde\phi_I, \tilde\phi_J]$ takes the same form as $[\phi_I, \phi_J]$, and integrality of $M$ and $M^{-1}$ preserves the periodicities $\phi_I \sim \phi_I + 2\pi$.
In other words, $M$ is an automorphism of the lattice $\Lambda = \ZZ^{2n}$ with the form $\<\cdot, \cdot\>$.
Vertex operators and the charge vector transform as
\begin{align}
	\vec l \mapsto \tilde{\vec l} = M^{-T} \vec l, \qquad \vec q \mapsto \tilde{\vec q} = M^{-T} \vec q ,
\end{align}
so that $\vec l \cdot \vec \phi$, the charges $\<\vec l, \vec q\>$ and the spins $\frac{1}{2} \<\vec l, \vec l\>$ are invariant.
The Hamiltonian keeps its form with $\vec q \to \tilde{\vec q}$, while the velocity matrix becomes $\tilde \Vmat = M^{-T} \Vmat M^{-1}$.

\minisec{Solving a 1+1-dimensional chiral gauge theory via T-duality}
Suppose a T-duality maps the charge vector to
\begin{align}
	\tilde{\vec q} = (1, 0, \dotsc, 0 \,|\, 1, 0, \dotsc, 0) .
\end{align}
In this dual frame, the gauge field couples only to the first pair of chiral bosons, which form a Schwinger model with unit charge.
The Schwinger mechanism gaps the gauge-invariant combination $\tilde{\vec q} \cdot \tilde{\vec \phi} = \vec q \cdot \vec \phi$.
The other $2(n-1)$ chiral bosons do not couple to the gauge field; they mix with the first pair only through off-diagonal entries of $\tilde \Vmat$, which keeps the Hamiltonian quadratic, so they remain massless.

The infrared spectrum is given by the lattice $\Lambda_\perp$ of gauge-invariant vertex operators.
It contains $\vec q$ itself, because $\<\vec q, \vec q\> = 0$. The vertex operator $V_{\vec q} = e^{i \vec q \cdot \vec \phi}$ is the gapped Schwinger mode, so the massless sector is described by the quotient $\Lambda_\perp / \ZZ \vec q$, of rank $2(n-1)$.
In the dual frame this is manifest since $\tilde\Lambda_\perp = \{ \tilde{\vec l} : \tilde l_{L,1} = \tilde l_{R,1} \} = \ZZ \tilde{\vec q} \oplus \ZZ^{2(n-1)}$, where the second factor is spanned by the decoupled bosons.

\minisec{\TFFZ{} model}
Take the $2n$-dimensional charge vector to be $\vec q = (3,4 | 5, 0)$.
A basis for the perpendicular lattice can be found using the row-based Hermite normal form of the column vector $A = \Kinv \vec q$, which gives a $2n \times 2n$ unimodular matrix $U$ and a $2n \times 1$ matrix $H$ such that $U A = H$.
For a column vector, the Hermite normal form is $H= g e_1$, where $g$ is the gcd of its entries. Here $g = 1$, so $U \Kinv \vec q = e_1$, which means that a basis for the perpendicular lattice is given by the last $2n-1$ rows of $U$:
\begin{align}
  l_1 &= \begin{bmatrix} -4 & 3 & | & 0 & 0 \end{bmatrix}, \\
  l_2 &= \begin{bmatrix} 15 & -10 & | & 1 & 0 \end{bmatrix}, \\
  l_3 &= \begin{bmatrix} 0 & 0 & | & 0 & 1 \end{bmatrix}.
\end{align}
The conformal weights of the operators are determined by the projected left and right momenta of these operators. For $l = (l_L | l_R )$
\begin{align}
	\pi(l) = (\pi_L | \pi_R), \qquad \pi_{L,R} = l_{L,R} - \frac{l_q}{Q^2} q_{L,R},
\end{align}
where we have defined $l_q = l_R \cdot q_R = l_L \cdot q_L$ and $Q^2 = q_L^2 = q_R^2$. The conformal weights are given by
\begin{align}
	h = \frac{1}{2} \pi_L^2 , \quad \bar{h} = \frac{1}{2} \pi_R^2.
\end{align}
For the basis vectors above, we have
\begin{align}
  \pi(l_1) &= \begin{bmatrix} -4 & 3 & | & 0 & 0 \end{bmatrix}, \\
  \pi(l_2) &= -\tfrac{18}{5} \begin{bmatrix} -4 & 3 & | & 0 & 0 \end{bmatrix}, \\
  \pi(l_3) &= \begin{bmatrix} 0 & 0 & | & 0 & 1 \end{bmatrix}.
\end{align}
A left-moving Weyl fermion has $(h, \bar h) = (\frac{1}{2}, 0) $ and a right-moving Weyl fermion has $(h,\bar h) = (0, \frac{1}{2})$.

Therefore, we can immediately read off that $l_3 \cdot \phi $ corresponds to a right-moving Weyl fermion.
Is there a left-moving Weyl fermion? The subspace  with $\bar{h}=0$ is spanned by $l_1, l_2$.
The conformal weights of a vertex operator corresponding to $l = \alpha l_1 + \beta l_2$ with $\alpha, \beta \in \ZZ$ are $h = \frac{1}{2} (5 \alpha - 18 \beta)^2$ and $\bar{h} = 0$.
A left-moving Weyl fermion needs $h = \frac{1}{2}$, which is possible since $\gcd(5,18) = 1 $.
One solution is $l' = 11 l_1 + 3 l_2 = (1,3 | 3, 0) $. Therefore, $l' \cdot \phi$ corresponds to a left-moving Weyl fermion, a composite particle made of elementary left- and right-moving Weyl fermions.
The (fermionic) \TFFZ{} model therefore has one free left- and one right-moving Weyl fermion in the \ac{IR}.

\section{Category theoretic construction of chiral gauge theories}
\label{sec:abstract-nonsense}

As an independent verification of the various gauged modified Villain models, we can use recently developed technology for gauging non-anomalous symmetries on the lattice \cite{cuiper2025gauging}. The idea is that gauging amounts to a change in the representation of local symmetric operators \cite{lootens2023dualities}, and that all such representations lead to isospectral Hamiltonians, provided that one is careful with boundary conditions \cite{lootens_dualities_2024}. Any correlation function can be computed in any representation, with the caveat that local operators can become nonlocal, which can be easily accommodated in the tensor network formulation. When adding a non-symmetric electric field term, enforcing its locality restricts the possible representations. For the Schwinger model, this representation is unique; for models with more fermion flavours, there are more possibilities, and we are free to choose one that best suits the numerics.

\subsection{(1-1)-model (Schwinger)}
As before, we start from the $\mathbb Z_{N^2}$ clock model:
\begin{equation}
    H = - \sum_i (J_1 X_i X_{i+1}^\dag + J_2 Z_i) + \text{h.c.} 
\end{equation}
with
\begin{equation}
    (X_i)^{N^2} = (Z_i)^{N^2} = \mathbb 1, \quad ZX = e^{\frac{2\pi i}{N^2}} XZ.
\end{equation}
This model has a $\mathbb Z_{N^2}$ symmetry generated by $\prod_i Z_i$. Following \cite{lootens2023dualities}, this model can be understood by considering the algebra of symmetric operators governed by a fusion category $\mc D = \Rep(\mathbb Z_{N^2})$ and picking a representation by choosing the module category $\mc R = \Vect$ over this fusion category. The symmetries of this model are then given by $\mc C := \End_{\mc D}(\mc R) \simeq \Vect_{\mathbb Z_{N^2}}$, which is indeed the on-site $\mathbb Z_{N^2}$ symmetry of the clock model. More general operators in this model are labeled by simple objects in $\mc Z(\mc D)$ \cite{Ji:2019jhk,Kong:2020cie,Freed:2022qnc}, which in this case is a modular fusion category with simple objects $\{e^a m^b, \enspace a,b \in \mathbb Z_{N^2}\}$. These are the anyons of the $\mathbb Z_{N^2}$ toric code, with $\mathbb Z_{N^2} \times \mathbb Z_{N^2}$ fusion rules, trivial associators, with $R$-matrices and topological spins given by
\begin{equation}
    R(e^a m^b,e^c m^d) = e^{\frac{2\pi i ad}{N^2}}, \quad \theta(e^a m^b) = e^{\frac{2\pi i ab}{N^2}}.
\end{equation}
The representations of these more general operators depend on the choice of $\mc R$ through the Morita equivalence defined by $\mc R$ as an invertible $(\mc C,\mc D)$ bimodule category. In the language of \cite{kitaev_models_2012}, the choice of $\mc R$ defines a functor from $\mc Z(\mc D)$ to $\mc C$ by composing the braided equivalence $\mc Z(\mc D) \simeq \mc Z(\mc C)$ induced by Morita equivalence with the forgetful functor from $\mc Z(\mc C)$ to $\mc C$. The operators in $\mc Z(\mc D)$ that map to the identity object in $\mc C$ are local; they form the Lagrangian algebra $L_{\mc R} \in \mc Z(\mc D)$ associated with the $\mc D$-module category $\mc R$ \cite{Davydov:2010kfz}. For the choice $\mc R = \Vect$, the Lagrangian algebra is
\begin{equation}
    L_\mc R = \langle e \rangle \simeq \mathbb C[\mathbb Z_{N^2}],
\end{equation}
corresponding to the lattice realization $\rho^{\mc R}_i$ on site $i$:
\begin{equation}
    \rho_i^{\mc R}(e^a m^b) = \prod_{j<i} Z_j^b X_i^a.
\end{equation}
From this, we indeed see that the operators $e^a \in L_{\mc R}$ are local. In terms of the Lagrangian algebra $L_{\mc R}$, the symmetry $\mc C$ can be understood as $\Mod_{\mc Z(\mc D)}(L_{\mc R})$ \cite{Kong:2013aya}, with monoidal structure $\otimes_{L_\mc R}$ given by the tensor product over $L_\mc R$. The underlying fusion ring of $\mc Z(\mc D)$ is simply $\mathbb Z^e_{N^2} \times \mathbb Z^m_{N^2}$, and so $L_\mc R$ modules are labelled by cosets
\begin{equation}
    \frac{\mathbb Z^e_{N^2} \times \mathbb Z^m_{N^2}}{\mathbb Z^e_{N^2}} = \mathbb Z^m_{N^2}.
\end{equation}
Picking representatives $m^a \otimes L_{\mc R}, a \in \mathbb Z^m_{N^2}$ for the cosets, the monoidal structure is given by
\begin{equation}
    m^a \otimes L_{\mc R} \otimes_{L_{\mc R}} m^b \otimes L_{\mc R} \simeq m^{a+b} \otimes L_{\mc R},
\end{equation}
and we indeed find that $\Mod_{\mc Z(\mc D)}(L_\mc R) \simeq \Vect_{\mathbb Z^m_{N^2}}$ with trivial associators.

\subsubsection{Villain model}

To obtain the $\mathbb Z_N$ modified Villain model, we gauge the $\mathbb Z^m_N$ subgroup of the $\mathbb Z^m_{N^2}$ symmetric clock model. Following \cite{lootens2023dualities,cuiper2025gauging}, this can be done by changing the module category $\mc R = \Vect$ that defines the representations of the algebra of symmetric operators to $\mc R' = \Rep(\mathbb Z_N)$. This yields the dual Hamiltonian
\begin{equation}
    H' = - \sum_i (J_1 X_i \tilde X_{i+\frac{1}{2}} X_{i+1}^\dag + J_2 Z_i) + \text{h.c.},
\end{equation}
subject to the Gauss constraints
\begin{equation}
    \tilde Z_{i-\frac{1}{2}} Z^N_i \tilde Z_{i+\frac{1}{2}}^\dag = \mathbb 1, \quad \text{with} \quad \tilde Z \tilde X = e^{\frac{2\pi i}{N}} \tilde X \tilde Z.
\end{equation}
The dual symmetry is now $\End_{\mc D}(\mc R')$, which we expect to be $\Vect_{\mathbb Z_N^e \times \mathbb Z_N^m}^\omega$, with $\omega$ a type-II cocycle \cite{Dijkgraaf:1989pz,deWildPropitius:1995cf}. The easiest way to confirm this is to compute the dual symmetry as $\Mod_{\mc Z(\mc D)}(L_{\mc R'})$. The new Lagrangian algebra $L_{\mc R'}$ after gauging should contain anyons of the type $e^a m^{Nb}, \enspace b \in \mathbb Z_N$ for some $a \in \mathbb Z_{N^2}$, as these correspond to the symmetry to be gauged in the original theory. These anyons should have trivial topological spin \cite{Bais:2008ni,Kong:2013aya}, so the new Lagrangian algebra is given by
\begin{equation}
    L_{\mc R'} = \langle e^N, m^N \rangle \simeq \mathbb C[\mathbb Z_N \times \mathbb Z_N],
\end{equation}
and its modules are the cosets
\begin{equation}
    \frac{\mathbb Z^e_{N^2} \times \mathbb Z^m_{N^2}}{N \mathbb Z^e_{N^2} \times N \mathbb Z^m_{N^2}} = \mathbb Z^e_{N} \times \mathbb Z^m_{N}.
\end{equation}
Picking representatives $e^a m^b \otimes L_{\mc R'}, \enspace a,b \in \mathbb Z_N$ for the cosets, the monoidal structure is given by
\begin{align}
     &e^{a_1} m^{b_1} \otimes L_{\mc R'} \otimes_{L_{\mc R'}}  e^{a_2} m^{b_2} \otimes L_{\mc R'}\nonumber\\
     &\simeq e^{a_1 +_N a_2} m^{b_1 +_N b_2} \otimes \kappa(e^{a_1} m^{b_1}, e^{a_2} m^{b_2}) \otimes L_{\mc R'},\nonumber\\
     &\simeq e^{a_1 +_N a_2} m^{b_1 +_N b_2} \otimes L_{\mc R'},
\end{align}
with $+_N$ addition modulo $N$ and $\kappa: \mc Z(\mc D) \times \mc Z(\mc D) \rightarrow L_{\mc R'}$ defined as
\begin{equation}
    \kappa(e^{a_1} m^{b_1}, e^{a_2} m^{b_2}) = e^{N \lfloor(a_1 + a_2)/N\rfloor} m^{N \lfloor(b_1 + b_2)/N\rfloor}.
\end{equation}
To compute the associator for $e^{a_1} m^{b_1} \otimes L_{\mc R'} \otimes_{L_{\mc R'}} e^{a_2} m^{b_2} \otimes L_{\mc R'} \otimes_{L_{\mc R'}} e^{a_3} m^{b_3} \otimes L_{\mc R'}$, we find that we have to braid $\kappa(e^{a_1}m^{b_1},e^{a_2}m^{b_2})$ past $e^{a_3}m^{b_3}$ to compare the two brackets. This gives the associator
\begin{equation}
    \omega(e^{a_1}m^{b_1},e^{a_2}m^{b_2},e^{a_3}m^{b_3}) = e^{\frac{2\pi i}{N}\lfloor \frac{a_1 + a_2}{N}\rfloor b_3}
\end{equation}
which is indeed a type-II cocycle in $H^3(\mathbb Z_N^e \times \mathbb Z_N^m, U(1))$ denoting a mixed anomaly between $\mathbb Z_N^e$ and $\mathbb Z_N^m$. This is precisely the single-flavor modified Villain model of \cref{sec:chiral}: under the identification $X_i = \Xm{i}$, $Z_i = \Zm{i}$, $\tilde X_{i+\frac12} = \Zl{i+\frac12}^\dagger$ and $\tilde Z_{i+\frac12} = \Xl{i+\frac12}^N$ (the $\Zl{}^N = 1$ subspace of the link), $H'$ is $\hkin + \hhop$ and the Gauss constraint is $\Qm{i}^N = 1$. In particular, the symmetry generated by $m$, $\prod_i Z_i = \prod_i \Qm{i}$, is the momentum symmetry, and the symmetry generated by $e$, $\prod_i \tilde X_{i+\frac12}^\dagger = \prod_i \Ql{i+\frac12}$, is the winding symmetry. 

\subsubsection{Gauged Villain model}

Having established the construction of the $\mathbb Z_N$ modified Villain model, we are now in a position to gauge a non-anomalous $\mathbb Z_N$ subgroup of its $\mathbb Z_N^{\times 2}$ symmetry. A generic $\mathbb Z_N$ subgroup is generated by $g = (\nw,\nm) \in \mathbb Z_N^{\times 2}$ such that $\text{gcd}(\nw,\nm,N) = 1$. As before, gauging this subgroup requires the new Lagrangian algebra $L_{\mc R''}$ to contain anyons of the form $\alpha = e^{Na + \nw} m^{Nb + \nm}$, for some $a,b \in \mathbb Z_N$. The topological spin of these anyons is
\begin{equation}
    \theta(\alpha) = e^{\frac{2\pi i \nw \nm}{N^2}} e^{\frac{2 \pi i(a \nm + b \nw)}{N}},
\end{equation}
which should be trivial for some $a,b \in \mathbb Z_N$ if we are to condense them. In particular, this requires $\theta(\alpha)^N = 1$, meaning we need $\nw \nm = 0 \mod N$ for this to be possible at all. This is precisely the condition that the subgroup we are gauging is non-anomalous, and since we are trying to approximate the $U(1)$ modified Villain model as $N \rightarrow \infty$, we will assume $\nw \nm = 0$. The topological spin of $\alpha$ then becomes
\begin{equation}
    \theta(\alpha) = e^{\frac{2 \pi i(a \nm + b \nw)}{N}},
\end{equation}
so the anyons of the form $e^{Na + \nw} m^{Nb + \nm}$ that condense to the generator of the subgroup symmetry that we are trying to gauge satisfy $a \nm + b \nw = 0 \mod N$. This equation has $N$ solutions; one obvious solution is $(a,b) = (0,0)$, corresponding to the boson $e^{\nw} m^{\nm}$, which indeed generates a $\mathbb Z_N$ symmetry after condensation since $e^{\nw N} m^{\nm N}$ is a condensed boson in $L_{\mc R'}$. In fact, because $\text{gcd}(\nw,\nm,N) = 1$, the algebra generated by this boson is Lagrangian and we write
\begin{equation}
    L_{\mc R''} = \langle e^{\nw}m^{\nm} \rangle \simeq \mathbb C [\mathbb Z_{N^2}].
\end{equation}

To write down the corresponding lattice model, we want to identify the module category $\mc R''$ over $\Rep(\mathbb Z_{N^2})$ that corresponds to this Lagrangian algebra. In general, such a module category $\Rep^\psi(H)$ is labeled by a subgroup $H$ of $\mathbb Z_{N^2}$ together with a torsion $\psi \in H^2(H,U(1))$ \cite{Ostrik:2002ohv,naidu2007morita}. Here, all the subgroups are cyclic, which have trivial second cohomology so the module category is completely determined by a choice of subgroup $H$. Taking $h \in \mathbb Z_{N^2}$ as the generator of this subgroup, the corresponding Lagrangian algebra is
\begin{equation}
    L_{\Rep(H)} = \langle e^{N^2/h}, m^h\rangle.
\end{equation}
From this, we see that taking $H$ to be generated by $h = 1$ if $\nw = 0$ and $h = N^2$ if $\nm = 0$ recovers the Lagrangian algebra $L_{\mc R''}$. Putting everything together, we find that the dual model labeled by $\mc R'' = \Vect$ obtained by choosing $\nm = 0$ is given by
\begin{equation}
    H'' = - \sum_i (J_1 X_i X_{i+1}^\dag + J_2 Z_i) + \text{h.c.}
\end{equation}
The local electric field term in this model is the operator $\rho_i^{\mc R''}(e^{\nw}) + \text{h.c.}$, which works out as
\begin{equation}
    H_E'' = \sum_i (X_i^{\nw} + \text{h.c.}).
\end{equation}
Choosing instead $\nw = 0$, we find the dual model labeled by $\mc R'' = \Rep(\mathbb Z_{N^2})$ is
\begin{equation}
    H'' = - \sum_i (J_1 X_i + J_2 Z_i Z_{i+1}^\dag) + \text{h.c.}
\end{equation}
with electric field term the operator $\rho_i^{\mc R''}(m^{\nm}) + \text{h.c.}$ working out to
\begin{equation}
    H_E'' = \sum_i (Z_i^{\nm} + \text{h.c.}).
\end{equation}

\subsection{(34-50)-model}
We can repeat the above derivation for two copies of the $\mathbb Z_N$ modified Villain model. The clock model is now built from the input category $\mc D = \Rep(\mathbb Z_{N^2} \times \mathbb Z_{N^2})$ with the choice of $\mc R = \Vect$ as module category and symmetry given by $\Vect_{\mathbb Z_{N^2} \times \mathbb Z_{N^2}}$. The Hamiltonian is given by
\begin{multline}
    H = - \sum_i (J_{1,1} X_{i,1} X_{i+1,1}^\dag + J_{2,1} Z_{i,1}\\
    + J_{1,2} X_{i,2} X_{i+1,2}^\dag + J_{2,2} Z_{i,2}) + \text{h.c.}.
\end{multline}
The operators in this theory are described by the anyons of the $\mathbb Z_{N^2} \times \mathbb Z_{N^2}$ toric code, which we write as $\{e_1^a m_1^b e_2^c m_2^d, \enspace a,b,c,d \in \mathbb Z_{N^2}\}$. They have $\mathbb Z_{N^2}^{\times 4}$ fusion rules, trivial associators, with $R$-matrices and topological spins given by
\begin{align}
    R(e_1^a m_1^b e_2^c m_2^d, e_1^e m_1^f e_2^g m_2^h) &= e^{\frac{2\pi i (af + ch)}{N^2}},\nonumber\\ 
    \theta(e_1^a m_1^b e_2^c m_2^d) &= e^{\frac{2\pi i(ab+cd)}{N^2}}.
\end{align}
For $\mc R = \Vect$, the corresponding Lagrangian algebra is
\begin{equation}
    L_{\mc R} = \langle e_1, e_2 \rangle \simeq \mathbb C[\mathbb Z_{N^2} \times \mathbb Z_{N^2}].
\end{equation}

\subsubsection{Villain model}

The Villain model is obtained by gauging the $\mathbb Z_N \times \mathbb Z_N$ subgroup of the clock model, corresponding to changing to the module category $\mc R' = \Rep(\mathbb Z_N \times \mathbb Z_N)$ and Lagrangian algebra
\begin{equation}
    L_{\mc R'} = \langle e_1^N, m_1^N, e_2^N, m_2^N \rangle \simeq \mathbb C[\mathbb Z_N^{\times 4}].
\end{equation}
We denote the resulting symmetry as $\mathbb Z_N^{e_1} \times \mathbb Z_N^{m_1} \times \mathbb Z_N^{e_2} \times \mathbb Z_N^{m_2}$ with generators $e_1, m_1, e_2$ and $m_2$ respectively. The corresponding fusion category is $\Vect_{\mathbb Z_N^{\times 4}}^\omega$, with $\omega \in H^3(\mathbb Z_N^{\times 4},U(1))$ capturing the mixed anomalies between $\mathbb Z_N^{e_i}$ and $\mathbb Z_N^{m_i}$. 

\subsubsection{Gauged Villain model}

We now consider a $\mathbb Z_N$ subgroup of the $\mathbb Z_N^{\times 4}$ symmetry of the Villain model generated by $(\nw[1],\nm[1], \nw[2], \nm[2]) \in \mathbb Z_N^{\times 4}$, which again requires $\text{gcd}(\nw[1],\nm[1],\nw[2],\nm[2],N) = 1$. Gauging this subgroup requires the new Lagrangian algebra $L_{\mc R''}$ to contain anyons of the form $\alpha = e_1^{Na + \nw[1]} m_1^{Nb + \nm[1]} e_2^{Nc + \nw[2]} m_2^{Nd + \nm[2]}$ for some $a,b,c,d \in \mathbb Z_N$. The topological spin of these anyons is
\begin{equation}
    \theta(\alpha) = e^{\frac{2\pi i (\nw[1] \nm[1] + \nw[2] \nm[2])}{N^2}} e^{\frac{2\pi i(a \nm[1] + b \nw[1] + c \nm[2] + d \nw[2])}{N}},
\end{equation}
and requiring at least some of these to be condensable bosons requires $\nw[1] \nm[1] + \nw[2] \nm[2] = 0 \mod N$. This is the condition that the $\mathbb Z_N$ subgroup we are gauging is anomaly free, and again we will assume $\nw[1] \nm[1] + \nw[2] \nm[2] = 0$ since we are interested in the $N \rightarrow \infty$ behavior. The spin of the anyons $\alpha$ reduces to
\begin{equation}
    \label{eq:spin1}
    \theta(\alpha) = 
    \exp\left(\frac{2\pi i (a \nm[1] + b \nw[1] + c \nm[2] + d \nw[2])}{N}\right).
\end{equation}
meaning we are looking for $a,b,c,d$ such that $a \nm[1] + b \nw[1] + c \nm[2] + d \nw[2] = 0 \mod N$. Since $\text{gcd}(\nm[1],\nw[1],\nm[2],\nw[2],N) = 1$, this equation has $N^3$ solutions; the map $(a,b,c,d) \rightarrow a \nm[1] + b \nw[1] + c \nm[2] + d \nw[2] \mod N$ has image $\mathbb Z_N$, meaning its kernel has $N^3$ elements. As before, we can generate $N^2$ bosons by considering the anyons generated by fusion of $e_1^{\nw[1]} m_1^{\nm[1]} e_2^{\nw[2]} m_2^{\nm[2]}$ with itself, generating an algebra
\begin{equation}
    A = \langle e_1^{\nw[1]} m_1^{\nm[1]} e_2^{\nw[2]} m_2^{\nm[2]} \rangle \simeq \mathbb C[\mathbb Z_{N^2}]
\end{equation}
of dimension $N^2$.  This is not yet a Lagrangian algebra, which is expected since it does not yet contain all the local operators of the modified Villain model that are neutral under the $\mathbb Z_N$ symmetry we are gauging and therefore remain local. Since this symmetry $\mathbb Z_N$ is generated by $e_1^{\nw[1]} m_1^{\nm[1]} e_2^{\nw[2]} m_2^{\nm[2]}$ and the local operators in the modified Villain model are precisely $L_{\mc R'}$, we are looking for those operators in $L_{\mc R'}$ that braid trivially with this generator. The full set of anyons that braids trivially with this generator is denoted as $A^{\perp}$, so we write
\begin{equation}
    L_{\mc R''} = A + (L_{\mc R'} \cap A^{\perp}).
\end{equation}
We note that this is not a direct sum, since $A$ contains some anyons of $L_{\mc R'}$ already. In terms of the $R$-matrices, a full braid of two anyons $\alpha$ and $\beta$ is given by $B(\alpha,\beta) = R(\alpha,\beta)R(\beta,\alpha)$, so the braiding of a general anyon in $L_{\mc R'}$ with the generator of $A$ is given by
\begin{multline}
    B(e_1^{\nw[1]} m_1^{\nm[1]} e_2^{\nw[2]} m_2^{\nm[2]}, e_1^{aN} m_1^{bN} e_2^{cN} m_2^{dN}) = \\
    \exp\left(\frac{2\pi i (a \nm[1] + b \nw[1] + c \nm[2] + d \nw[2])}{N}\right).
\end{multline}
Requiring this to be trivial is the same condition as eq.~\eqref{eq:spin1}, and it again has $N^3$ solutions. Of these, $N$ are given by $(a,b,c,d) = k(\nw[1], \nm[1], \nw[2], \nm[2]), k \in \mathbb Z_N$ due to the anomaly cancellation condition, and these solutions are already contained in $A$ leaving us with $N^2$ additional bosons. This shows that $L_{\mc R''}$ is $N^4$ dimensional, so it is indeed Lagrangian. We find
\begin{multline}
    L_{\mc R''} = \langle e_1^{\nw[1]} m_1^{\nm[1]} e_2^{\nw[2]} m_2^{\nm[2]} \rangle + \{e_1^{aN} m_1^{bN} e_2^{cN} m_2^{dN},\\ a \nm[1] + b \nw[1] + c \nm[2] + d \nw[2] = 0 \mod N\} \simeq \mathbb C[\mathbb Z_{N^2} \times \mathbb Z_N^{\times 2}].
\end{multline}
The symmetry operators for this choice of Lagrangian algebra are labeled by the cosets
\begin{equation}
\label{eq:gauged_villain_coset}
    \frac{\mathbb Z_{N^2}^{\times 4}}{\mathbb Z_{N^2} \times \mathbb Z_N^{\times 2}} = \mathbb Z_{N^2} \times \mathbb Z_N^{\times 2},
\end{equation}
corresponding to a fusion category $\Vect_{\mathbb Z_{N^2}} \boxtimes \Vect_{\mathbb Z_N^{\times 2}}^\omega$; the appearance of a group extension in the dual symmetry is the familiar consequence of gauging a subgroup that has a mixed anomaly with the rest of the symmetry \cite{Tachikawa:2017gyf}.

To write down the corresponding lattice model, we need the module category $\mc R'' = \Rep^{\psi}(H)$ with $H \subset \mathbb Z_{N^2}^{\times 2}$ and $\psi \in H^2(H,U(1))$ that gives rise to the Lagrangian algebra $L_{\mc R''}$. Since this second cohomology class is no longer necessarily trivial, we need the general form of a Lagrangian algebra in $\mc Z(\Rep(\mathbb Z_{N^2}^{\times 2}))$ for a choice of $H,\psi$ given by \cite{naidu2008lagrangian,davydov2010modular}
\begin{multline}
    L_{\Rep^\psi(H)} = \{e_1^a m_1^b e_2^c m_2^d\enspace : \enspace (b,d) \in H,\\ \enspace e^{\frac{2\pi i (ab' + cd')}{N^2}} = \beta((b,d),(b',d')) \enspace \forall \enspace (b',d') \in H\},
\end{multline}
with $\beta(h,h') = \psi(h,h')/\psi(h',h)$. The subgroup $H \subset \mathbb Z_{N^2} \times \mathbb Z_{N^2}$ can therefore be read off from the flux content of $L_{\mc R''}$. We write $(\nm[1],\nm[2]) = g_m (\hat n_{1}, \hat n_{2})$ with $g_m = \text{gcd}(\nm[1],\nm[2])$, so that $(\hat n_{1},\hat n_{2})$ is primitive. Anomaly cancellation then implies $(\nw[1],\nw[2]) = \lambda (\hat n_{2}, -\hat n_{1})$ for some integer $\lambda$, and $\text{gcd}(\nw[1],\nm[1],\nw[2],\nm[2],N)=1$ implies $\text{gcd}(\lambda, g_N) = 1$, where $g_N = \text{gcd}(g_m,N)$. The fluxes in $L_{\mc R''}$ are $k(\nm[1],\nm[2])$ from $A$, together with $N(b,d)$ for which the condition $a \nm[1] + b \nw[1] + c \nm[2] + d \nw[2] = 0 \mod N$ can be solved for $(a,c)$, i.e.\ $b \hat n_{2} - d \hat n_{1} = 0 \mod g_N$. Hence the subgroup $H$ is generated by
\begin{align}
    H &= \langle h_1, h_2 \rangle \simeq \mathbb Z_{N^2/g_N} \times \mathbb Z_{N/g_N},\nonumber\\
    h_1 &= g_N(\hat n_{1}, \hat n_{2}), \quad h_2 = g_N(a_2 N, - a_1 N),
\end{align}
with $a_1 \hat n_{1} + a_2 \hat n_{2} = 1$. If $g_N = 1$, $h_1$ generates the same subgroup as $(\nm[1],\nm[2])$; for the 34-50 model, $g_N = \text{gcd}(4,N)$, so this is the case for odd $N$ only. To determine $\beta$, it suffices to consider one pair of charges $(a,c)$ for every $(b,d) \in H$ such that $e_1^a m_1^b e_2^c m_2^d \in L_{\mc R''}$, where consistency is guaranteed by $L_{\mc R''}$ being Lagrangian. Additionally, $\beta$ as extracted from the Lagrangian algebra definition is bilinear, so it suffices to specify it on generators. We can pair the generator $h_1$ with the charges $s(\nw[1],\nw[2])$, where $s g_m = g_N \mod N$ (the flux of $\alpha^s$ differs from $h_1$ by a multiple of $N(\hat n_{1},\hat n_{2})$, which can be paired with vanishing charge), and we find
\begin{align}
    \beta(h_1,h_1) &= e^{\frac{2\pi i s g_N (\nw[1] \hat n_{1} + \nw[2] \hat n_{2})}{N^2}} = 1, \nonumber\\
    \beta(h_1,h_2) &= e^{\frac{2\pi i s g_N (a_2 \nw[1] - a_1 \nw[2])}{N}}.
\end{align}
The generator $h_2 = g_N(a_2 N, -a_1 N)$ can be paired with a charge $(x_1 N,x_3 N)$ such that
\begin{equation}
    x_1 \nm[1] + x_3 \nm[2] + g_N (a_2 \nw[1] - a_1 \nw[2]) = 0 \mod N,
\end{equation}
which has a solution since $a_2 \nw[1] - a_1 \nw[2] = \lambda$ and $x_1 \nm[1] + x_3 \nm[2]$ ranges over $g_N \mathbb Z_N$, and we find
\begin{align}
    \beta(h_2,h_1) &= e^{\frac{2\pi i g_N (x_1 \hat n_{1} + x_3 \hat n_{2})}{N}},\nonumber\\
    \beta(h_2,h_2) &= e^{2 \pi i g_N (x_1 a_2 - x_3 a_1)} = 1. 
\end{align}
The corresponding lattice model has $\Rep^\psi(H)$-valued gauge degrees of freedom on the links. This somewhat complicates the structure of the local Hilbert space, since the number and dimension of the $\psi$-projective representations of $H$ is now sensitive to $N$.

\subsubsection{A different frame}

These subtleties can be remedied by choosing a different Lagrangian algebra that still contains $A$, which corresponds to gauging additional symmetry in the Villain model. Making the choice $H = \mathbb Z_{N^2}$ gives a translation invariant local Hilbert space and additionally forces $\psi = 1$ since it has trivial second cohomology. With $g_m$ and $(\hat n_{1},\hat n_{2})$ as above, the group $H$ generated by the primitive vector $(\hat n_{1},\hat n_{2}) \in \mathbb Z_{N^2}^{\times 2}$ is $\mathbb Z_{N^2}$ for any $N$. With $\psi = 1$, the condition that $ab' + cd' = 0 \mod N^2, \enspace \forall \enspace (b',d') \in H$ forces $(a,c) = \langle (\hat n_{2},-\hat n_{1}) \rangle$. The resulting Lagrangian algebra is then
\begin{equation}
    L_{\Rep(H)} = \langle e_1^{\hat n_{2}} e_2^{-\hat n_{1}}, m_1^{\hat n_{1}} m_2^{\hat n_{2}}\rangle \simeq \mathbb C[\mathbb Z_{N^2}^{\times 2}].
\end{equation}
Due to the anomaly cancellation we have $\nw[1]\nm[1] + \nw[2]\nm[2] = 0$ and since $(\nw[1],\nw[2]) = \lambda (\hat n_{2},-\hat n_{1})$, the electric anyon $e_1^{\hat n_{2}}e_2^{-\hat n_{1}}$ does generate the electric anyon $e_1^{\nw[1]} e_2^{\nw[2]}$, while $m_1^{\nm[1]} m_2^{\nm[2]} = (m_1^{\hat n_{1}} m_2^{\hat n_{2}})^{g_m}$, and so $A \subset L_{\mc R''}$ for $\mc R'' = \Rep(H)$. 

The symmetry operators corresponding to this choice of Lagrangian algebra are labeled by the cosets
\begin{equation}
    \frac{\mathbb Z_{N^2}^{\times 4}}{\mathbb Z_{N^2}^{\times 2}} = \mathbb Z_{N^2}^{\times 2},
\end{equation}
corresponding to the fusion category $\Vect_{\mathbb Z_{N^2}^{\times 2}}$. From this, we see that this model is related to the previous one by gauging a $\mathbb Z_N$ subgroup of its $\mathbb Z_{N^2}$ symmetry, which provides the mapping between the correlation functions of both theories.

Putting everything together, we obtain the resulting model by gauging the subgroup generated by $(\hat n_{1}, \hat n_{2})$ of the $\mathbb Z_{N^2}^{\times 2}$ symmetry of the clock model and adding the local electric field term $\rho_i^{\mc R''}(e_1^{\nw[1]}e_2^{\nw[2]}m_1^{\nm[1]}m_2^{\nm[2]}) + \text{h.c.}$. The gauged clock model Hamiltonian is
\begin{multline}
    H'' = - \sum_i (J_{1,1} X_{i,1} \tilde X_{i+\frac{1}{2}}^{\hat n_{1}} X_{i+1,1}^\dag + J_{2,1} Z_{i,1}\\
    + J_{1,2} X_{i,2} \tilde X_{i+\frac{1}{2}}^{\hat n_{2}} X_{i+1,2}^\dag + J_{2,2} Z_{i,2}) + \text{h.c.},
\end{multline}
and the electric term is
\begin{equation}
    H''_E = \sum_i \prod_{j<i} Z_{j,1}^{\nm[1]} Z_{j,2}^{\nm[2]} X_{i,1}^{\nw[1]} X_{i,2}^{\nw[2]} + \text{h.c.},
\end{equation}
both subject to the Gauss constraint
\begin{equation}
    \tilde Z_{i-\frac{1}{2}} Z_{i,1}^{\hat n_{1}} Z_{i,2}^{\hat n_{2}} \tilde Z_{i+\frac{1}{2}}^\dag = 1, \quad \text{with} \quad \tilde Z \tilde X = e^{\frac{2\pi i}{N^2}} \tilde X \tilde Z.
\end{equation}
The Gauss constraint fixes the $H$-charge of the matter degrees of freedom as the difference of the neighboring gauge charges. Defining $r_1 \hat n_{1} + r_2 \hat n_{2} = 1$, which has integer solutions for any $N$, we can write
\begin{alignat}{2}
    U_{i,1} &= X_{i,1}^{r_1} X_{i,2}^{r_2}, &\quad V_{i,1} &= Z_{i,1}^{\hat n_{1}} Z_{i,2}^{\hat n_{2}},\nonumber\\ 
    U_{i,2} &= X_{i,1}^{\hat n_{2}} X_{i,2}^{-\hat n_{1}}, &\quad V_{i,2} &= Z_{i,1}^{r_2} Z_{i,2}^{-r_1},
\end{alignat}
such that $V_{i,b} U_{i,a} = e^{\delta_{a,b} \frac{2\pi i}{N^2}} U_{i,a} V_{i,b}$. Inverting these relations, we find
\begin{alignat}{2}
    X_{i,1} &= U_{i,1}^{\hat n_{1}} U_{i,2}^{r_2}, &\quad Z_{i,1} &= V_{i,1}^{r_1} V_{i,2}^{\hat n_{2}},\nonumber\\
    X_{i,2} &= U_{i,1}^{\hat n_{2}} U_{i,2}^{-r_1}, &\quad Z_{i,2} &= V_{i,1}^{r_2} V_{i,2}^{-\hat n_{1}}.
\end{alignat}
Imposing the Gauss constraints now amounts to setting
\begin{equation}
    V_{i,1} = \tilde Z_{i-\frac{1}{2}}^\dag \tilde Z_{i+\frac{1}{2}}, \quad U_{i,1} = 1,
\end{equation}
and the gauged clock model becomes
\begin{multline}
    H'' = - \sum_i (J_{1,1} U_{i}^{r_2} \tilde X_{i+\frac{1}{2}}^{\hat n_{1}} U_{i+1}^{-r_2} + J_{2,1} \tilde Z_{i-\frac{1}{2}}^{-r_1} V_{i}^{\hat n_{2}} \tilde Z_{i+\frac{1}{2}}^{r_1}\\
    + J_{1,2} U_{i}^{-r_1} \tilde X_{i+\frac{1}{2}}^{\hat n_{2}} U_{i+1}^{r_1} + J_{2,2} \tilde Z_{i-\frac{1}{2}}^{-r_2} V_{i}^{-\hat n_{1}} \tilde Z_{i+\frac{1}{2}}^{r_2}) + \text{h.c.},
\end{multline}
where we define $U_i := U_{i,2}$ and $V_i := V_{i,2}$. As expected, the electric term becomes local; the string $\prod_{j<i} Z_{j,1}^{\nm[1]} Z_{j,2}^{\nm[2]} = \prod_{j<i} V_{j,1}^{g_m}$ telescopes to $\tilde Z_{i-\frac12}^{g_m}$:
\begin{equation}
    H_E'' = \sum_i \tilde Z_{i-\frac{1}{2}}^{g_m} U_{i}^{\nw[1] r_2 - \nw[2] r_1} + \text{h.c..}
\end{equation}

The $34-50$ model corresponds to choosing the winding and momentum charges $(\nw[1],\nw[2],\nm[1],\nm[2]) = (-1,2,8,4)$ of \cref{eq:qm-qw}, since $e_a$ and $m_a$ generate the winding and momentum symmetries of flavor $a$. This gauges the same symmetry $\Qchi_{j+\frac12}$ as \cref{sec:chiral}. We then have $g_m = 4$, $(\hat n_{1},\hat n_{2}) = (2,1)$, and we can take $r_1 = 0$, $r_2 = 1$. The Hamiltonian becomes
\begin{align}
    H'' &+ g H_E = - \sum_i (J_{1,1} U_{i} \tilde X_{i+\frac{1}{2}}^{2} U_{i+1}^\dag + J_{2,1} V_{i}\nonumber\\
    &+ J_{1,2} \tilde X_{i+\frac{1}{2}} + J_{2,2} \tilde Z_{i-\frac{1}{2}}^\dag V_{i}^{-2} \tilde Z_{i+\frac{1}{2}}\nonumber\\
    &+ g \tilde Z_{i-\frac{1}{2}}^{4} U_{i}^\dag) + \text{h.c.}.
\end{align}

\subsubsection{Yet another frame}

The above Hamiltonian does not match the one derived in the main text, since this one corresponds to yet another frame in which the electric term is local. We will now show that the Hamiltonian from the main text can be obtained by gauging the remaining $\mathbb Z_{N^2}$ symmetry. The electric field term commutes with
\begin{equation}
    G = \prod_i g_i, \quad g_i = \tilde X_{i-\frac{1}{2}} V_i^{-4},
\end{equation}
and it can readily be checked that this $\mathbb Z_{N^2}$ symmetry commutes with the rest of the Hamiltonian terms as well. After blocking, it is an on-site symmetry, so it can readily be gauged using standard methods \cite{Haegeman:2014maa}. We have $g_i U_i g_i^\dag = \omega^{-4} U_i$ and $g_i \tilde Z_{i-\frac{1}{2}} g_i^\dag = \omega^{-1} \tilde Z_{i-\frac{1}{2}}$, with $V$ and $\tilde X$ neutral. Denoting the corresponding gauge fields as $(\xm{i},\zm{i})$ with $\zm{i}\xm{i} = e^\frac{2\pi i}{N^2} \xm{i}\zm{i}$, the gauged Hamiltonian becomes
\begin{align}
    \tilde H'' &+ g \tilde H_E = - \sum_i (J_{1,1} U_{i} \zm{i}^{4} \tilde X_{i+\frac{1}{2}}^{2} U_{i+1}^\dag + J_{2,1} V_{i}\nonumber\\
    &+ J_{1,2} \tilde X_{i+\frac{1}{2}} + J_{2,2} \tilde Z_{i-\frac{1}{2}}^\dag V_{i}^{-2} \zm{i}^\dag \tilde Z_{i+\frac{1}{2}}\nonumber\\
    &+ g \tilde Z_{i-\frac{1}{2}}^{4} U_{i}^\dag) + \text{h.c.},
\end{align}
subject to the Gauss constraint
\begin{equation}
    \xm{i-1} \tilde X_{i - \frac{1}{2}} V_i^{-4} \xm{i}^\dag = 1.
\end{equation}
Note that the electric field term remains unchanged, since it is already gauge invariant. To solve the Gauss constraint, we define
\begin{alignat}{2}
    a_{i} &= \tilde X_{i-\frac{1}{2}} V_i^{-4}, &\quad b_{i} &= \tilde Z_{i-\frac{1}{2}}^\dag,\nonumber\\
    \Zm[1]{i} &= V_i^\dag, &\quad \Xm[1]{i} &= \tilde Z_{i - \frac{1}{2}}^{4} U_i^\dag,
\end{alignat}
such that $a_{i} b_{i} = e^{\frac{2\pi i}{N^2}} b_{i} a_{i}$ and $\Zm[1]{i} \Xm[1]{i} = e^{\frac{2\pi i}{N^2}} \Xm[1]{i} \Zm[1]{i}$, with the two pairs commuting. Inverting these relations, we find
\begin{alignat}{2}
    V_i &= \Zm[1]{i}^\dag, &\quad U_i &= \Xm[1]{i}^\dag b_{i}^{-4},\nonumber\\
    \tilde X_{i-\frac{1}{2}} &= a_{i} \Zm[1]{i}^{-4}, &\quad \tilde Z_{i - \frac{1}{2}} &= b_{i}^\dag.
\end{alignat}
Imposing the Gauss constraints amounts to setting
\begin{equation}
    a_{i} = \xm{i-1}^\dag \xm{i}, \quad b_{i} = 1,
\end{equation}
where in the $J_{1,1}$ term the gauge fixing $b_{i} = 1$ is done after moving $b_{i+1}^{4}$ (from $U_{i+1}^\dag$) to the left of $a_{i+1}^{2}$ (from $\tilde X_{i+\frac12}^2$), producing a phase $\omega^{8}$. Writing the $J_{1,2}$ term as its Hermitian conjugate, the gauged model becomes
\begin{align}
    \tilde H'' &+ g \tilde H_E = - \sum_i (J_{1,1}\, \omega^{8}\, \Xm[1]{i}^\dag \zm{i}^{4} (\xm{i}^\dag \xm{i+1})^2 \Zm[1]{i+1}^{-8} \Xm[1]{i+1} \nonumber\\
    &+ J_{2,1} \Zm[1]{i}^\dag + J_{1,2} \xm{i} \xm{i+1}^\dag \Zm[1]{i+1}^{4} \nonumber\\
    &+ J_{2,2} \Zm[1]{i}^{2} \zm{i}^\dag + g\, \Xm[1]{i}) + \text{h.c.}
\end{align}
This is exactly the Hamiltonian of \cref{sec:chiral} in its final, unconstrained frame, with $(\xm{i},\zm{i})$ the site fields of \cref{sec:chiral}.

\subsection{Three-flavor models}

As shown in the main text, the (34-50) model is no longer chiral in the IR after gauging. To get something that is chiral after gauging, one can repeat the derivation for three-flavor models, which is particularly straightforward in the categorical formulation. Below, we present two models: the (333-511) model, which is expected to be non-chiral after gauging, and the (041-223) model, which is. While these models are numerically more challenging than their two-flavor counterparts, they still lie within the capabilities of current tensor network methods, and we hope to explore these further in future works. 

\subsubsection{(333-511)-model}

The winding and momentum charges of the (333-511) model are $(\nw[1],\nw[2],\nw[3],\nm[1],\nm[2],\nm[3]) = (-1,1,1,8,4,4)$. As in the different frame of the 34-50 model, we gauge the subgroup generated by the primitive vector $(\hat n_{1},\hat n_{2},\hat n_{3}) = (2,1,1)$, with $\nm[a] = g_m \hat n_{a}$ and $g_m = 4$. The gauged clock model Hamiltonian is
\begin{multline}
    H'' = - \sum_i \sum_{a = 1}^3 (J_{1,a} X_{i,a} \tilde X_{i+\frac{1}{2}}^{\hat n_{a}} X_{i+1,a}^\dag + J_{2,a} Z_{i,a}) + \text{h.c.},
\end{multline}
and the electric term is
\begin{equation}
    H''_E = \sum_i \prod_{j<i} \prod_{a=1}^3 Z_{j,a}^{\nm[a]} X_{i,a}^{\nw[a]} + \text{h.c.},
\end{equation}
both subject to the Gauss constraint
\begin{equation}
    \tilde Z_{i-\frac{1}{2}} \prod_{a=1}^3 Z_{i,a}^{\hat n_{a}} \tilde Z_{i+\frac{1}{2}}^\dag = 1, \quad \text{with} \quad \tilde Z \tilde X = e^{\frac{2\pi i}{N^2}} \tilde X \tilde Z.
\end{equation}
To solve the Gauss constraint for the 333-511 model, we define
\begin{alignat}{2}
    U_{i,1} &= X_{i,3}, &\quad V_{i,1} &= Z_{i,1}^2 Z_{i,2} Z_{i,3},\nonumber\\ 
    U_{i,2} &= X_{i,1} X_{i,3}^{-2}, &\quad V_{i,2} &= Z_{i,1},\nonumber\\
    U_{i,3} &= X_{i,2} X_{i,3}^\dag, &\quad V_{i,3} &= Z_{i,2},\nonumber
\end{alignat}
again such that $V_{i,b} U_{i,a} = e^{\delta_{a,b} \frac{2\pi i}{N^2}} U_{i,a} V_{i,b}$. Inverting these relations, we find
\begin{alignat}{2}
    X_{i,1} &= U_{i,1}^2 U_{i,2}, &\quad Z_{i,1} &= V_{i,2},\nonumber\\
    X_{i,2} &= U_{i,1} U_{i,3}, &\quad Z_{i,2} &= V_{i,3},\nonumber\\
    X_{i,3} &= U_{i,1}, &\quad Z_{i,3} &= V_{i,1} V_{i,2}^{-2} V_{i,3}^\dag.
\end{alignat}
Imposing the Gauss constraints now amounts to setting
\begin{equation}
    V_{i,1} = \tilde Z_{i-\frac{1}{2}}^\dag \tilde Z_{i+\frac{1}{2}}, \quad U_{i,1} = 1,
\end{equation}
and the gauged clock model becomes
\begin{multline}
    H'' = - \sum_i (J_{1,1} U_{i,2} \tilde X_{i+\frac{1}{2}}^2 U_{i+1,2}^\dag
    + J_{2,1} V_{i,2}\\
    + J_{1,2} U_{i,3} \tilde X_{i+\frac{1}{2}} U_{i+1,3}^\dag
    + J_{2,2} V_{i,3}
    + J_{1,3} \tilde X_{i+\frac{1}{2}}\\
    + J_{2,3} \tilde Z_{i-\frac{1}{2}}^\dag V_{i,2}^{-2} V_{i,3}^\dag \tilde Z_{i+\frac{1}{2}}) + \text{h.c.}.
\end{multline}
As expected, the electric term becomes local:
\begin{equation}
    H_E'' = \sum_i \tilde Z_{i-\frac{1}{2}}^4 U_{i,2}^\dag U_{i,3} + \text{h.c..}
\end{equation}

\subsubsection{(041-223)-model}
The winding and momentum charges of the (041-223) model are $(\nw[1],\nw[2],\nw[3],\nm[1],\nm[2],\nm[3]) = (-1,1,-1,2,6,4)$, and we gauge the subgroup generated by $(\hat n_{1},\hat n_{2},\hat n_{3}) = (1,3,2)$, with $g_m = 2$. The gauged clock model Hamiltonian is
\begin{multline}
    H'' = - \sum_i \sum_{a = 1}^3 (J_{1,a} X_{i,a} \tilde X_{i+\frac{1}{2}}^{\hat n_{a}} X_{i+1,a}^\dag + J_{2,a} Z_{i,a}) + \text{h.c.},
\end{multline}
and the electric term is
\begin{equation}
    H''_E = \sum_i \prod_{j<i} \prod_{a=1}^3 Z_{j,a}^{\nm[a]} X_{i,a}^{\nw[a]} + \text{h.c.},
\end{equation}
both subject to the Gauss constraint
\begin{equation}
    \tilde Z_{i-\frac{1}{2}} \prod_{a=1}^3 Z_{i,a}^{\hat n_{a}} \tilde Z_{i+\frac{1}{2}}^\dag = 1, \quad \text{with} \quad \tilde Z \tilde X = e^{\frac{2\pi i}{N^2}} \tilde X \tilde Z.
\end{equation}
To solve the Gauss constraint for the 041-223 model, we define
\begin{alignat}{2}
    U_{i,1} &= X_{i,1}, &\quad V_{i,1} &= Z_{i,1} Z_{i,2}^3 Z_{i,3}^2,\nonumber\\ 
    U_{i,2} &= X_{i,1}^{-3} X_{i,2}, &\quad V_{i,2} &= Z_{i,2},\nonumber\\
    U_{i,3} &= X_{i,1}^{-2} X_{i,3}, &\quad V_{i,3} &= Z_{i,3},
\end{alignat}
again such that $V_{i,b} U_{i,a} = e^{\delta_{a,b} \frac{2\pi i}{N^2}} U_{i,a} V_{i,b}$. Inverting these relations, we find
\begin{alignat}{2}
    X_{i,1} &= U_{i,1}, &\quad Z_{i,1} &= V_{i,1} V_{i,2}^{-3} V_{i,3}^{-2},\nonumber\\
    X_{i,2} &= U_{i,1}^3 U_{i,2}, &\quad Z_{i,2} &= V_{i,2},\nonumber\\
    X_{i,3} &= U_{i,1}^2 U_{i,3}, &\quad Z_{i,3} &= V_{i,3}.
\end{alignat}
Imposing the Gauss constraints now amounts to setting
\begin{equation}
    V_{i,1} = \tilde Z_{i-\frac{1}{2}}^\dag \tilde Z_{i+\frac{1}{2}}, \quad U_{i,1} = 1,
\end{equation}
and the gauged clock model becomes
\begin{multline}
    H'' = - \sum_i (J_{1,1} \tilde X_{i+\frac{1}{2}}
    + J_{2,1} \tilde Z_{i-\frac{1}{2}}^\dag V_{i,2}^{-3} V_{i,3}^{-2} \tilde Z_{i+\frac{1}{2}}\\
    + J_{1,2} U_{i,2} \tilde X_{i+\frac{1}{2}}^3 U_{i+1,2}^\dag
    + J_{2,2} V_{i,2}\\
    + J_{1,3} U_{i,3} \tilde X_{i+\frac{1}{2}}^2 U_{i+1,3}^\dag
    + J_{2,3} V_{i,3}) + \text{h.c.}.
\end{multline}
As expected, the electric term becomes local:
\begin{equation}
    H_E'' = \sum_i \tilde Z_{i-\frac{1}{2}}^2 U_{i,2} U_{i,3}^\dag + \text{h.c..}
\end{equation}

\end{document}